\documentclass[conference]{IEEEtran}
\IEEEoverridecommandlockouts

\usepackage{tikz}
\usepackage{textcomp}
\usepackage{hyperref}
\usepackage{lipsum}
\usepackage{textcomp}
\usepackage{cite}
\usepackage{amsmath,amssymb,amsfonts}
\usepackage{algorithmic}
\usepackage{graphicx}
\usepackage{multicol,multirow}
\usepackage{array}
\usepackage{gensymb}
\usepackage{mathtools, cuted}
\usepackage{breqn}
\usepackage{lipsum}
\usepackage{xcolor}
\usepackage{siunitx}
\usepackage{subfloat}
\usepackage{subfig}

\usepackage[font=small]{caption}  

\newcolumntype{L}{>{\raggedright\arraybackslash}p}
\newcolumntype{C}{>{\centering\arraybackslash}m}

\newcommand\copyrighttext{%
  \footnotesize \textcopyright 2020 IEEE. Personal use of this material is permitted.
  Permission from IEEE must be obtained for all other uses, in any current or future
  media, including reprinting/republishing this material for advertising or promotional
  purposes, creating new collective works, for resale or redistribution to servers or
  lists, or reuse of any copyrighted component of this work in other works.
  DOI: 10.1109/TAP.2020.3008658}
\newcommand\copyrightnotice{%
\begin{tikzpicture}[remember picture,overlay]
\node[anchor=south,yshift=10pt] at (current page.south) {\fbox{\parbox{\dimexpr\textwidth-\fboxsep-\fboxrule\relax}{\copyrighttext}}};
\end{tikzpicture}%
}

\begin{document}

\title{A Simple Equivalent Circuit Approach for Anisotropic Frequency Selective Surfaces and Metasurfaces}

\author{Michele~Borgese,~\IEEEmembership{Member,~IEEE}
        and~Filippo~Costa,~\IEEEmembership{Senior Member,~IEEE}

\thanks{Filippo Costa is with University of Pisa, Dipartimento di Ingegneria dell'Informazione, Pisa, Italy.  (corresponding author, e-mail: filippo.costa@unipi.it).}
\thanks{Digital Object Identifier 10.1109/\discretionary{}{}{}TAP.2019.xxx}

}

\maketitle
\copyrightnotice

\begin{abstract}
\boldmath
An equivalent circuit model for Frequency Selective Surfaces (FSS) comprising anisotropic elements is presented. The periodic surface is initially simulated with an arbitrary azimuthal incidence angle and its surface impedance matrix is derived. The impedance matrix is subsequently rotated by an angle $\varphi^{rot}$ on the crystal axes $\chi_1$, $\chi_2$ thus nullifying its extra diagonal terms. The rotation angle $\varphi^{rot}$ is derived according to the spectral theorem by using the terms of the matrix initially extracted. The diagonal terms of the rotated matrix, that is, the impedances $Z_{\chi_1}$ and $Z_{\chi_2}$, are finally matched with simple LC networks. The circuit model representation of the anisotropic element can be used to analyse anisotropic FSSs rotated by a generic azimuth angle. The methodology provides a compact description of generic FSS elements with only five  parameters: the lumped parameters of the LC network $L_{\chi_1}$, $C_{\chi_1}$, $L_{\chi_2}$, $C_{\chi_2}$ and the rotation angle $\varphi^{rot}$. The circuit model can take into account the presence of dielectric substrates close to the FSS or a variation of the FSS periodicity without additional computational efforts. The equivalent circuit model is finally applied to the design of two transmitting polarization converts based on anisotropic metasurfaces.     
\end{abstract}

\begin{IEEEkeywords}
anisotropic FSS, Equivalent Circuit model, Frequency Selective Surface (FSS), Metamaterials, Metasurfaces, Transmission Line (TL).

\end{IEEEkeywords}

\section{Introduction}
\label{sec:introduction}

The equivalent circuit representation of Frequency Selective Surfaces (FSSs) \cite{munk2000frequency} and metasurfaces \cite{Holloway_metasurface, costametamaterials} represents a powerful approach both for the design and the physical understanding of these bidimensional spatial filters. When an electromagnetic wave strikes a metasuraface, a local perturbation of the electric and magnetic field distribution is observed. The strong reactive fields caused by the metasurface discontinuity can be represented as a summation of spatial harmonics \cite{rodriguez2015analytical, costa2016wideband}. A convenient approach for analysing these planar periodic structures is to resort to a Transmission Line (TL) model where the metasurface is represented by a shunt complex impedance $ Z=R+jX $ \cite{costa2014overview}. The real part of the equivalent sheet impedance $ R $ \cite{costa2012circuit}, takes into account ohmic losses in the metal and in dielectric layers and the reactance $ X $ includes the effect of the high order harmonics which represents the perturbation of the electromagnetic (EM) field locally \cite{Mesa_ECA_IEEE_magazine, Berral_MTT}. 
The equivalent circuit approach for the analysis of FSSs, dates back to the beginning of $20th$ century \cite{Marcuvitz}. 

In general, the analysis of planar periodic surfaces can be separated in three different fundamental regions: effective media region, resonant region, Floquet-Bloch region \cite{Holloway_metasurface, costa2014overview}. At low frequencies, where the FSS periodicity $D$ is much smaller than the operating wavelength $\lambda$, the periodic surface can be analysed by using the homogenization theory \cite{lukkonen_simple_model}. In the intermediate region, where the periodicity is smaller but comparable with the operating wavelength, periodic surfaces become resonant and simple closed form expressions are not available except a few empirical formulas for loops and Jerusalem crosses  \cite{Langley_loops,Langley_double_loops, Anderson_Jcross}. Even if closed-form expressions are generally not available, periodic surfaces can still be modelled by using the TL circuit theory in the intermediate region. Indeed, the metasurface impedance can be represented by simple LC circuits whose values of the lumped parameters need to be retrieved by using a full-wave simulations and an inversion procedure \cite{costa_efficient_2012}. In the Floquet-Bloch region, where the operating frequency exceeds the cut-off frequency of high order Floquet modes, FSS elements (also single resonant ones) need to be represented by using a multi-mode network \cite{Palocz,Dubrovka}. 
In the large majority of works dedicated to the circuit analysis of FSSs, the shape of the element is considered symmetric so that its behaviour is independent of the azimuth angle and the coupling between TE and TM response can be neglected \cite{Maci_pole_zero_matching, caminita_bianisotropic}. 
However, in practice, there are several applications where anisotropic FSS elements are employed to design innovative devices or antennas \cite{patel_TMTT_2014_devices,selvanayagam_2014}. Some examples are the polarization converting surfaces \cite{Mao2017a}, surface wave waveguides \cite{gregoire_SW_waveguides,quarfoth_2013,mencagli_2015_SWdispersion,gomez2015hyperbolic} and metasurface antennas \cite{fong_sievenpiper_2010,casaletti_2017,Minatti_TAP_2015}. Tensorial metasurfaces are frequently included within equivalent transmission line formulation \cite{patel_TAP_2013,selvanayagam_2011} by pre-emptively deriving the full impedance matrix of the FSS element as a function of frequency from a full-wave simulation. However, a circuit model for anisotropic FSS elements is not available in the literature.
 
The aim of this paper is to present a simple LC model of anisotropic elements and to show that even unconventional FSS elements can be characterized in terms of a set of lumped parameters ($L_{\chi_1}$ $C_{\chi_1}$, $L_{\chi_2}$ $C_{\chi_2}$) and the rotation angle $\varphi^{rot}$.

The objective of the paper is to analyse the FSS impedance matrix and to extract an equivalent circuit topology to represent a specific element shape. The present work is not focalized on a specific unit cell shape but it is aimed at presenting a general approach for the analysis of generic FSS geometries. The proposed approach can be subsequently used for the synthesis of periodic surfaces in different scenarios with respect to the one used  for the extraction of the LC network (freestanding case). For instance, the unit cell period can be rescaled by scaling LC parameters and the response of the FSS within different dielectric layers, can be calculated with a correction of the FSS capacitance \cite{costa_efficient_2012}.

The paper is organized as follows. Section II describes step by step the procedure to derive the equivalent circuit model of anisotropic FSS elements.  Section III describes how the transmission and reflection coefficients are computed once that the LC network is derived. In Section IV, some representative examples aimed at clarifying the novelty and the usefulness of the proposed method are presented. Section V shows how to use the derived LC networks for the synthesis of devices. In particular, the designs of two transmission-type polarization converters are presented. Conclusions are reported in Section VI.  

\section{Calculation of the FSS lumped parameters} 
\label{sec_Method} 

\begin{figure} 
    \centering
    \hfill
  \subfloat[]{%
       \includegraphics[width=4cm]{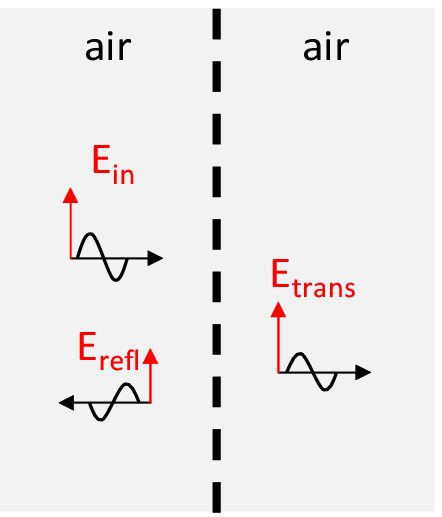}}
    \hfill
  \subfloat[]{%
        \includegraphics[width=4cm]{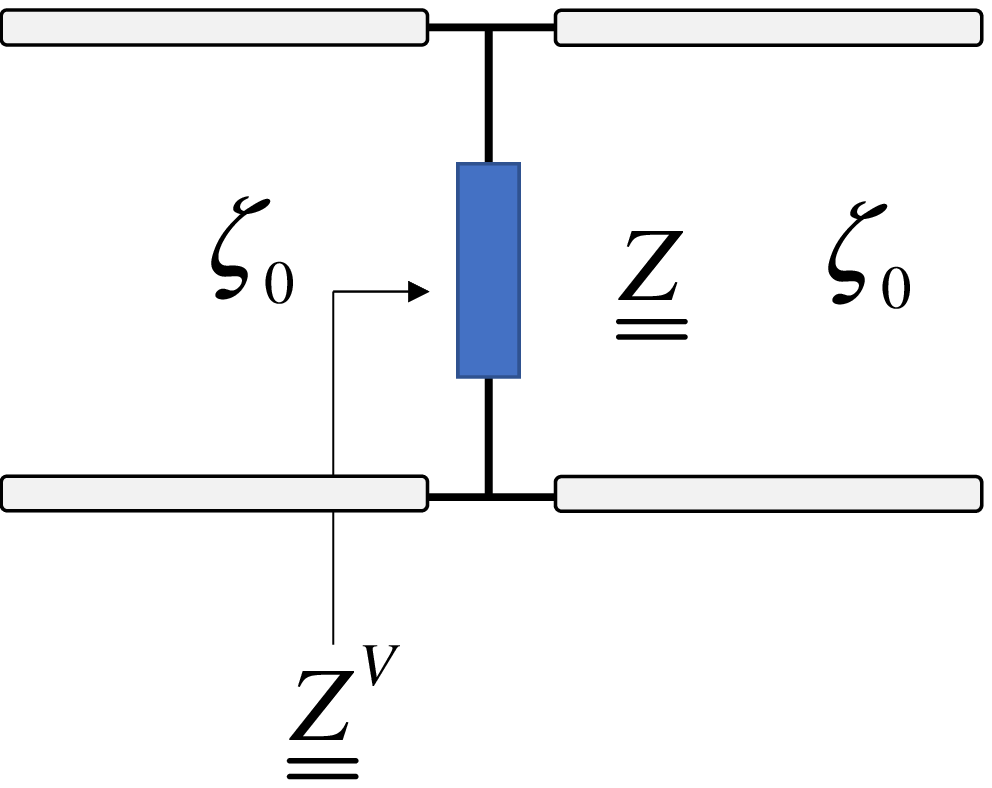}}\\              
\caption{Freestanding FSS: (a) stack-up; (b) equivalent transmission line model.}
  \label{fig_eq_circuit} 
  \vspace{0.2cm}
\end{figure}

Let us consider the canonical problem represented in Fig.~\ref{fig_eq_circuit}(a) where a plane wave strikes on a planar periodic surface at normal incidence, $\theta^{inc}=0^\circ$ and with a generic plane of incidence, $\varphi^{inc}$. The reflected and transmitted fields can be decomposed in TE and TM polarizations. The electric field for TM polarization is aligned with the plane of incidence $\varphi^{inc}$, that is \textit{x}-axis if $\varphi^{inc}=0^\circ$.  The electric field for TE polarization is aligned with the normal to the plane of incidence, that is \textit{y}-axis if $\varphi^{inc}=0^\circ$. 

The EM problem can be analysed by resorting to the equivalent circuit of the cell geometry reported in Fig.~\ref{fig_eq_circuit}(b). 
When the FSS element is symmetric, the equivalent circuit is independent of the polarization of the impinging wave. However, this approach is very restrictive since the large part of FSS unit cells employed for the design of practical devices, are polarization dependent \cite{gao2015ultrawideband,lin2016dual, Pfeiffer_PRA, gregoire_SW_waveguides, quarfoth_2013, casaletti_2017, patel_TMTT_2014_devices}, or anisotropic. 

\begin{figure} 
    \centering 
            \includegraphics[width=8cm]{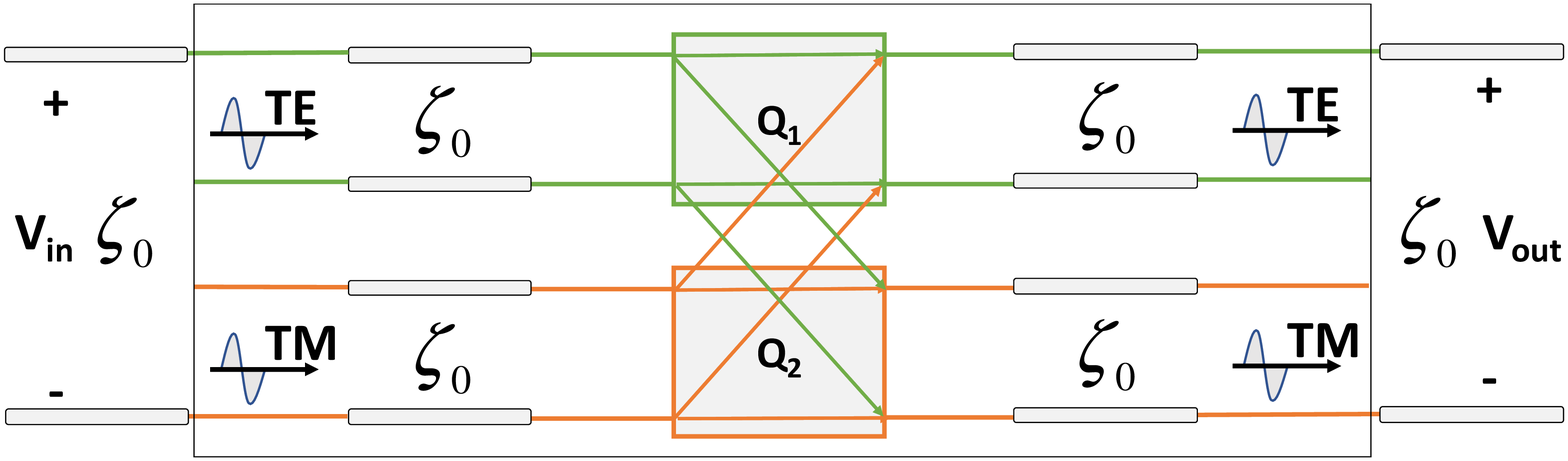}           
\caption{Equivalent circuit representation of anisotropic FSS. The quadrupoles Q1 and Q2  represent the anisotropic FSS.}
  \label{fig_LC_circuit_full} 
\end{figure}

A more general visualization of the EM problem in terms of equivalent circuit is represented in Fig.~\ref{fig_LC_circuit_full} where the impinging wave is decomposed in TE and TM polarizations. The reflection coefficient of the anisotropic metasurface can be defined as follow: 

\begin{equation} \label{eq_Gamma}
 \underline{\underline{\Gamma}} = \begin{bmatrix}
                    \Gamma_{TE-TE} & \Gamma_{TE-TM} \\[5pt]
                    \Gamma_{TM-TE} & \Gamma_{TM-TM}  \\

                  				\end{bmatrix} \\
\end{equation}

As shown in Fig.~\ref{fig_LC_circuit_full}, the representation of a certain FSS in terms of frequency dependent impedance matrix can be visualized as a set of two coupled quadrupoles \cite{Minatti_TAP_2015} whose transfer function is the impedance matrix extracted from a full wave simulation. 

According to the transmission line model, the reflection coefficient in equation \eqref{eq_Gamma} is obtained from the parallel connection of the impedance of the FSS $ \underline{\underline{Z}} $ and the free-space impedance $ \zeta_0 $.

Considering the Cartesian reference system, the FSS impedance relates the tangential components of the electric and magnetic fields according to the following expression:

\begin{equation}
\begin{bmatrix}
                    E_{x}  \\[5pt]
                    E_{y} \\

                  				\end{bmatrix}   \\ = \begin{bmatrix}
                    Z_{xx} & Z_{xy} \\[5pt]
                    Z_{yx} & Z_{yy}  \\

                  				\end{bmatrix}   \\  \begin{bmatrix}
                    -H_{y}  \\[5pt]
                    H_{x} \\
                  				\end{bmatrix}   \\
\end{equation}

The impedance matrix $ \underline{\underline{Z}}$ comprises four terms ($Z_{xx}$, $Z_{xy}$, $Z_{yx}$, $Z_{yy}$) and the interaction between polarizations must be considered for the correct calculation of the reflection/transmission coefficient of the spatial filter. Although there are cases in which $ Z_{xy} $ and $ Z_{yx} $ are equal to zero, in general $ Z_{xy}= Z_{yx} \neq 0 $. For this reason, in the equivalent circuit of Fig.~\ref{fig_eq_circuit}(b), the metasurface impedance must be represented in a matrix form. 

\begin{figure} 
    \centering
    \hfill
  \subfloat[]{%
       \includegraphics[width=4cm]{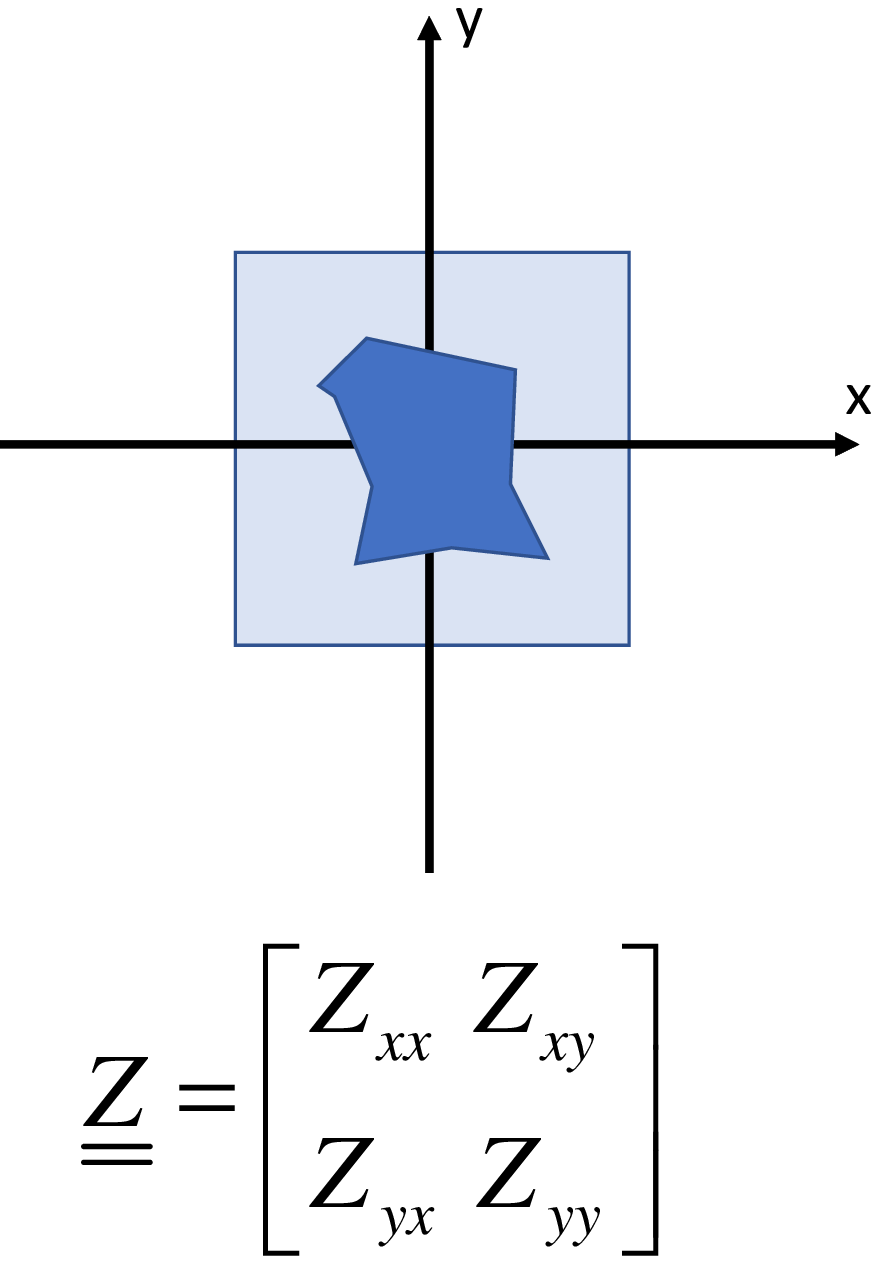}}
    \hfill
  \subfloat[]{%
        \includegraphics[width=4cm]{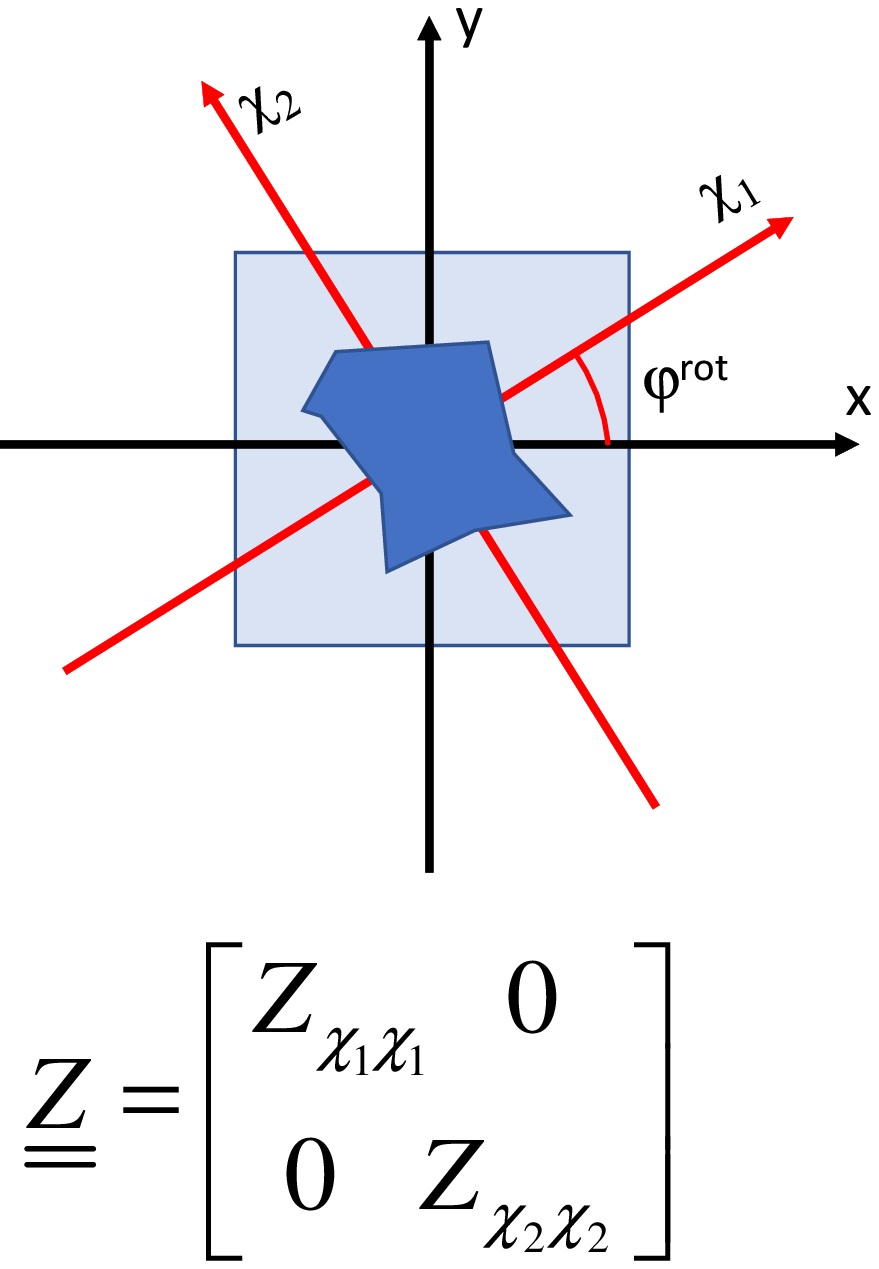}}        
        \\      
          \subfloat[]{%
        \includegraphics[width=8cm]{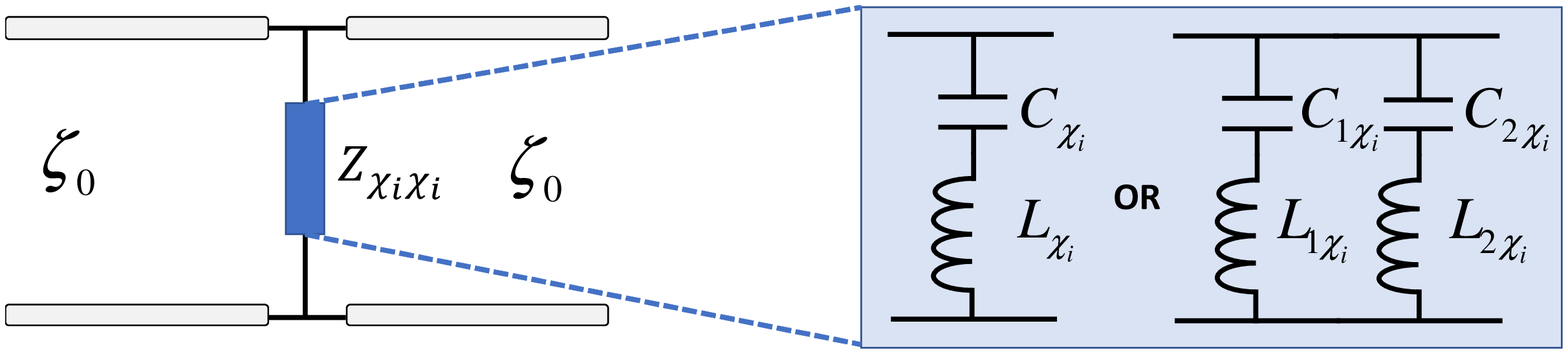}} 
\caption{(a) Generic FSS element on Cartesian axes and its impedance matrix; (b) Generic FSS matrix on crystal axes and its impedance matrix; (c) Equivalent circuit representation of the impedance on the two crystal axes. The equivalent circuit can be a simple LC circuit topology characterized by one zero and one pole or a more complicated parallel LC topology characterized by two zeros and two poles.}
  \label{fig_Stackup} 
\end{figure}

Let us consider for instance the generic unit cell element depicted in Fig.~\ref{fig_Stackup}(a) which is located on the $xy$-plane. The equivalent circuit representation of the anisotropic FSS cannot be calculated with the conventional approach \cite{costa2014overview} if  off-diagonal terms of $ \underline{\underline{Z}} $ are not zero. However, since a metasurface is a passive system, its matrix can be diagonalized by using the spectral theorem. The diagonalization is carried out trough a rotation of the coordinate system  from the Cartesian one $(x,y)$ to the crystal one  $(\chi_1,\chi_2)$ as shown in Fig.~\ref{fig_Stackup}(b). The rotation angle, $\varphi^{rot}$, is derived by using the terms of the initial matrix. The analytical derivations are reported in the Appendix \ref{Appendix-B}. Once  the matrix is diagonalized, the diagonal FSS impedance terms on the crystal axes can be represented in terms on LC equivalent circuit as shown in Fig.~\ref{fig_Stackup}(c). In particular, each FSS can be characterized with five parameters: $L_{\chi_1}$, $C_{\chi_1}$, $L_{\chi_2}$, $C_{\chi_2}$ which are the inductances and capacitances on the two crystal axes  and  $\varphi^{rot}$. The procedure for the calculation of these parameters is described in the following steps:

\begin{enumerate}
\setlength{\itemsep}{8pt}
\item Calculation of the impedance $ \underline{\underline{Z}} $ at normal incidence $ (\theta^{inc} = 0^{\circ},\varphi^{inc}) $.

The FSS impedance $ \underline{\underline{Z}} $ is computed by using the procedure described in detail in Appendix~\ref{Appendix-A}. The calculation starts from the reflection coefficient $\underline{\underline{\Gamma}} $.

\item Calculation of the rotation angle $ \varphi^{rot} $ which identifies the position of the crystal axis. The crystal angle is computed as follow:

\begin{equation} \label{eq:m_rot}
 \varphi^{\chi} = \varphi^{inc} - \varphi^{rot} 
\end{equation}

where the $\varphi^{rot}$ angle is computed according to the procedure described in Appendix \ref{Appendix-B}.

\item Calculation of the impedance $ \underline{\underline{Z}}^{\chi} $ on the crystal axis $ (\theta^{inc} = 0^{\circ}, \varphi^{inc} = \varphi^{\chi}) $. The FSS matrix $ \underline{\underline{Z}}^{\chi} $ is now diagonal;  

\item Calculation of the lumped parameters ($L_{\chi_1}$ $C_{\chi_1}$, $L_{\chi_2}$ $C_{\chi_2}$) of metasurface impedance on the crystal axes;

\item Calculation of the approximate impedance $ \underline{\underline{Z}}^{{\chi}_{LC}} $ from the parameters ($L_{\chi_1}$ $C_{\chi_1}$, $L_{\chi_2}$ $C_{\chi_2}$).

\end{enumerate}

\section{Calculation of reflection/transmission coefficients for a generic azimuth angle} 
\label{sec_Calculation of reflection coefficient for a generic azimuth angle}

Once that the five parameters characterizing a generic anisotropic FSS element are computed, they can be subsequently employed to calculate the impedance matrix  for a generic azimuth angle $\varphi^{inc}$. 
The approximate FSS impedance $ \underline{\underline{Z}}^{LC}(\varphi^{inc}) $ for a certain incident angle $ (\theta^{inc} = 0^{\circ},\varphi^{inc}) $ is computed as follow:

\begin{equation} \label{eq:XaxisRot}
\underline{\underline{Z}}^{LC}(\varphi^{inc})  = \underline{\underline{R}}^T \underline{\underline{Z}}^{{\chi}_{LC}} \underline{\underline{R}}
\end{equation}

where $ \underline{\underline{Z}}^{{\chi}_{LC}} $ is the approximate LC impedance calculated on the crystal axis $ (\theta^{inc} = 0^{\circ},\varphi^{inc} = \varphi^{\chi}) $ and $\underline{\underline{R}}$ is the rotation matrix:

\begin{equation} \label{eq:Rot}
 \underline{\underline{R}} = \begin{bmatrix}
                    \cos(-\varphi^{rot}) & -\sin(-\varphi^{rot}) \\[2pt]
                    \sin(-\varphi^{rot}) & \cos(-\varphi^{rot})  \\
                  				\end{bmatrix} \\
\end{equation}

Once the impedance matrix for the desired impinging direction is computed, the reflection coefficient can be calculated by using the $ABCD$ matrix approach \cite{Pfeiffer_PRA}:
          				
\begin{equation}\label{eq:S_matrix} 
\scalebox{0.95}[1]{$\begin{bmatrix}
                    \underline{\underline S} _{11} &\underline{\underline S} _{21} \\[5pt]
                    \underline{\underline S} _{12} & \underline{\underline S} _{22}  
                  				\end{bmatrix} = {\begin{bmatrix}
                     - \underline{\underline I}  &  \dfrac{\underline{\underline{B}} \, \underline{\underline{n}}}{\zeta _0}  + \underline{\underline{A}}  \\[12pt]
                    \dfrac{\underline{\underline{n}}}{\zeta _0}   & \dfrac{\underline{\underline{D}} \, \underline{\underline{n}}}{\zeta _0}  + \underline{\underline{C}}   
                  				\end{bmatrix}}^{-1} \begin{bmatrix}
                      \underline{\underline I}  &  \dfrac{\underline{\underline{B}} \, \underline{\underline{n}}}{\zeta _0}  - \underline{\underline{A}}  \\[12pt]
                    \dfrac{\underline{\underline{n}}}{\zeta _0}   & \dfrac{\underline{\underline{D}} \, \underline{\underline{n}}}{\zeta _0}  - \underline{\underline{C}}   
\end{bmatrix}$}
\end{equation}
		
where $\zeta _0$ is the free space impedance for TE or TM polarizations and
$ \underline{\underline{I}}=\begin{bmatrix}1&0\\0&1 \end{bmatrix}$ is the identity matrix and $ \underline{\underline{n}}=\begin{bmatrix}0&-1\\1&0 \end{bmatrix}$ is the $90^{\circ}$ rotation matrix. The term $\underline{\underline S}_{11}$ of the scattering matrix represents the reflection coefficient for the two polarizations from the left side of the screen $ (\underline{\underline \Gamma }=\underline{\underline S}_{11} )$ whereas the term $ \underline{\underline S}_{21}$ represents the transmission coefficient of the screen $ (\underline{\underline \tau }=\underline{\underline S} _{21} )$.
	
The terms of the $ABCD$ matrix for a freestanding FSS are computed as follows:
	
\begin{equation}
\begin{bmatrix}
 \underline{\underline {\text{A}}} & \underline{\underline B}  \hfill \\
  \underline{\underline C} & \underline{\underline D}
\end{bmatrix} =
\begin{bmatrix}
 \underline{\underline {\text{I}}} & \underline{\underline 0}  \hfill \\
  \underline{\underline {\text{n}}} \, {\underline{\underline {\text{Y}}} }&\underline{\underline I}
\end{bmatrix}  =   \underline{\underline{M}}^{FSS}
\end{equation}

where $\underline{\underline Y}$ is the FSS admittance matrix ($\underline{\underline Y}={\underline{\underline Z}}^{-1}$ ).

\section{Application of the Method: examples} 
\begin{figure}[h] 
    \centering
  \subfloat[]{%
       \includegraphics[width=1.7cm,height=1.7cm,angle =0]{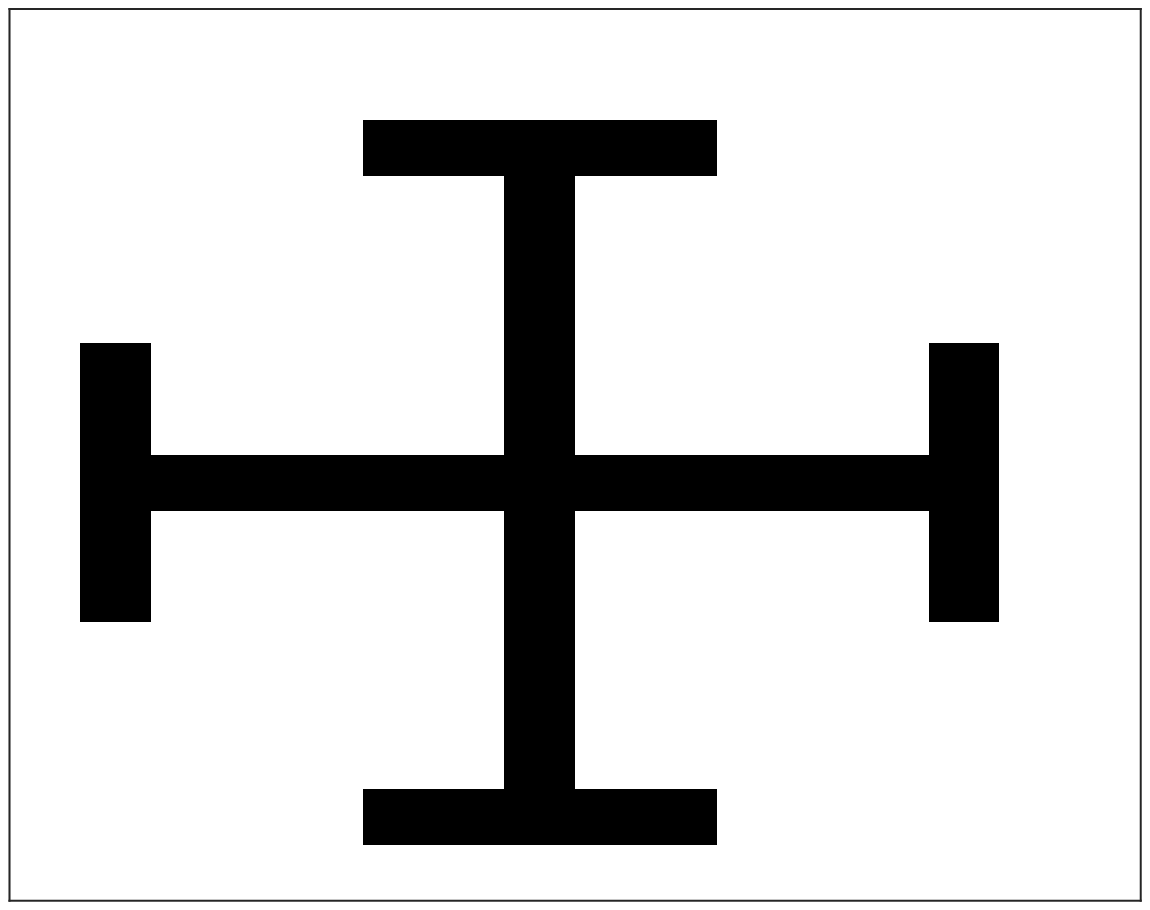}}
    \hfill
  \subfloat[]{%
        \includegraphics[width=1.7cm,height=1.7cm,angle =0]{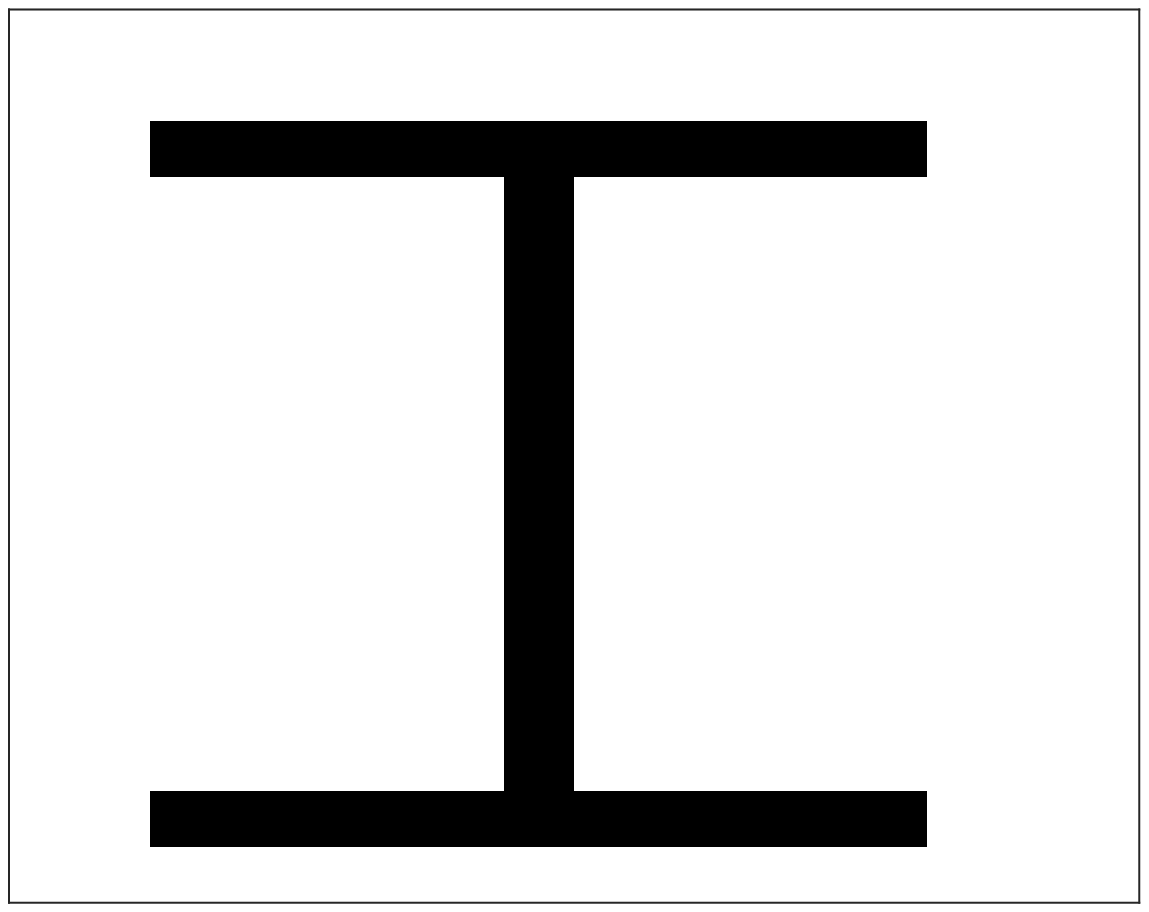}}    \hfill
  \subfloat[]{%
        \includegraphics[width=1.7cm,height=1.7cm,angle =0]{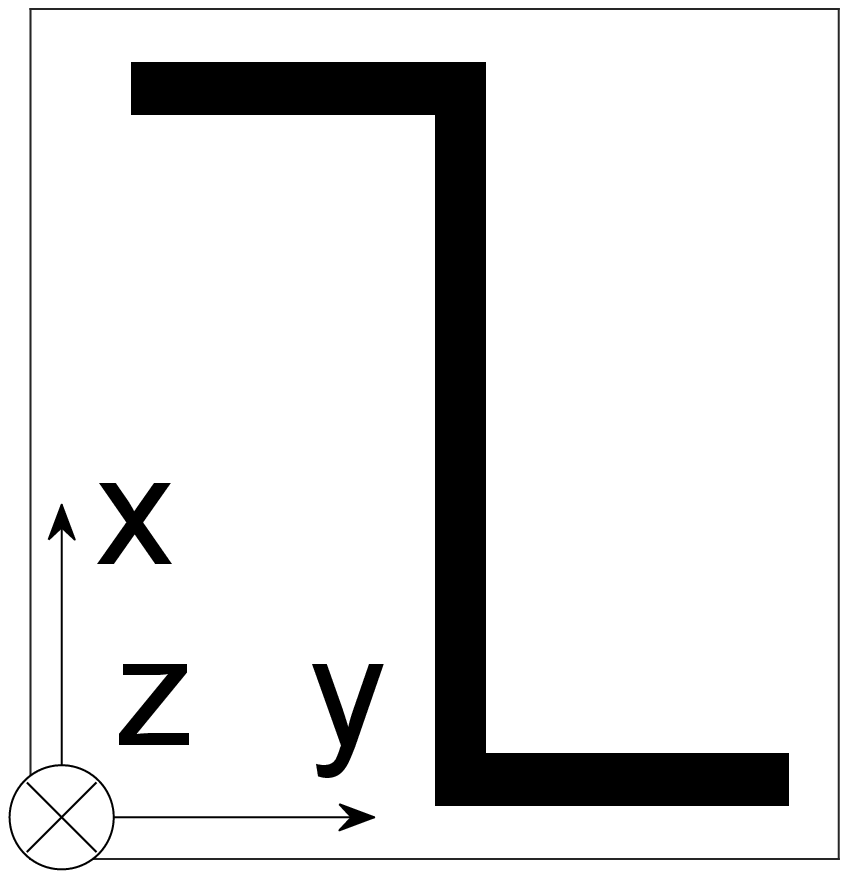}}
    \hfill
  \subfloat[]{%
        \includegraphics[width=1.7cm,height=1.7cm,angle =0]{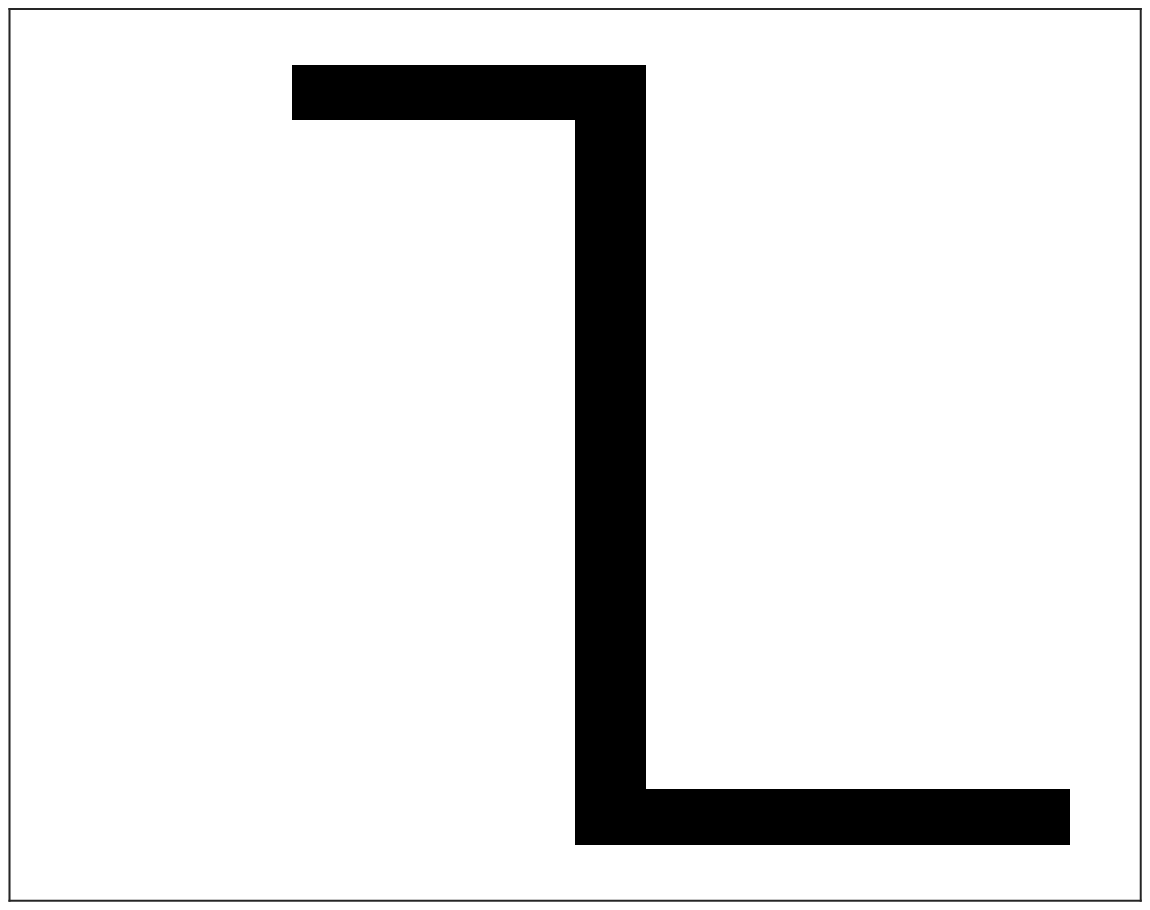}}
        \hfill
  \subfloat[]{%
        \includegraphics[width=1.7cm,height=1.7cm,angle =0]{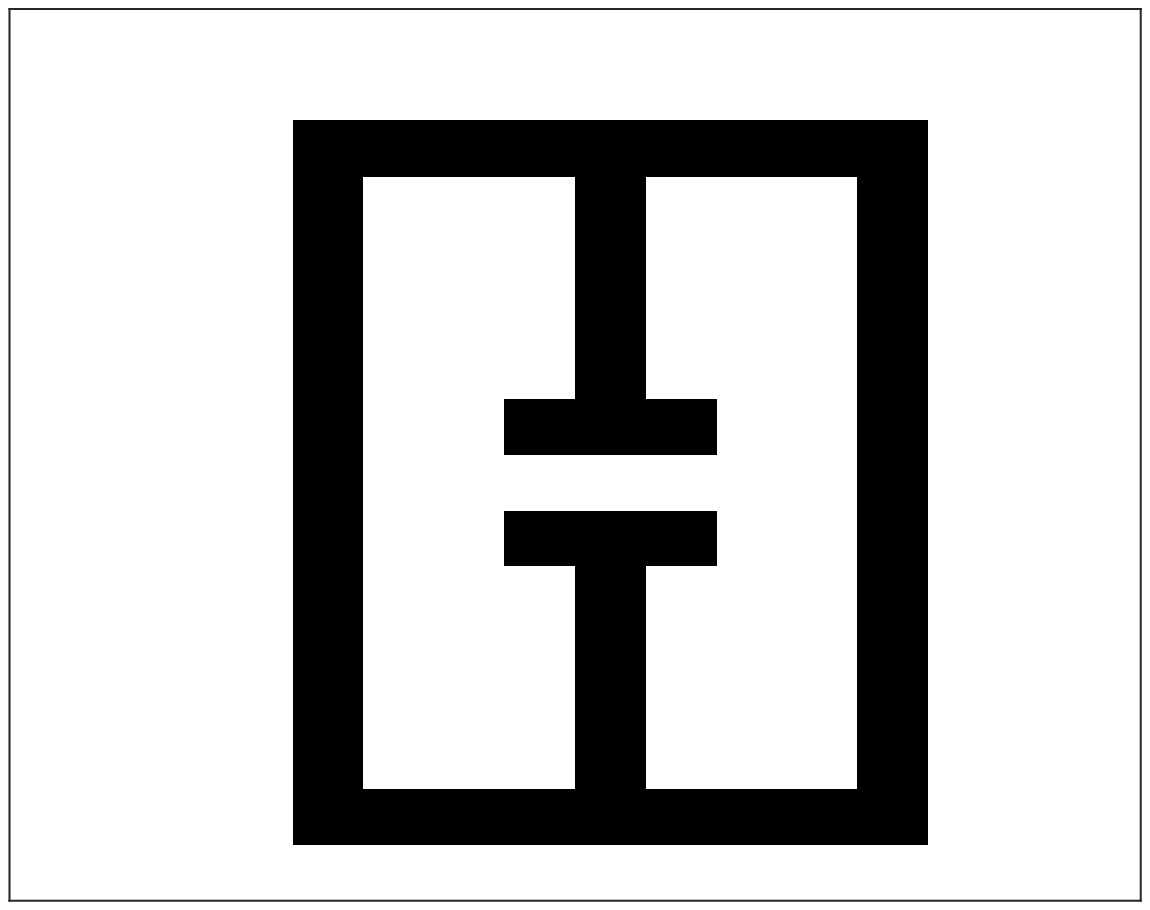}}
      \caption{Analyzed unit cells: (a) Jerusalem cross, (b) Dogbone, (c) loaded dipole, (d) asymmetric loaded dipole, (e) ref \cite{Chen2010}.}
  \label{fig_UnitCells}  
\end{figure}

Some representative examples to show the validity of the proposed equivalent circuit approach are presented in this section. Although both loops and cross type element categories \cite{munk2000frequency} are suitable examples, for the sake of brevity, we present only results for the topologies shown in Fig.~\ref{fig_UnitCells}. Initially, a fully symmetric unit cell element, i.e. the Jerusalem cross one, is presented in order to show the invariance of the impedance as a function of the azimuth angle \cite{yakovlev2009analytical}. 
Subsequently FSS elements with different degrees of anisotropy are analysed. It is worth underlining that, in particular cases, anisotropic elements might exhibit off-diagonal terms in the impedance matrix equal to zero. This happens if the crystal axes match the Cartesian axes. However, as the elements are analysed over another plane of incidence, the matrix becomes not diagonal and non-zero off-diagonal terms appear. 

\subsection{Jerusalem cross (mirror + $90^\circ$ rotational symmetry)}
\begin{figure} 
    \centering
  \subfloat[]{%
       \includegraphics[width=0.45\linewidth]{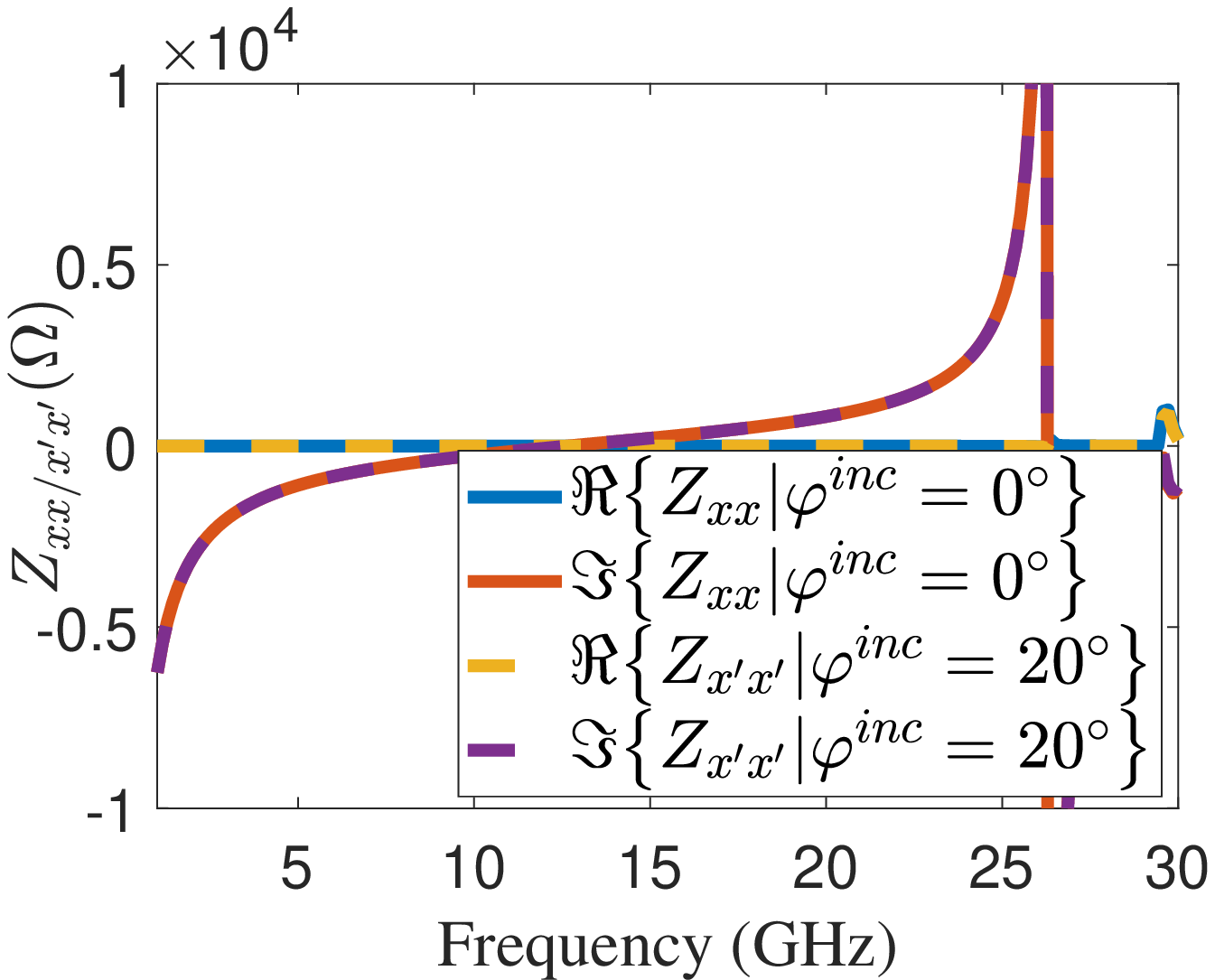}}
    \hfill
  \subfloat[]{%
        \includegraphics[width=0.45\linewidth]{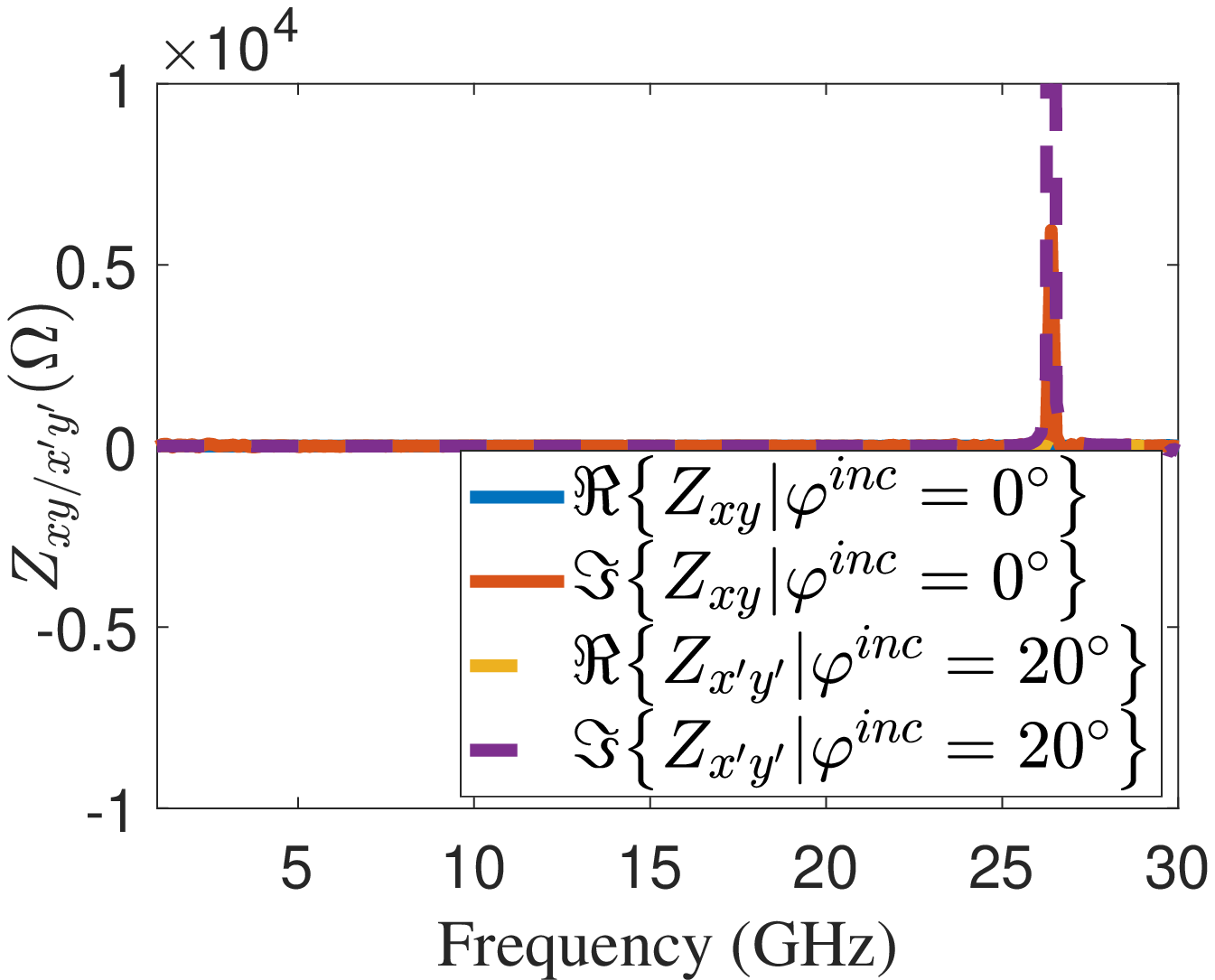}}
   \\
  \subfloat[]{%
        \includegraphics[width=0.45\linewidth]{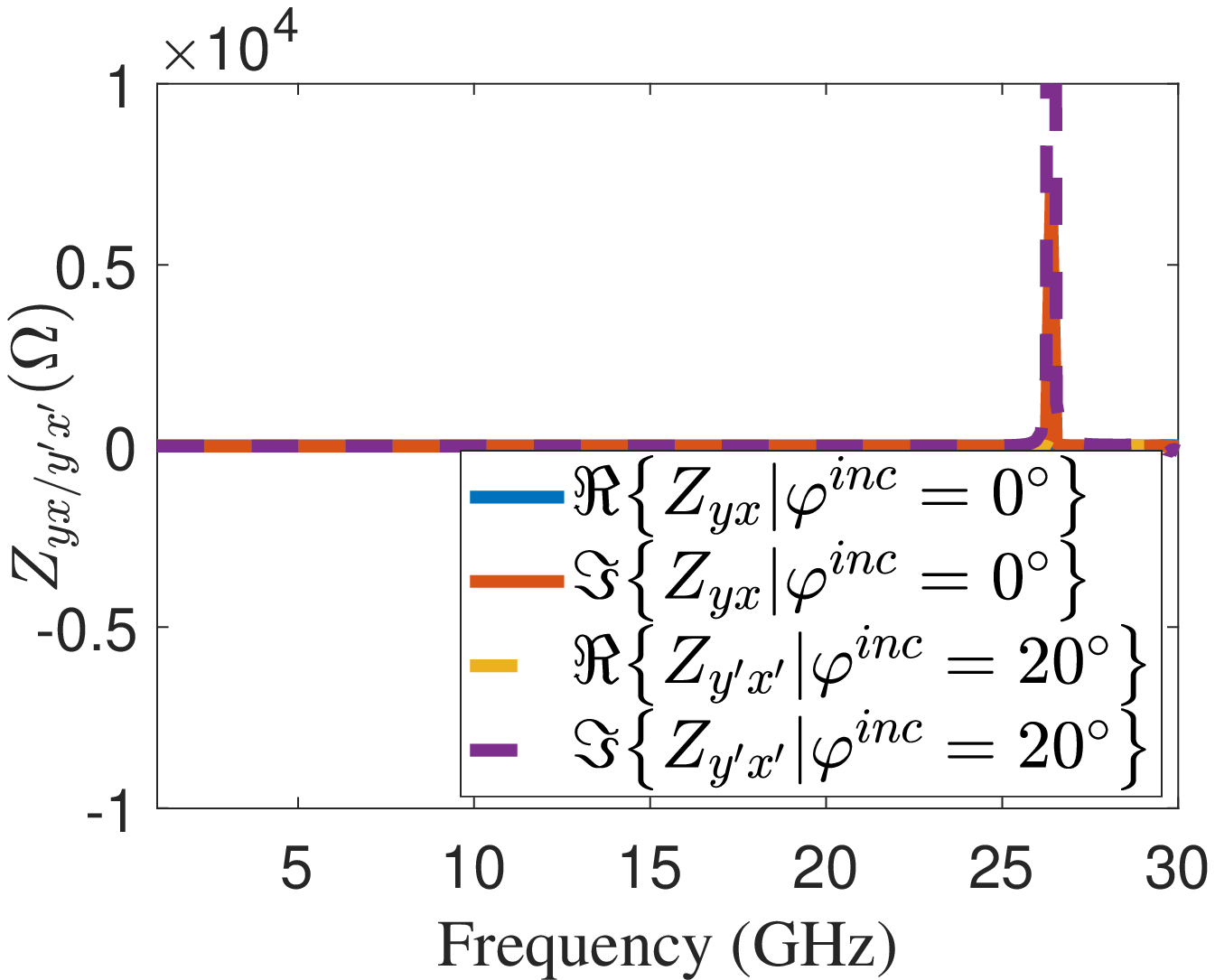}}
    \hfill
  \subfloat[]{%
        \includegraphics[width=0.45\linewidth]{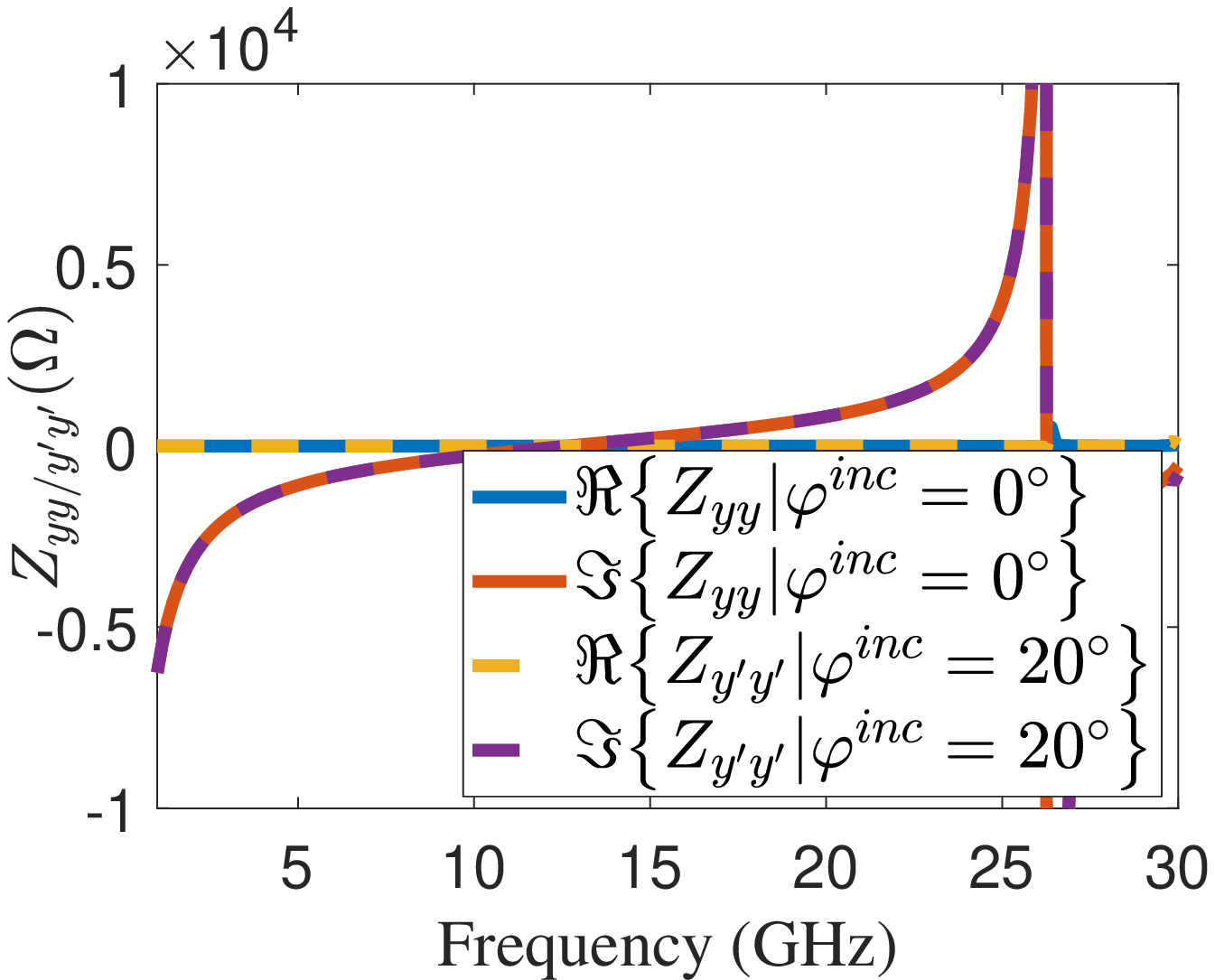}}
      \caption{$ \Re{\lbrace\underline{\underline{Z}}}\rbrace $ and $ \Im{\lbrace\underline{\underline{Z}}}\rbrace $ as a function of the frequency for the case of an FSS based on Jerusalem cross (JCross). The matrix $\underline{\underline{Z}}$ is represented on Cartesian axes $(x,y)$ and $(x^\prime,y^\prime)$ where the latter is the rotated reference system. The direction of the impinging electric field is $ (\theta^{inc} = 0^{\circ} ) $. The FSS is interrogated with two different azimuth angles $ (\varphi^{inc} = 0^{\circ}, 20^\circ)$; (a) $ Z_{xx/x^{\prime}x^{\prime}} $, (b) $ Z_{xy/x^{\prime}y^{\prime}} $, (c) $ Z_{yx/y^{\prime}x^{\prime}} $,(d) $ Z_{yy/y^{\prime}y^{\prime}} $. The periodicity is equal to \SI{1}{\centi\metre} along planar directions.}
  \label{fig_Z_Jcross} 
\end{figure}

The Jerusalem cross element is shown in Fig.~\ref{fig_UnitCells}(a). The element is characterized by a full symmetry along principal planes and diagonals ($1/8$ rotational symmetry). 
The four terms of the impedance matrix of the considered Jerusalem cross element are reported in Fig.~\ref{fig_Z_Jcross} for normal incidence and for two different azimuth angles ($\varphi^{inc} =0^\circ$ and $\varphi^{inc} = 20^\circ$). As is evident, in this case, the FSS impedance is diagonal for both the azimuth angles. Indeed, $Z_{xx} = Z_{yy}$ and the off-diagonal terms are equal to zero in the whole analysed frequency range independently of the $\varphi^{inc}$ interrogation angle. Due to the symmetry of the element, the equivalent circuit model is invariant with respect to azimuth rotations. It is therefore sufficient to derive the LC representation for the $Z_{xx}$ element of the matrix. As a consequence, symmetric elements can be represented by only two lumped parameters. A more accurate representation of the impedance can be obtained with a shunt connection of two series LC circuits in case of double-resonant shapes \cite{costa2014overview}. 

\subsection{Dogbone (mirror + $180^\circ$ rotational symmetry)}
\begin{figure} 
    \centering
  \subfloat[]{%
       \includegraphics[width=0.45\linewidth]{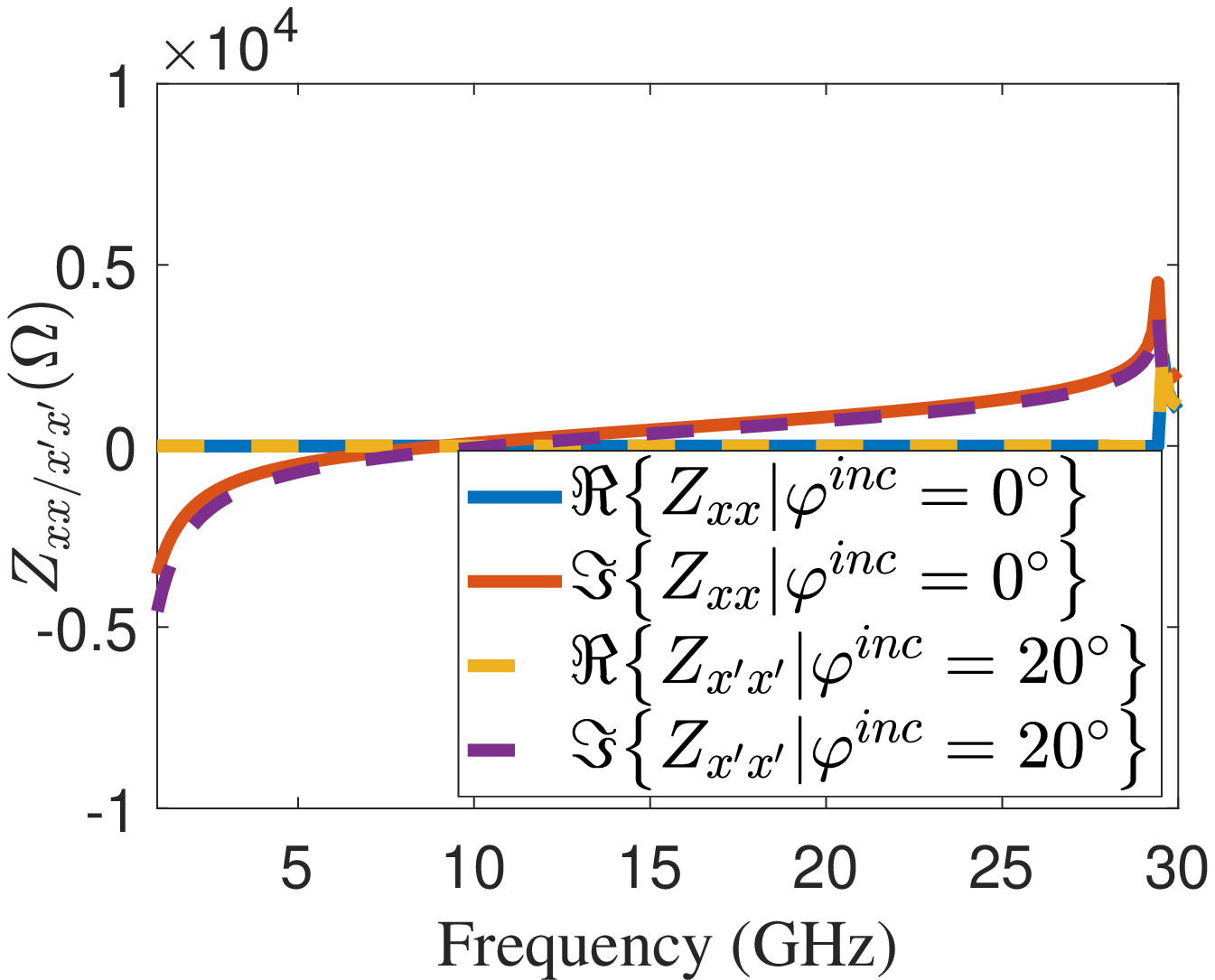}}
    \hfill
  \subfloat[]{%
        \includegraphics[width=0.45\linewidth]{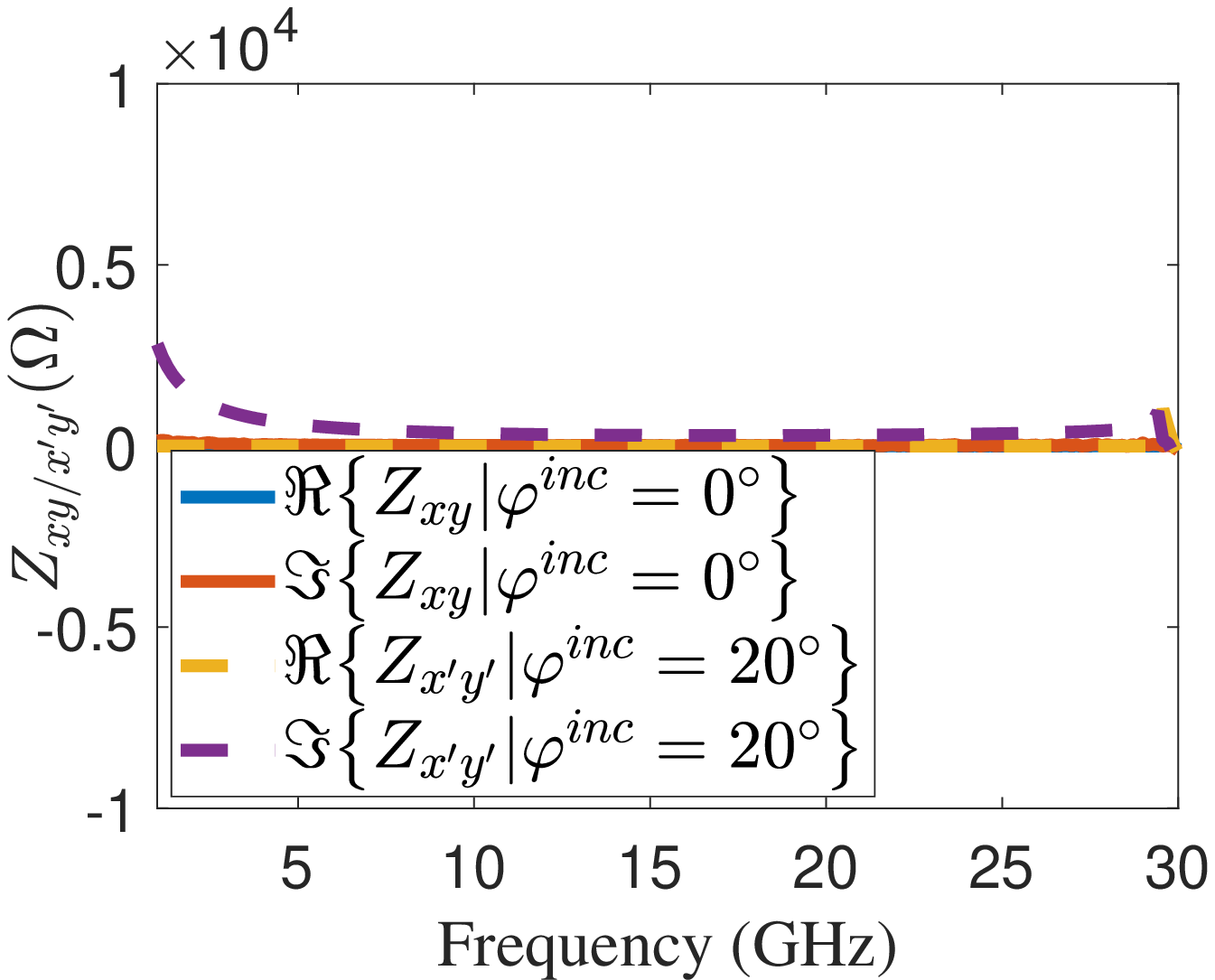}}
   \\
  \subfloat[]{%
        \includegraphics[width=0.45\linewidth]{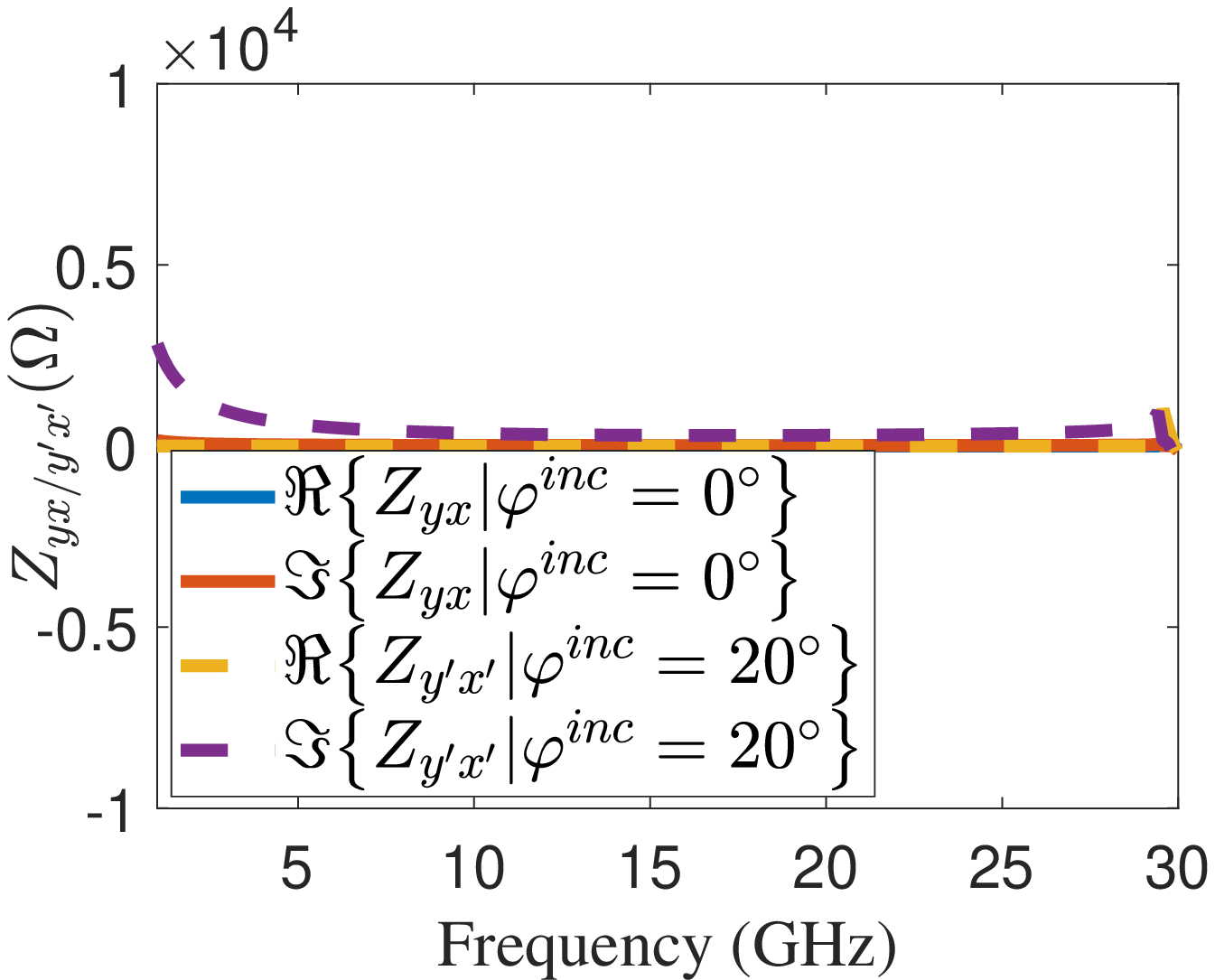}}
    \hfill
  \subfloat[]{%
        \includegraphics[width=0.45\linewidth]{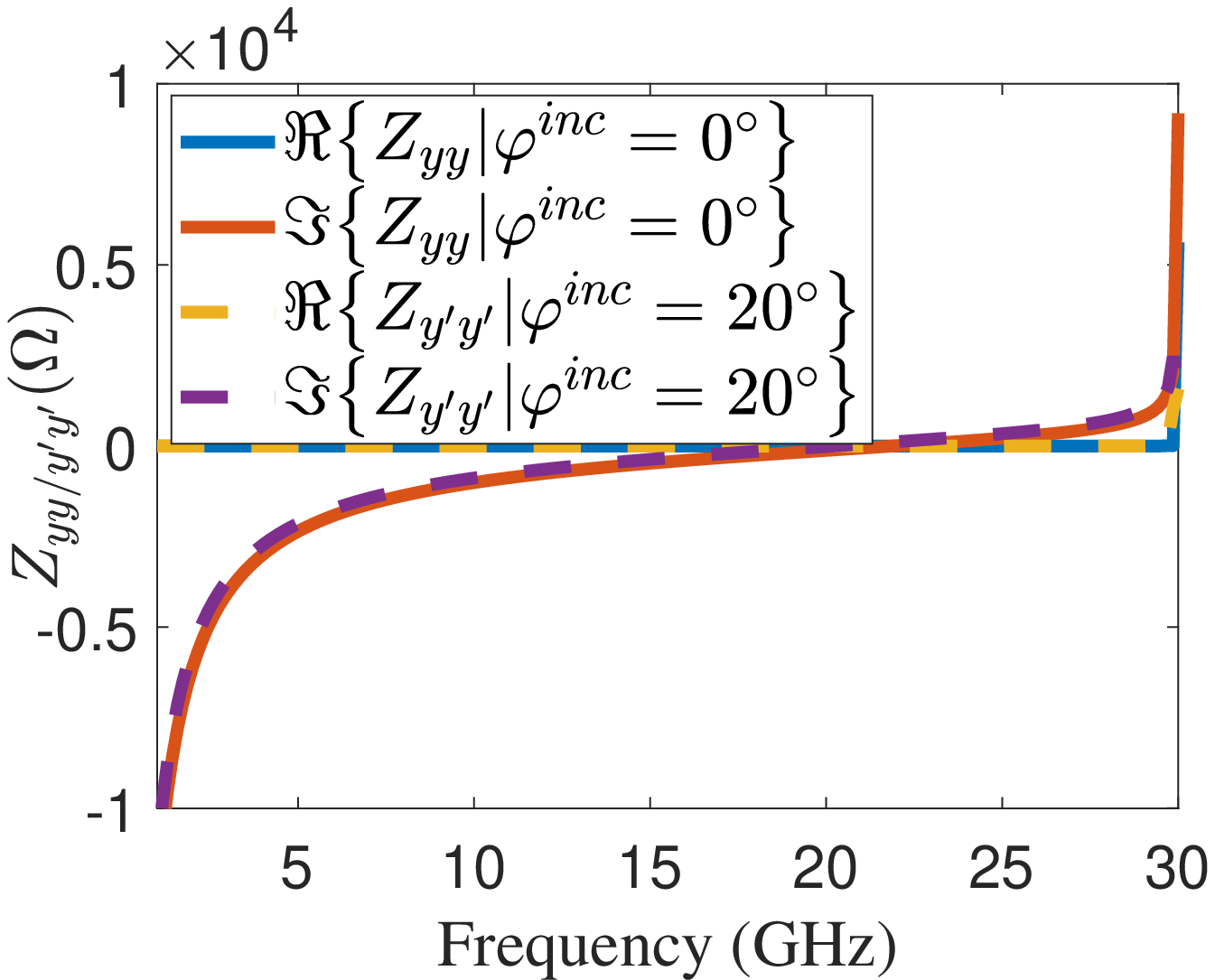}}
      \caption{$ \Re{\lbrace\underline{\underline{Z}}}\rbrace $ and $ \Im{\lbrace\underline{\underline{Z}}}\rbrace $ as a function of the frequency for the case of an FSS based on a dogbone shaped element. The matrix $\underline{\underline{Z}}$ is represented on Cartesian axes $(x,y)$ and $(x^\prime,y^\prime)$ where the latter is the rotated reference system. The direction of the impinging electric field is $ (\theta^{inc} = 0^{\circ} ) $. The FSS is interrogated with two different azimuth angles $ (\varphi^{inc} = 0^{\circ}, 20^\circ)$; (a) $ Z_{xx/x^{\prime}x^{\prime}} $, (b) $ Z_{xy/x^{\prime}y^{\prime}} $, (c) $ Z_{yx/y^{\prime}x^{\prime}} $,(d) $ Z_{yy/y^{\prime}y^{\prime}} $. The periodicity is equal to \SI{1}{\centi\metre} along planar directions.}
  \label{fig_ZFSS_dogbone_dipole} 
\end{figure}

\begin{figure} 
    \centering
  \subfloat[]{%
       \includegraphics[width=0.45\linewidth]{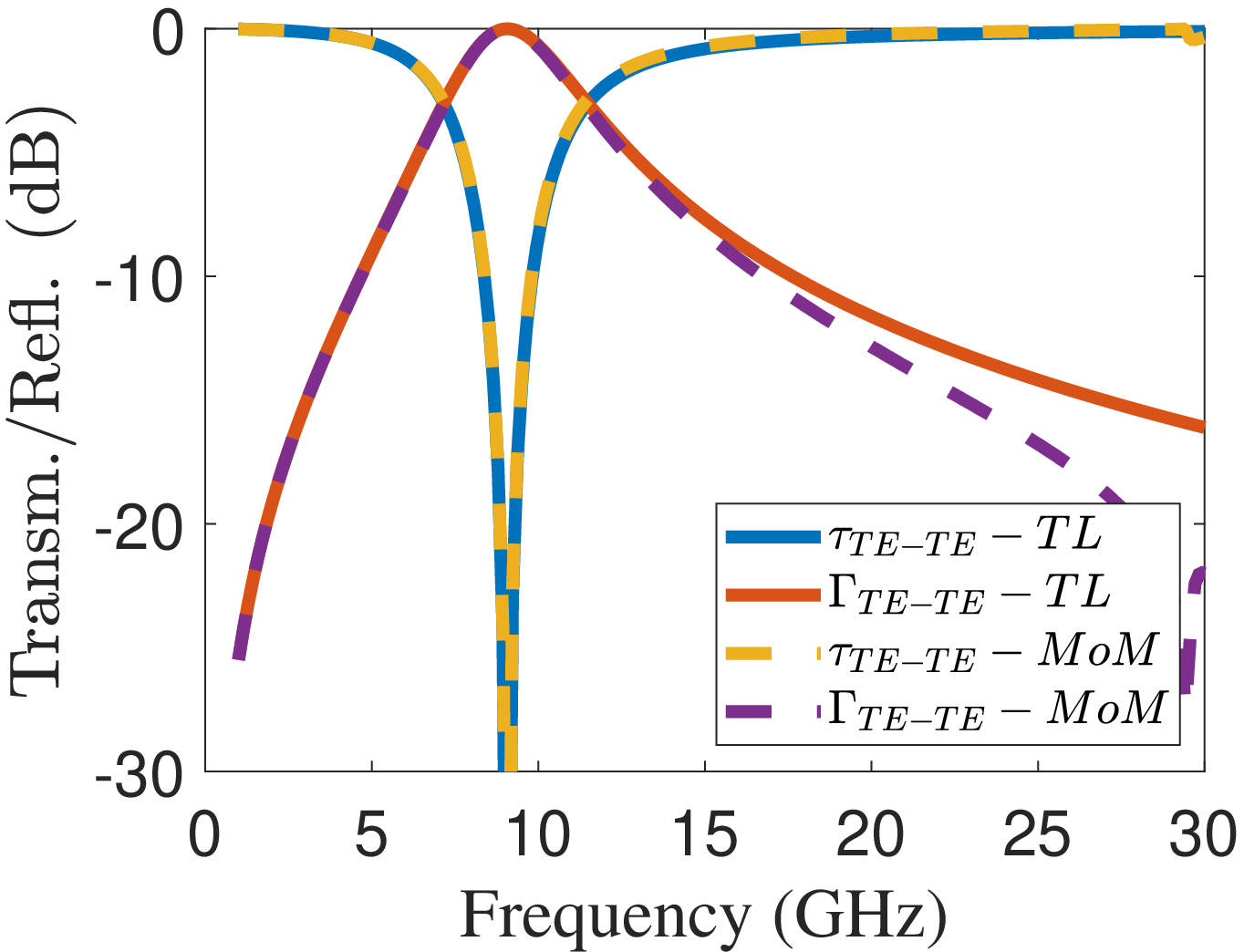}}
    \hfill
  \subfloat[]{%
        \includegraphics[width=0.45\linewidth]{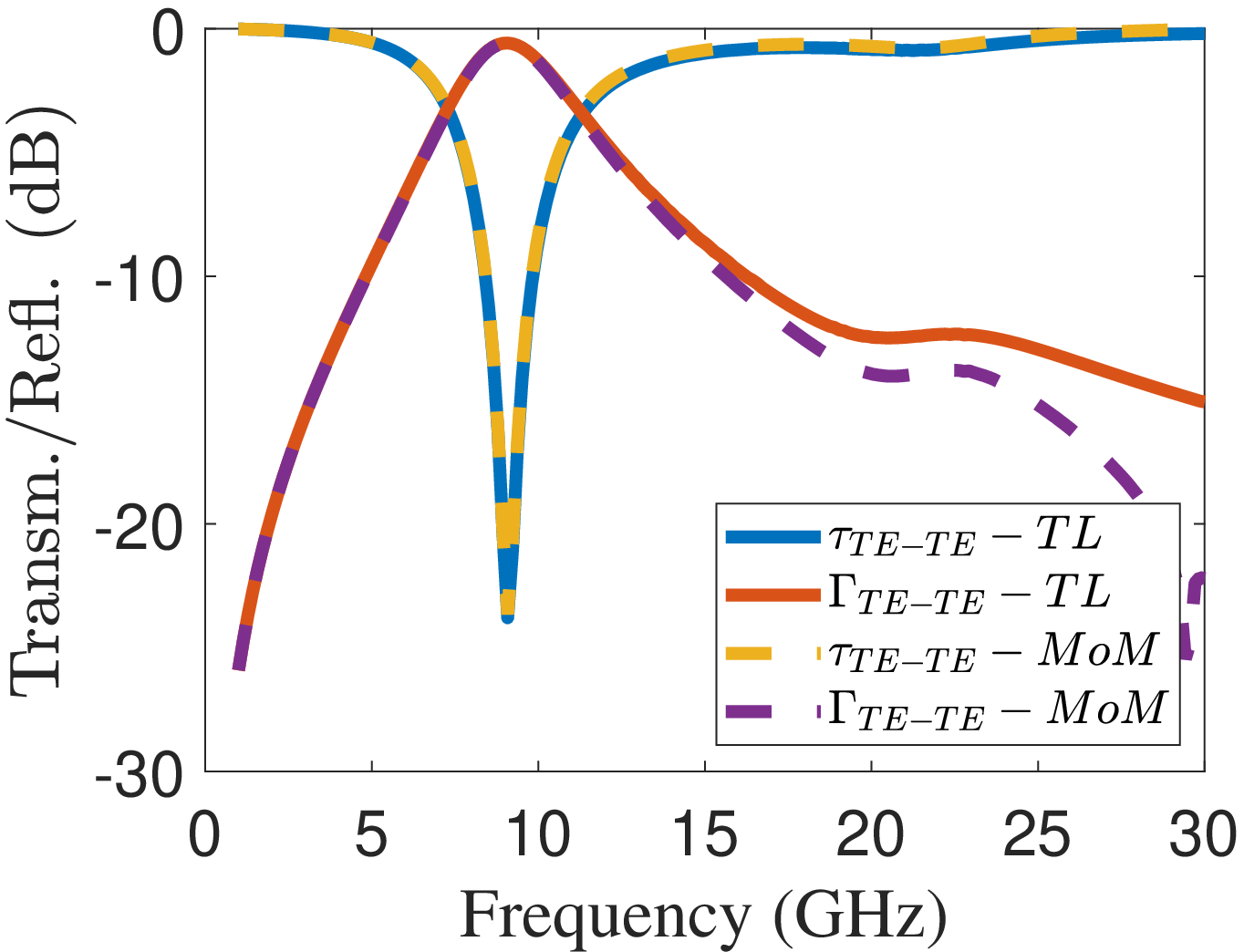}}
   \\
  \subfloat[]{%
        \includegraphics[width=0.45\linewidth]{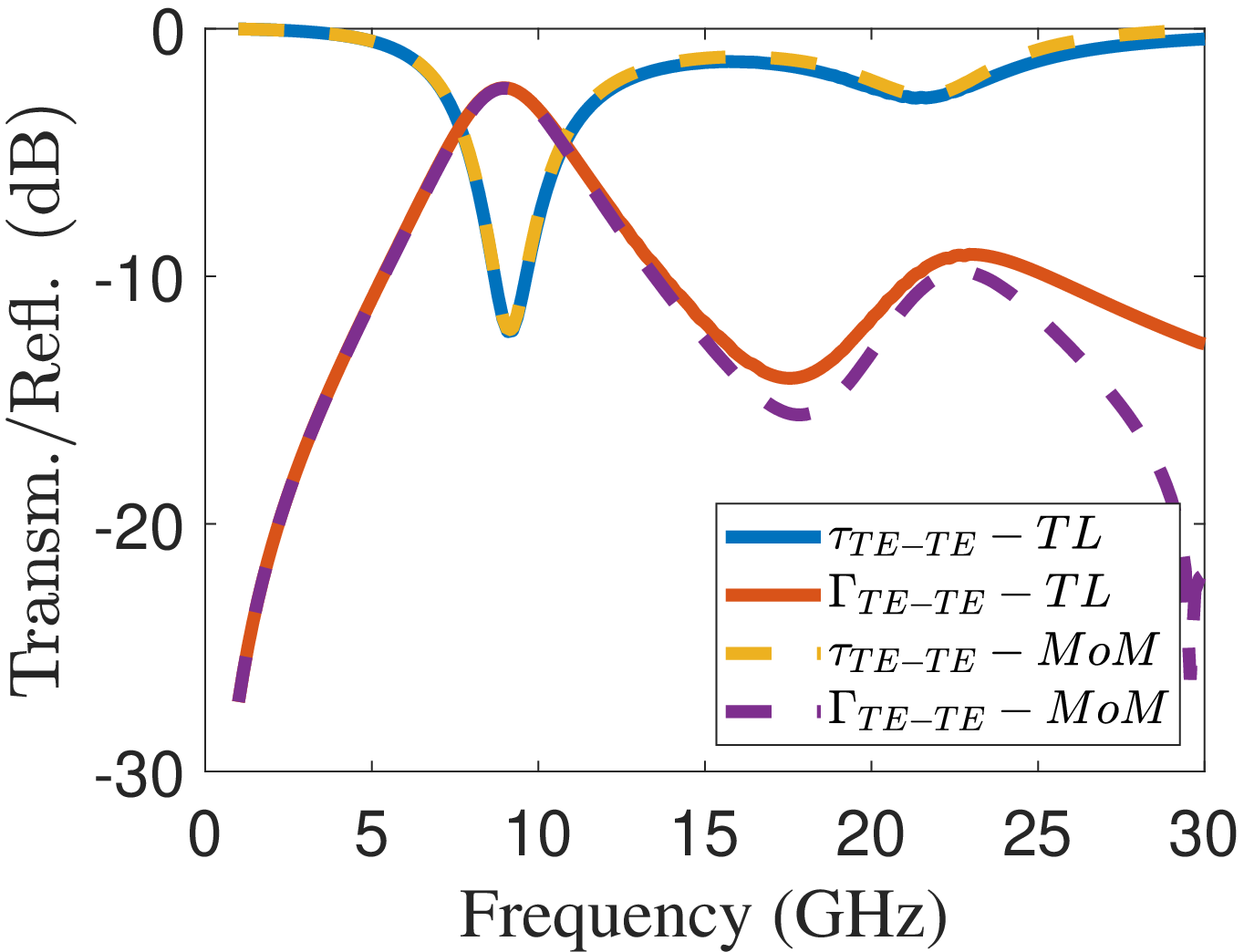}}
    \hfill
  \subfloat[]{%
        \includegraphics[width=0.45\linewidth]{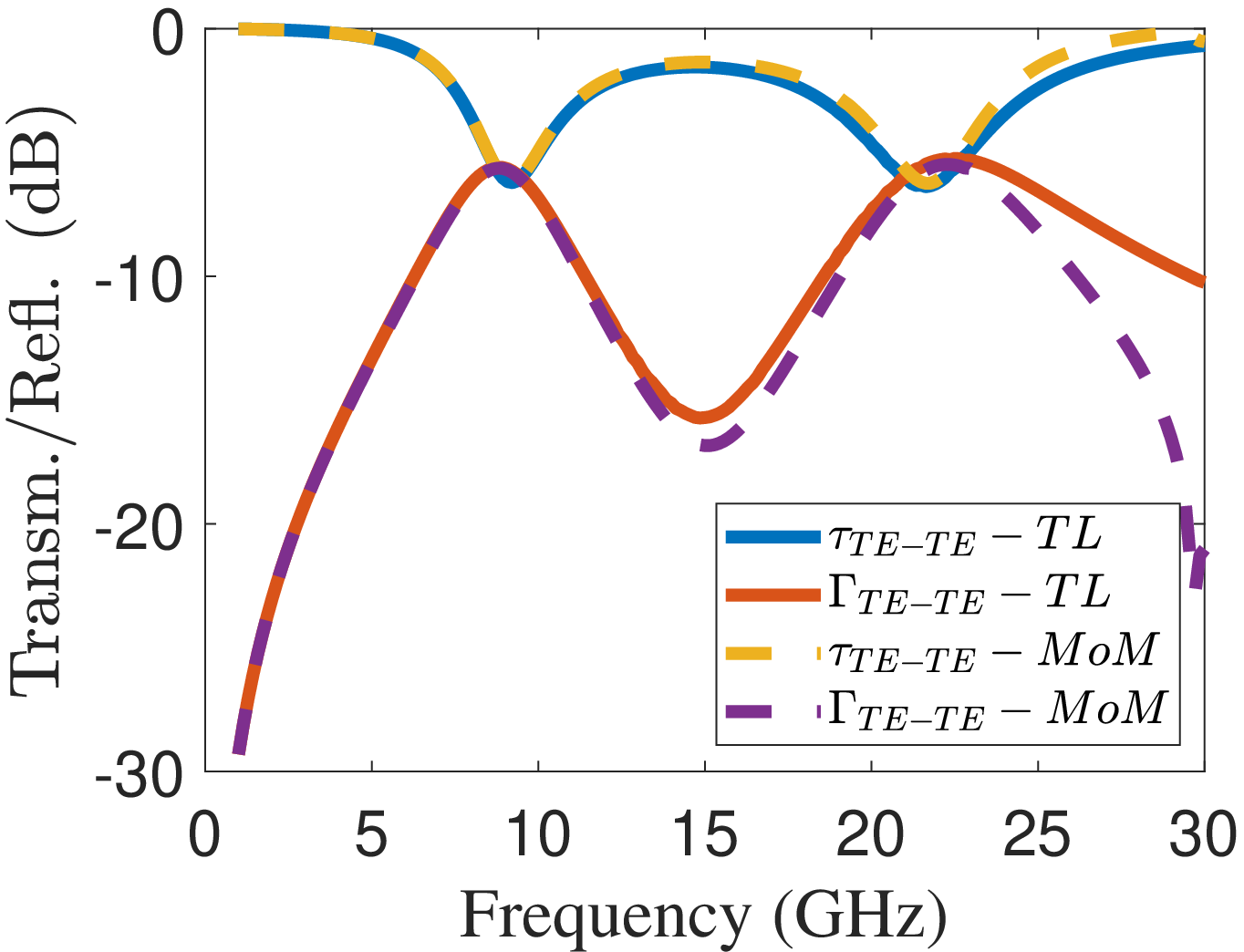}}
      \caption{Reflection and transmission coefficients of a dogbone shaped FSS for different polarization angles in freestanding configuration ((a) $\varphi^{inc} = 0^{\circ}$, (b) $\varphi^{inc} = 15^{\circ}$, (c) $\varphi^{inc} = 30^{\circ}$, (d) $\varphi^{inc} = 45^{\circ}$). The periodicity is equal to \SI{1}{\centi\metre} along planar directions. }
  \label{fig_dogbone_reflection_trasnmission_air} 
\end{figure}

The dogbone shape consists of one single arm of the Jerusalem cross element. If the end-loading is removed, a simple dipole element resonating at a half-wavelength is obtained. The end-loading arm allows for the reduction of the resonance frequency thanks to the increased capacitive coupling with the neighbour element in the periodic lattice \cite{costa2014overview}. In this case, the behaviour of the periodic surface is polarization dependent.  The impedance of the dogbone element is diagonal only if the impinging electric field is aligned with the dipole. Indeed, if the impedance is derived from a simulation carried out for a generic $\varphi^{inc}$, the impedance matrix is not diagonal. According to the formulation presented in Section~\ref{sec_Method}, at $\varphi^{inc} = 0^\circ$ the matrix  is diagonal since, with the dipole aligned with one of the Cartesian axes, the crystal axes are $\varphi^{inc} = 0^\circ$ and $\varphi^{inc} = 90^\circ$. The impedance matrix of the dogbone element, computed at $\varphi^{inc} = 0^\circ$ and $\varphi^{inc} = 20^\circ$ are reported in Fig.~\ref{fig_ZFSS_dogbone_dipole}. As is evident, the interrogation at $\varphi^{inc} = 20^\circ$ determines non zero off-diagonal terms. Therefore, it is evident that the only option for deriving an equivalent circuit model of the non-symmetric unit cell is to derive the impedance of the unit cell on the crystal axis. In this case, two LC circuits can be easily derived. Once the equivalent circuit (EC) parameters are calculated, the impedance along the other azimuth angles can be derived by a rotation of the matrix as described in Section \ref{sec_Calculation of reflection coefficient for a generic azimuth angle}. The reflection and transmission coefficients of the dogbone shaped FSS computed with the LC model and with a full-wave Periodic Method of Moments (MoM)\cite{Mittra_PMM}\footnote{ The simulator is based on a Periodic Method of Moments. The EFIE (Electric Field Integral Integration) is applied to the single cell together with the Floquet theorem, thus exploiting the infinite array model. The simulator discretizes each cell with a grid of $16 \times 16$ elements. The Method of Moments is used to compute current distribution on the element. From the currents, it is possible to calculate the electric field and subsequently the reflection and transmission coefficient.} simulations are compared in Fig.~\ref{fig_dogbone_reflection_trasnmission_air}. The comparison is carried out for four different azimuth angles. 

\subsection{Symmetric end-loaded dipole ($180^\circ$ rotational symmetry)}
\begin{figure} 
    \centering
  \subfloat[]{%
       \includegraphics[width=0.45\linewidth]{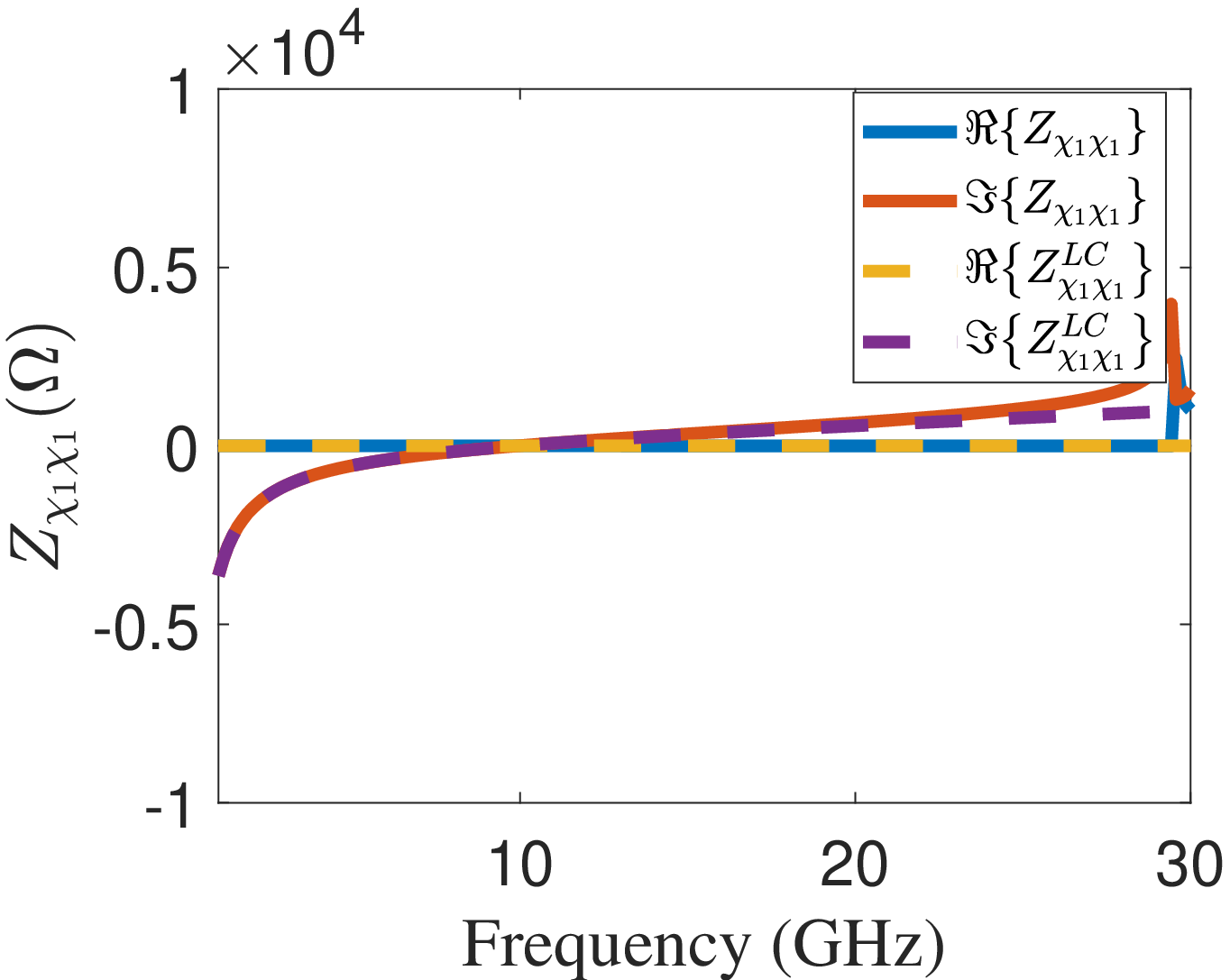}}
    \hfill
  \subfloat[]{%
        \includegraphics[width=0.45\linewidth]{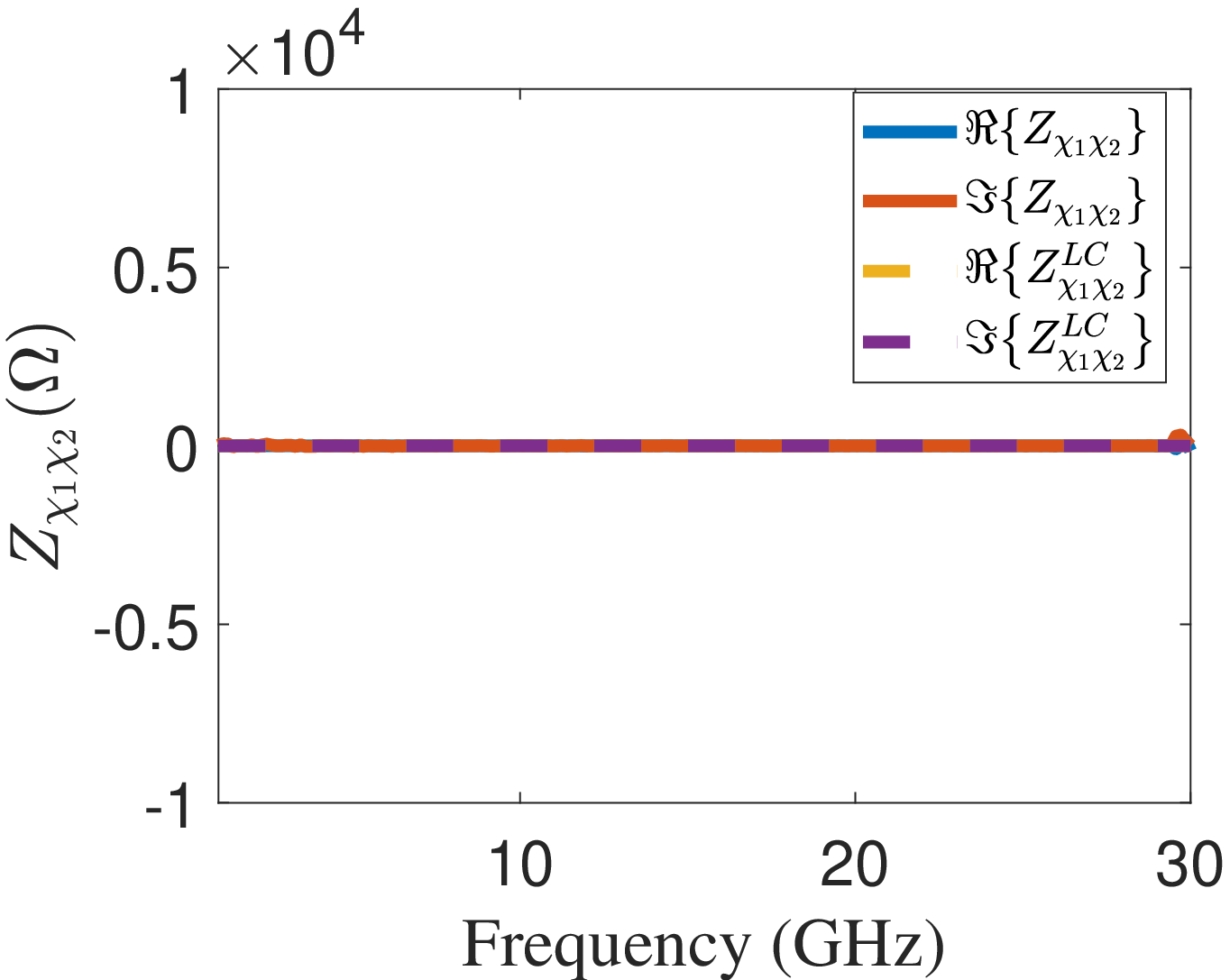}}
   \\
  \subfloat[]{%
        \includegraphics[width=0.45\linewidth]{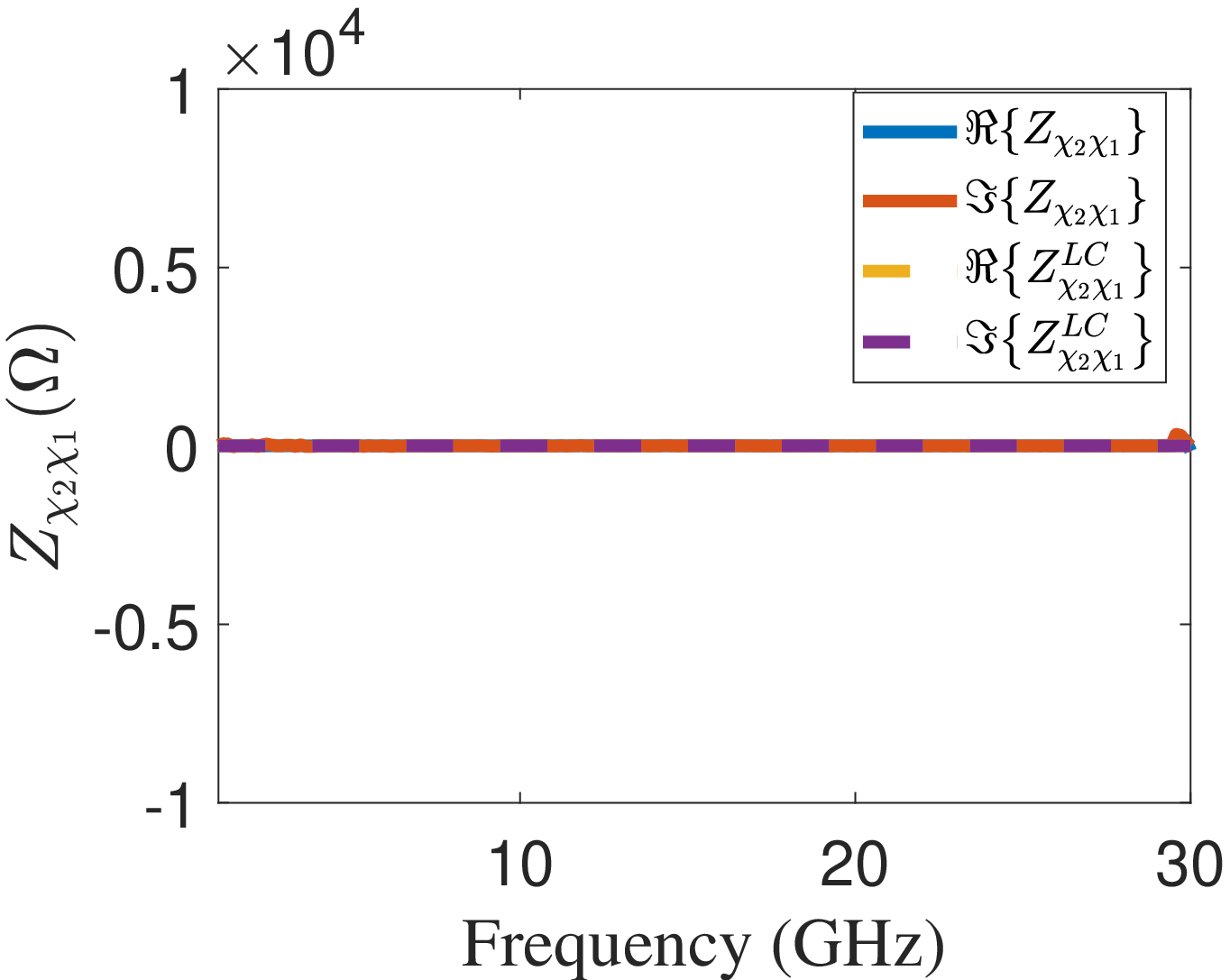}}
    \hfill
  \subfloat[]{%
        \includegraphics[width=0.45\linewidth]{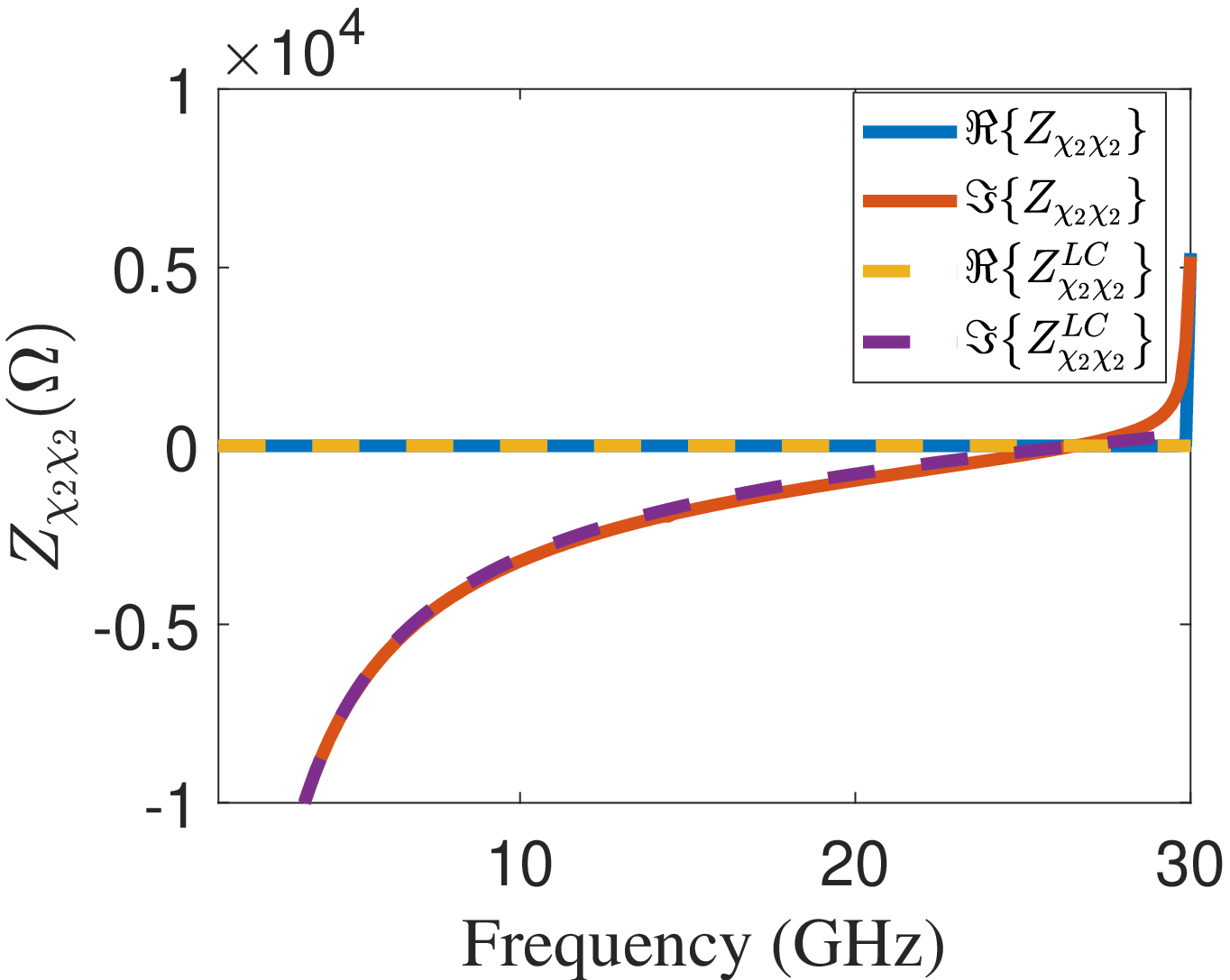}}
      \caption{$ \Re{\lbrace\underline{\underline{Z}}}\rbrace $ and $ \Im{\lbrace\underline{\underline{Z}}}\rbrace $ a function of the frequency for the case of an FSS based on a symmetric loaded dipole shaped element. The $\underline{\underline{Z}}$ matrix is represented on crystal axes. The direction of the impinging electric field is $ (\theta^{inc} = 0^{\circ},\varphi^{inc} = \varphi^{rot} ) $. (a) $ Z_{\chi_1 \chi_1} $, (b) $ Z_{\chi_1 \chi_2} $, (c) $ Z_{\chi_2 \chi_1} $,(d) $ Z_{\chi_2 \chi_2} $.  The periodicity is equal to \SI{1}{\centi\metre} along planar directions.}
  \label{fig_Zdipole_crystal} 
\end{figure}

\begin{figure} 
    \centering
  \subfloat[]{%
       \includegraphics[width=0.45\linewidth]{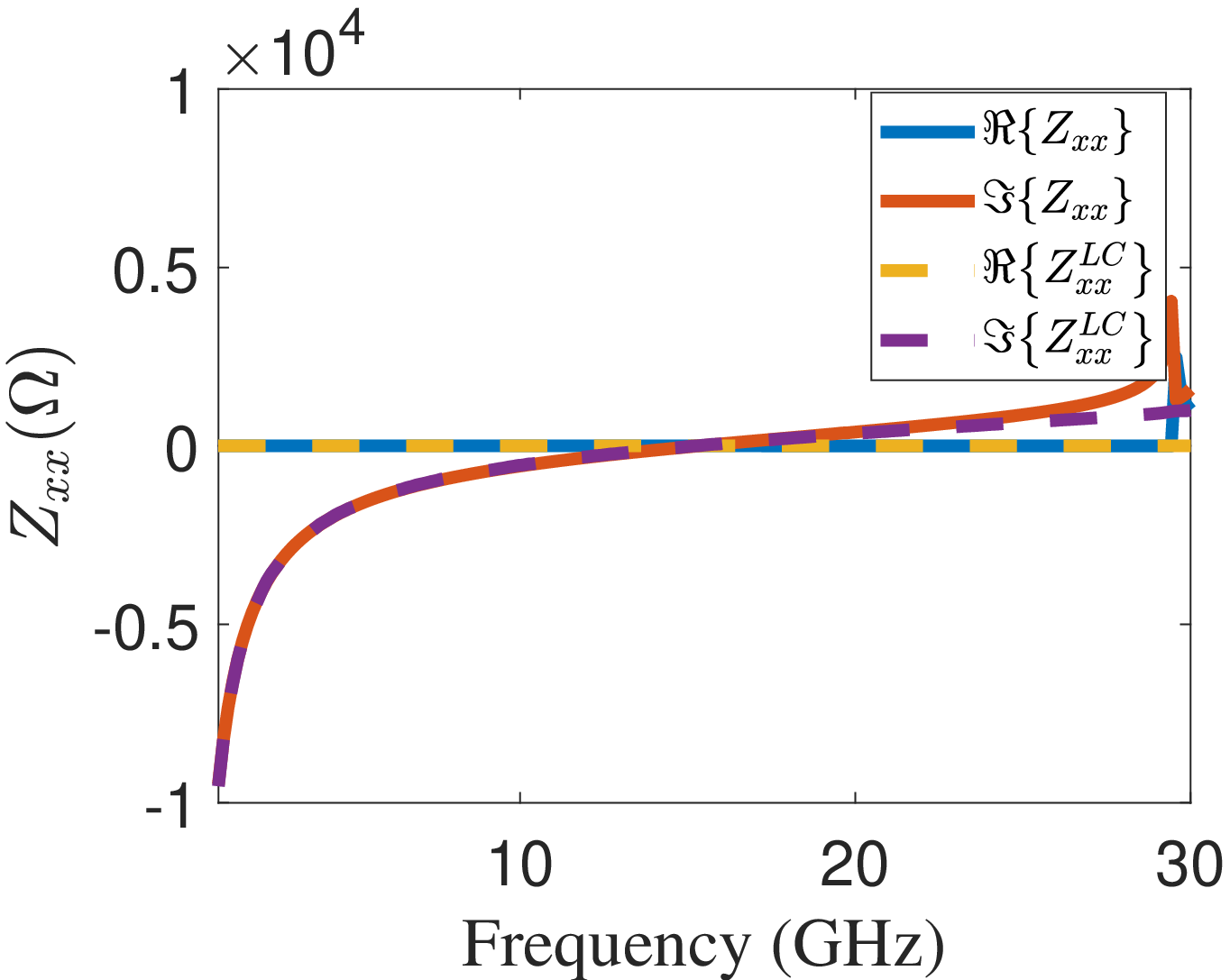}}
    \hfill
  \subfloat[]{%
        \includegraphics[width=0.45\linewidth]{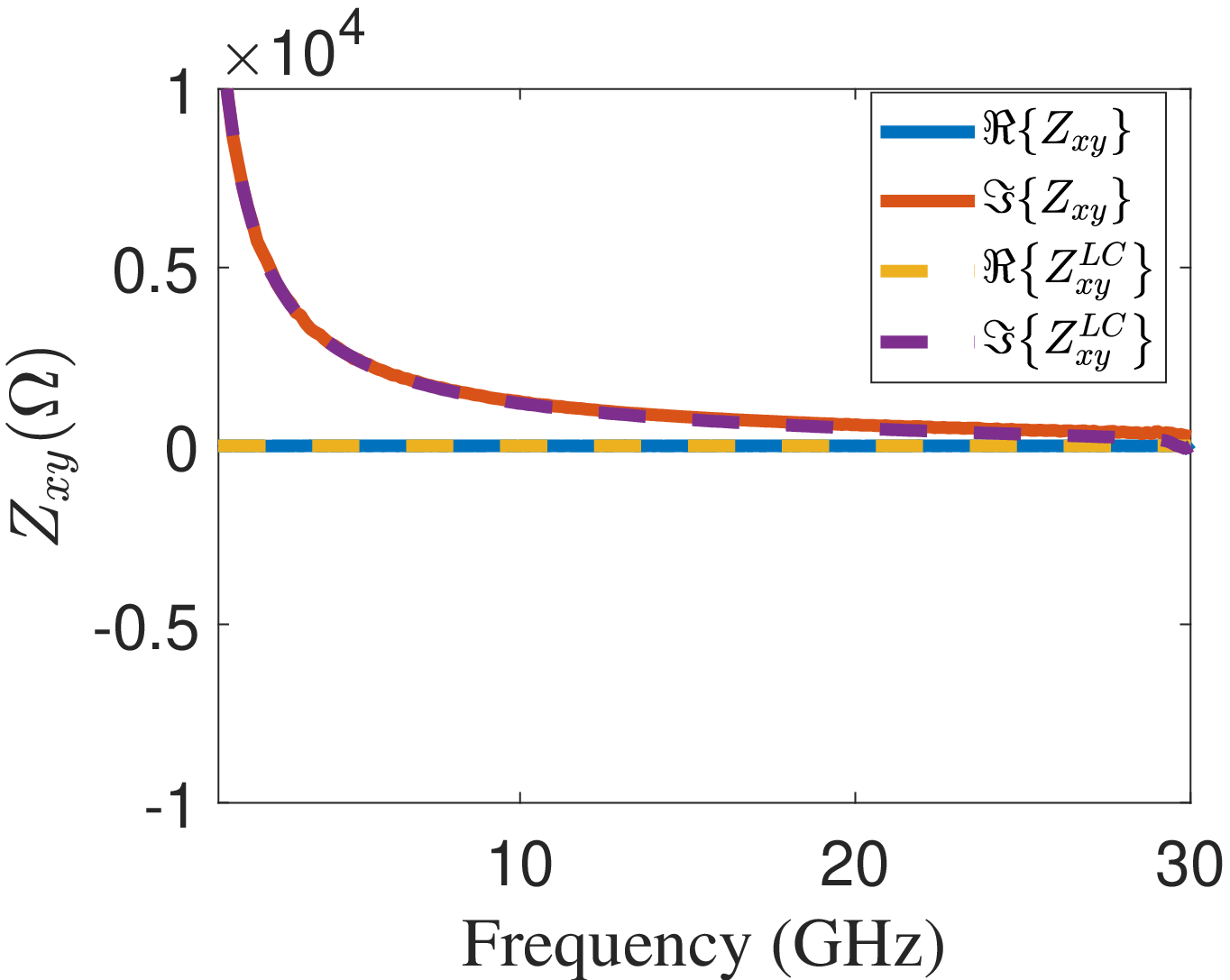}}
   \\
  \subfloat[]{%
        \includegraphics[width=0.45\linewidth]{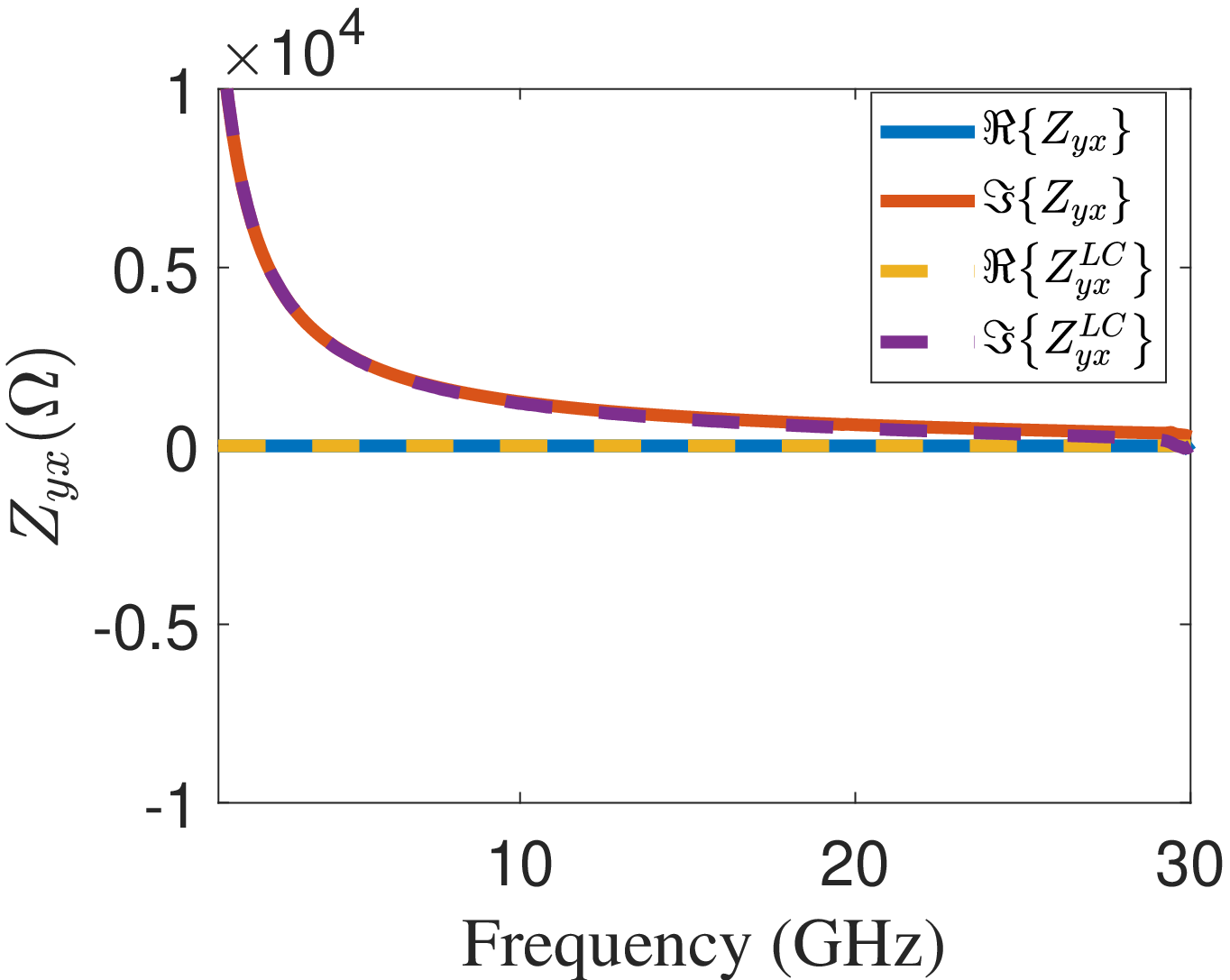}}
    \hfill
  \subfloat[]{%
        \includegraphics[width=0.45\linewidth]{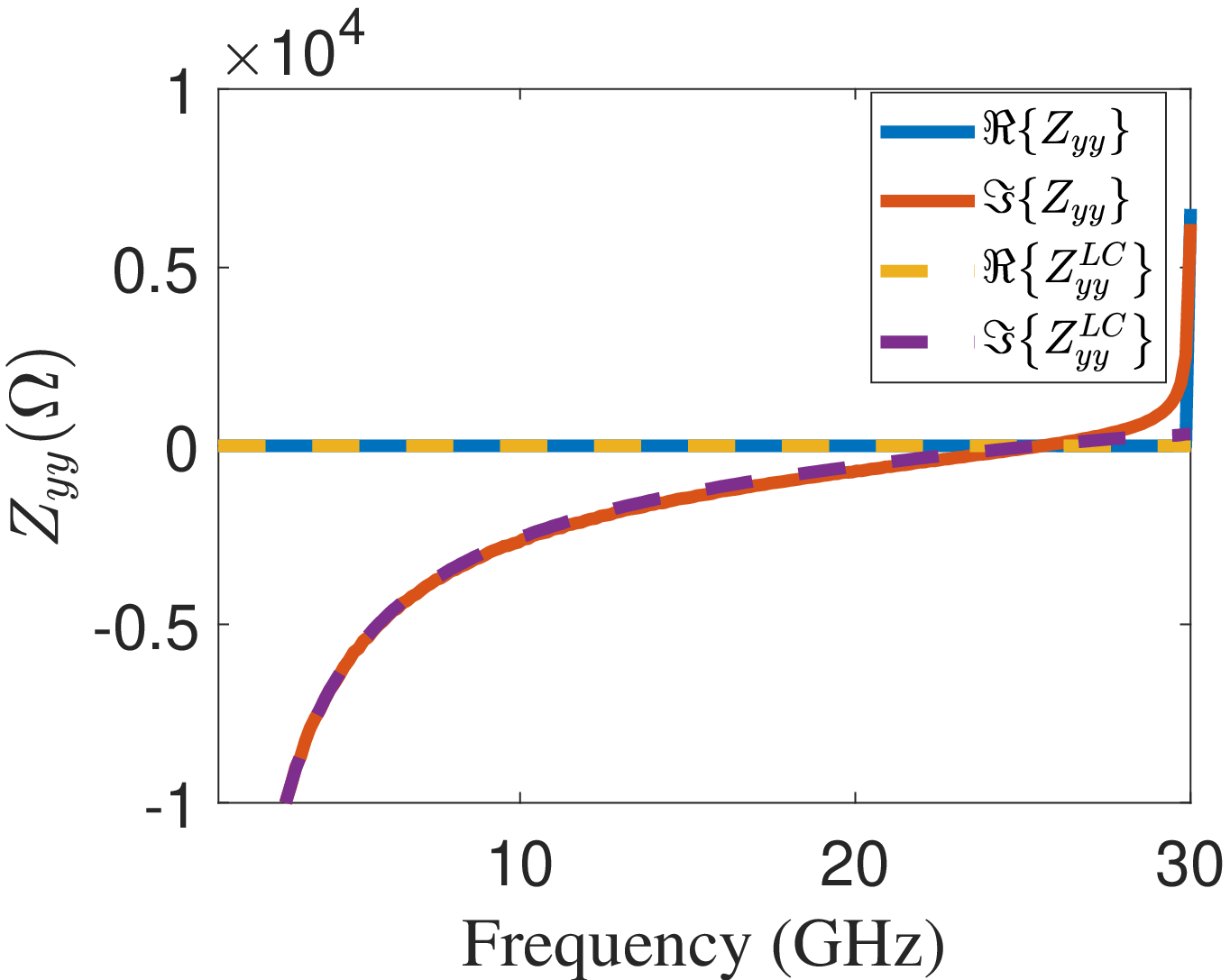}}
      \caption{$ \Re{\lbrace\underline{\underline{Z}}}\rbrace $ and $ \Im{\lbrace\underline{\underline{Z}}}\rbrace $ as a function of the frequency for the case of an FSS based on a symmetric loaded dipole shaped element. The matrix $\underline{\underline{Z}}$ is represented on Cartesian axes. The direction of the impinging electric field is $ (\theta^{inc} = 0^{\circ},\varphi^{inc} = 0^{\circ} ) $. (a) $ Z_{xx} $, (b) $ Z_{xx} $, (c) $ Z_{yx} $,(d) $ Z_{yy} $. The periodicity is equal to \SI{1}{\centi\metre} along planar directions.}
  \label{fig_Zdipole_cartesian} 
\end{figure}

If the end-loading of the dipole is obtained with two dipoles directed towards opposite directions, the $180^\circ$ rotational symmetry is maintained but not the mirror symmetry. In this case, the crystal axis does not coincide with the direction of the dipole and it is not possible to determine the crystal axis based on visual inspection as in the previous case. For a generic FSS element, the crystal axes direction can be derived according to Appendix \ref{Appendix-B}. The crystal axis can be then computed for all the desired frequency points. In the case of the symmetric end loaded dipole, the crystal axes remain stable for the whole investigated frequency range and is approximately $\chi_1 = -27^\circ$, $\chi_2 = 63^\circ$. According to the procedure described in Section \ref{sec_Method}, once the crystal axis is computed, a second simulation is performed along the crystal axis and the diagonalized impedance of the loaded dipole is derived. Finally, the $L_{\chi_1}$ $C_{\chi_1}$ and $L_{\chi_2}$ $C_{\chi_2}$ parameters are calculated for the symmetric end-loaded dipole element. The impedance matrix of the symmetric end-loaded dipole element computed along the crystal axes and its equivalent LC representation is shown in  Fig.~\ref{fig_Zdipole_crystal}. As the diagonalized impedance is rotated back to the initial simulation angle, the correct behaviour of the MoM impedance is obtained also for the off-diagonal elements. The impedance matrix of the symmetrically loaded dipole element, computed at the original simulation angle, say $\varphi^{inc} =0^\circ$, is reported in Fig.~\ref{fig_Zdipole_cartesian}. In conclusion, the equivalent circuit of the element can be again obtained with 5 parameters: ($L_{\chi_1}$ $C_{\chi_1}$, $L_{\chi_2}$ $C_{\chi_2}$ and $\varphi^{rot}$).

\subsection{Asymmetric end-loaded dipole (no symmetry)}
\begin{figure} 
    \centering
  \subfloat[]{%
       \includegraphics[width=0.45\linewidth]{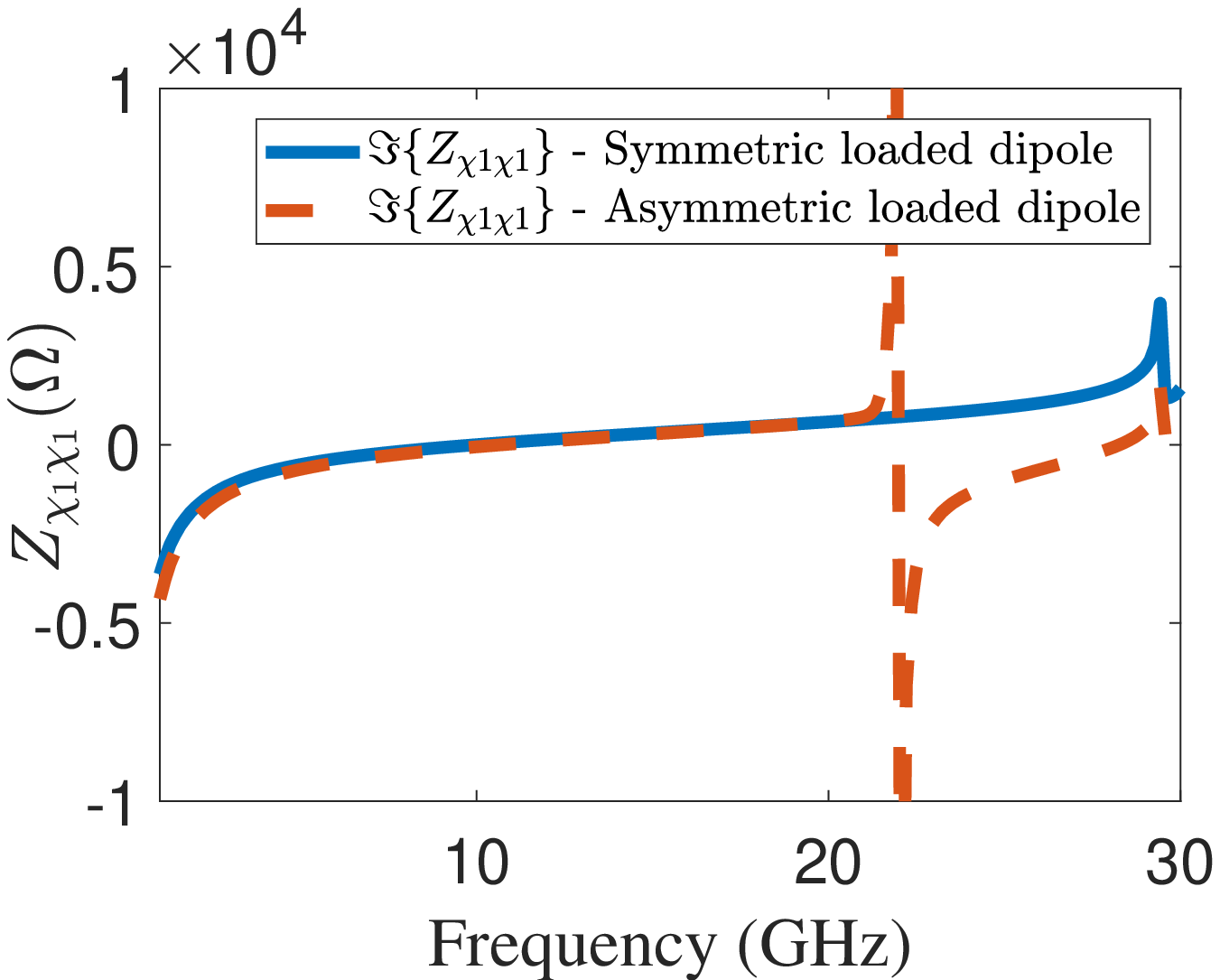}}
    \hfill
  \subfloat[]{%
        \includegraphics[width=0.45\linewidth]{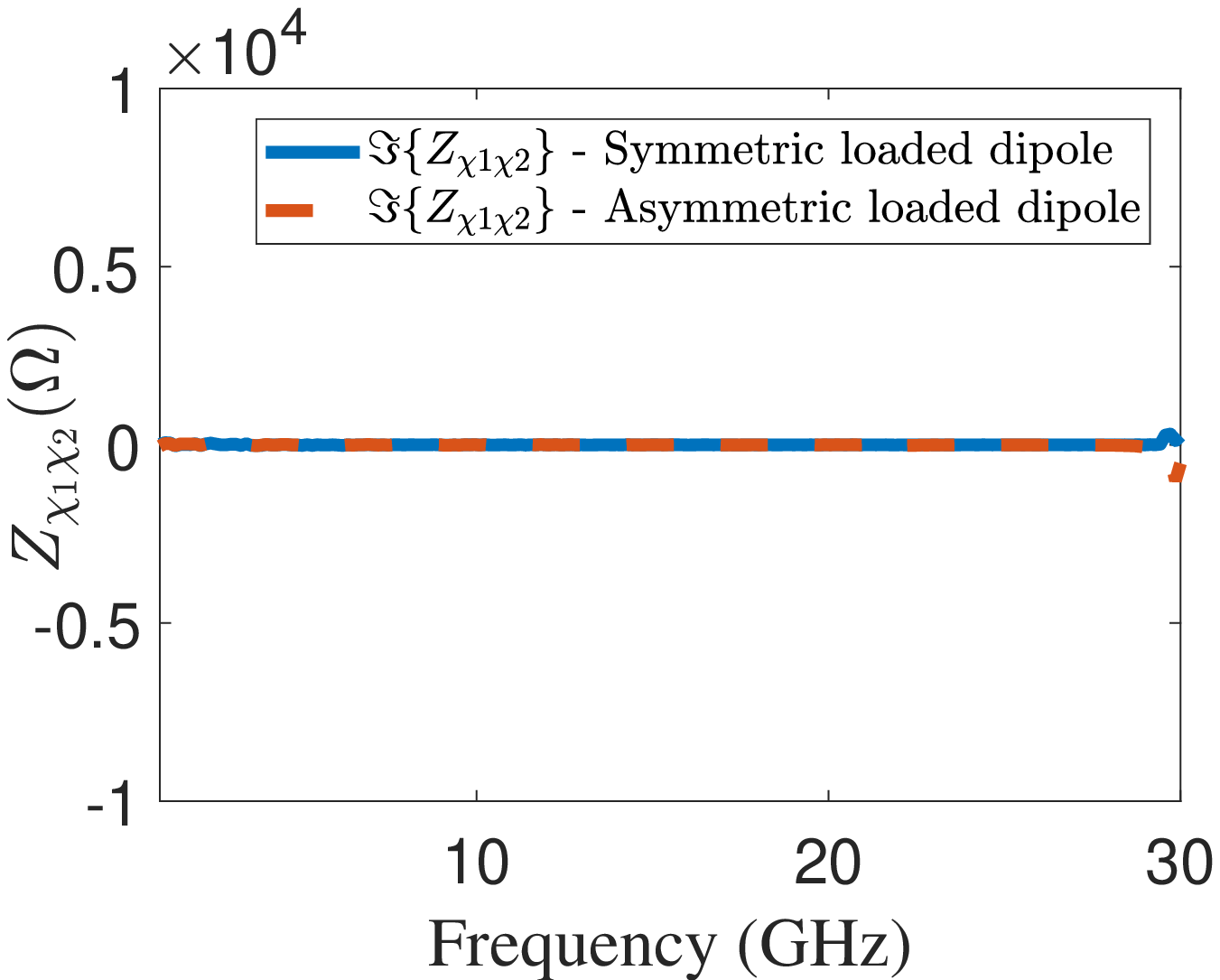}}
   \\
  \subfloat[]{%
        \includegraphics[width=0.45\linewidth]{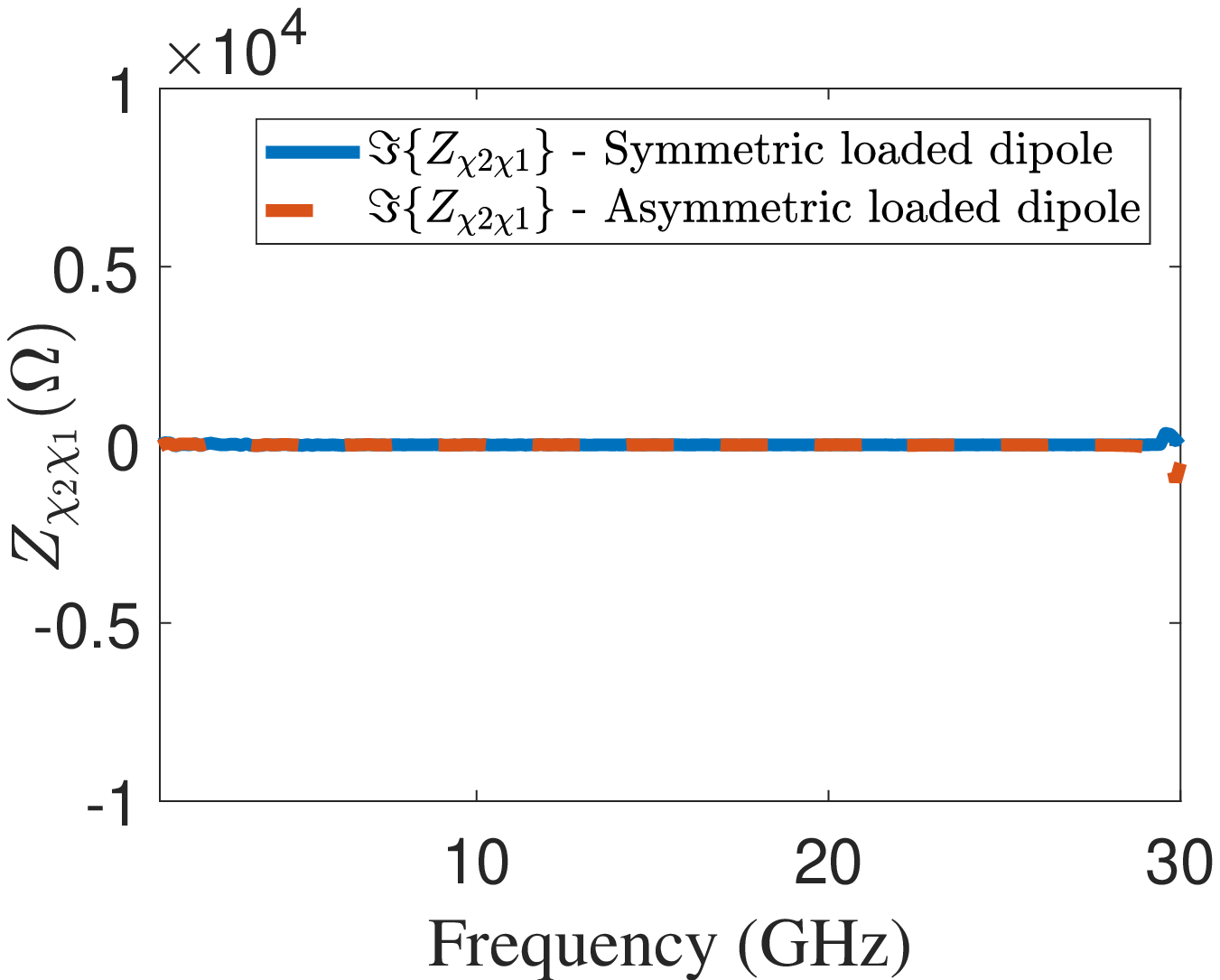}}
    \hfill
  \subfloat[]{%
        \includegraphics[width=0.45\linewidth]{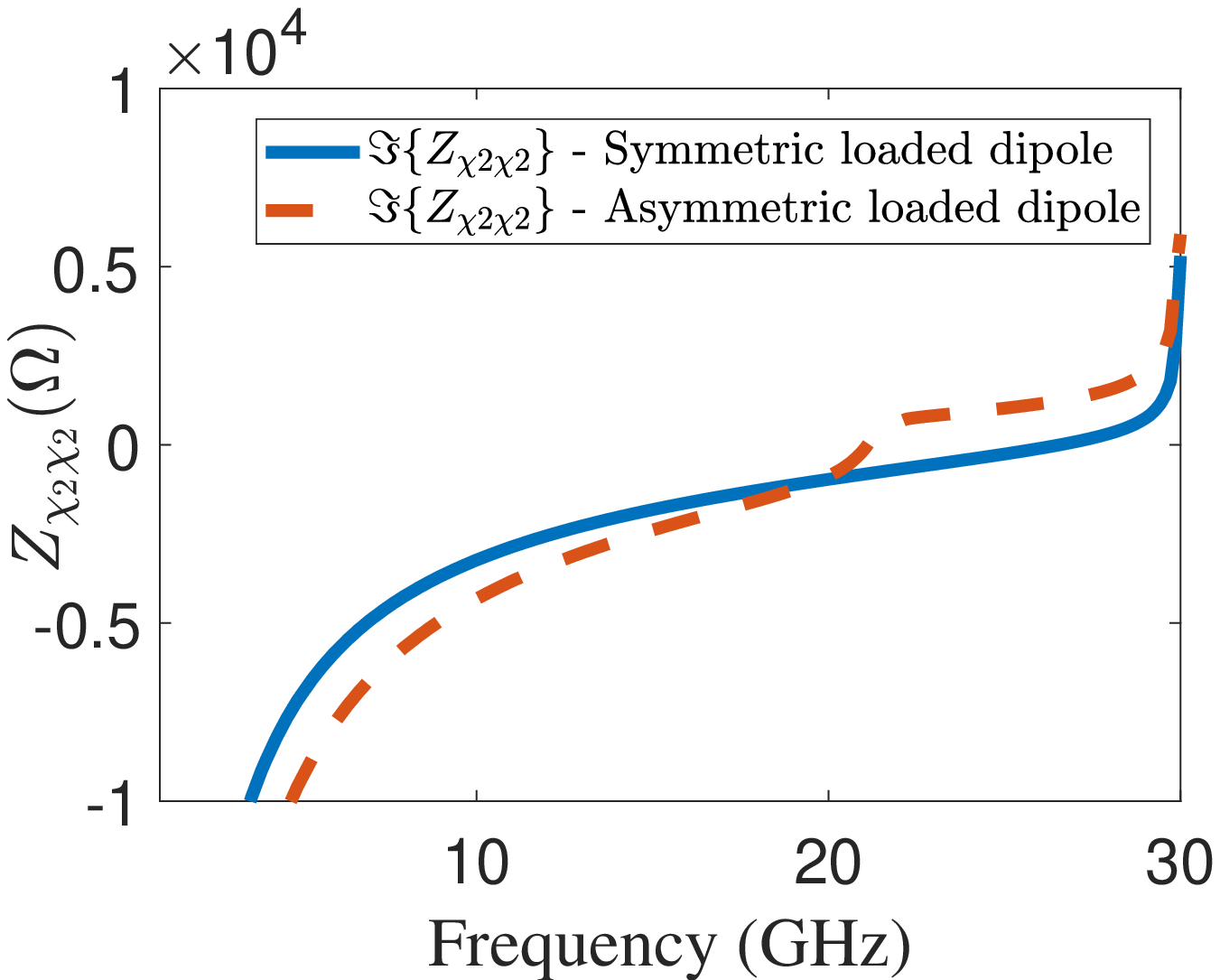}}
      \caption{$ \Im{\lbrace\underline{\underline{Z}}}\rbrace $ as a function of the frequency for a symmetric loaded dipole and the asymmetric loaded dipole . The matrix $\underline{\underline{Z}}$ is represented on crystal axes. The direction of the impinging electric field is $ (\theta^{inc} = 0^{\circ},\varphi^{inc} = \varphi^{rot} ) $. (a) $ Z_{\chi_1 \chi_1} $, (b) $ Z_{\chi_1 \chi_2} $, (c) $ Z_{\chi_2 \chi_1} $,(d) $ Z_{\chi_2 \chi_2} $. The periodicity is equal to \SI{1}{\centi\metre} along planar directions.}
  \label{fig_ZFSS_symm_loaded_vs_asymm_loaded} 
  \vspace{1.2cm}
\end{figure}

\begin{figure}
    \centering
        \includegraphics[width=0.9\linewidth]{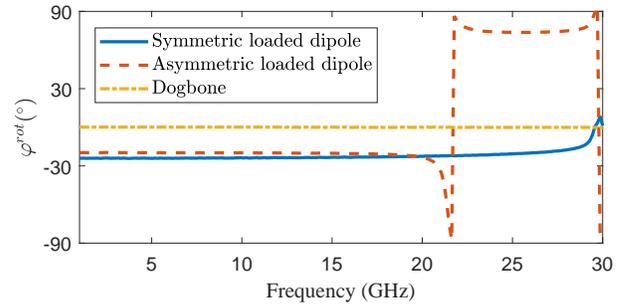}
      \caption{Rotation angle $ \varphi^{rot} $ as a function of the frequency for the symetric loaded dipole, the asymmetric loaded dipole and the dogbone shaped FSS depicted in Fig.~\ref{fig_UnitCells}(c).}
  \label{fig_Crystal_Axis} 
\end{figure}

The last representative example is the asymmetric loaded dipole (no symmetry is applicable) shown in Fig.~\ref{fig_UnitCells}(d). In this case, the steps followed in the previous example remain valid. The behaviour of the four terms of the FSS impedance on the two crystal axes is shown in in  Fig.~\ref{fig_ZFSS_symm_loaded_vs_asymm_loaded}. It is interesting to observe that the extracted crystal axis is not stable in  frequency. The behaviour of the crystal axis as a function of the frequency, compared with the one of the dogbone shape and the symmetric end loaded dipole, is shown in Fig.~\ref{fig_Crystal_Axis}.

While crystal axis is usually stable with frequency for conventional geometries characterized by some symmetry (e.g. dipole, dogbone, loop, Jerusalem cross etc.), it can vary with frequency for FSS elements without any symmetry. However, the main non-linearities arise close to the resonance frequency. In case of frequency variations, it is necessary to represent the frequency varying crystal angle with a polynomial expansion. The derived LC model is therefore valid in the whole analysed frequency range once we pre-emptively diagonalize the $\underline{\underline{Z}}$ matrix frequency by frequency with a different crystal angle. We employed a second order interpolation function in order to fit the behaviour of the crystal axis with frequency. Summarizing, in these particular cases, the anisotropic FSS representation will rely on 4  real numbers ($L_{\chi_1}$ $C_{\chi_1}$, $L_{\chi_2}$ $C_{\chi_2}$) and a second order interpolation function which accounts for the frequency dispersive behaviour of the crystal axis angle. Once these parameters are stored in a database, the behaviour of the anisotropic FSS can be simulated for a generic $\varphi^{rot}$ rotational angle and for a frequency range consistent with a first order equivalent circuit.  

\subsection{Arbitrary unit cell shapes} 
\begin{figure} 
    \centering
  \subfloat[]{%
       \includegraphics[width=0.45\linewidth]{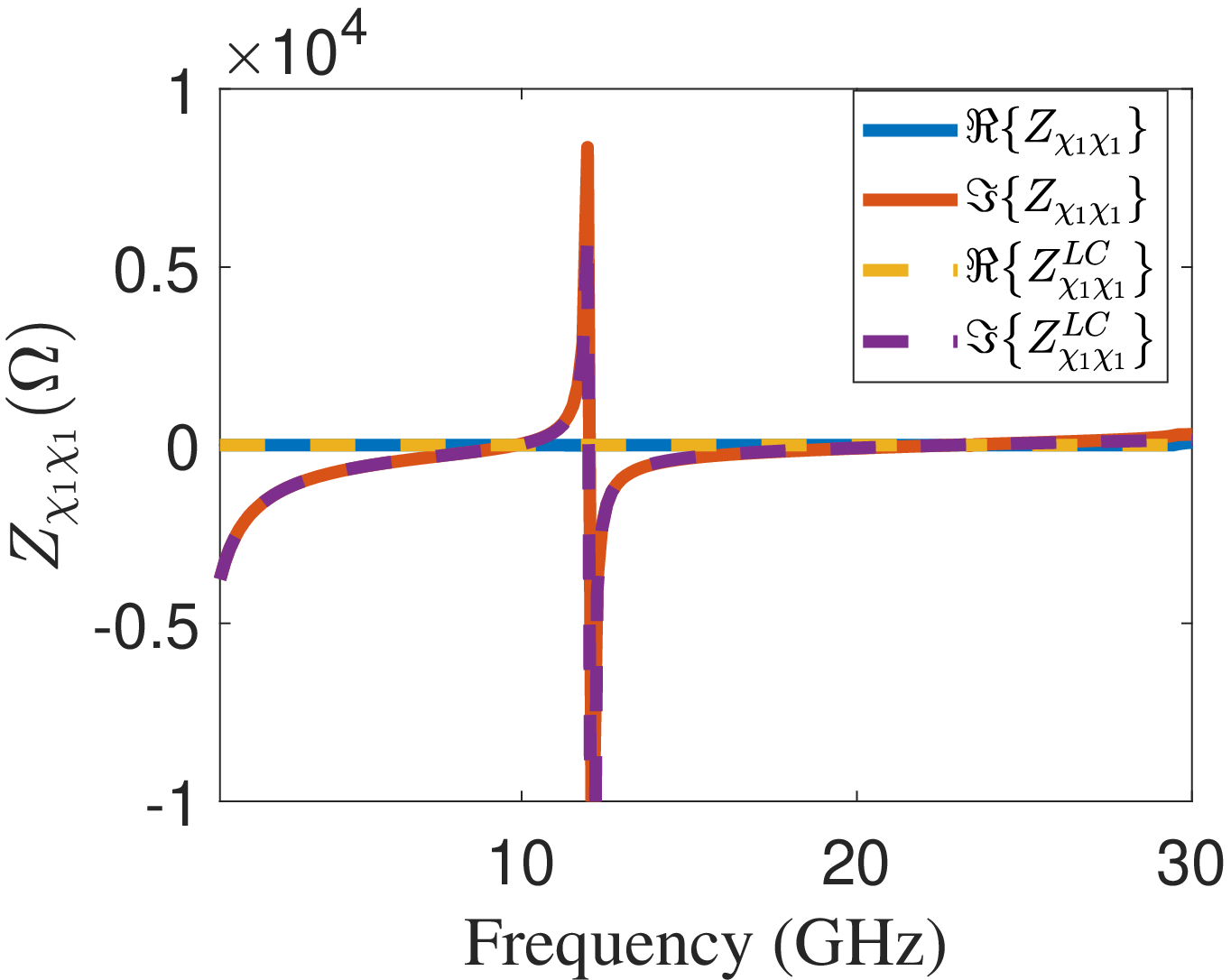}}
    \hfill
  \subfloat[]{%
        \includegraphics[width=0.45\linewidth]{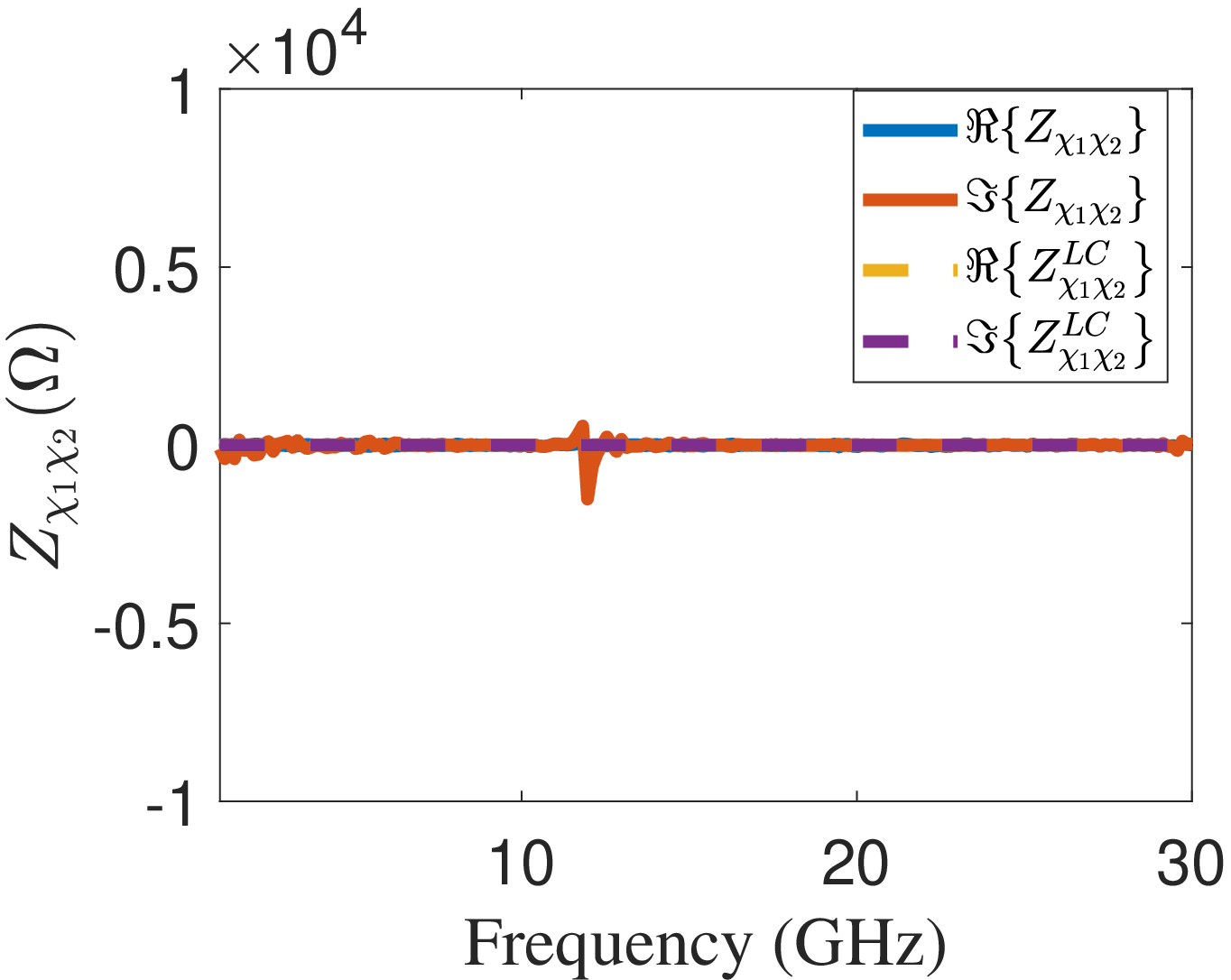}}
   \\
  \subfloat[]{%
        \includegraphics[width=0.45\linewidth]{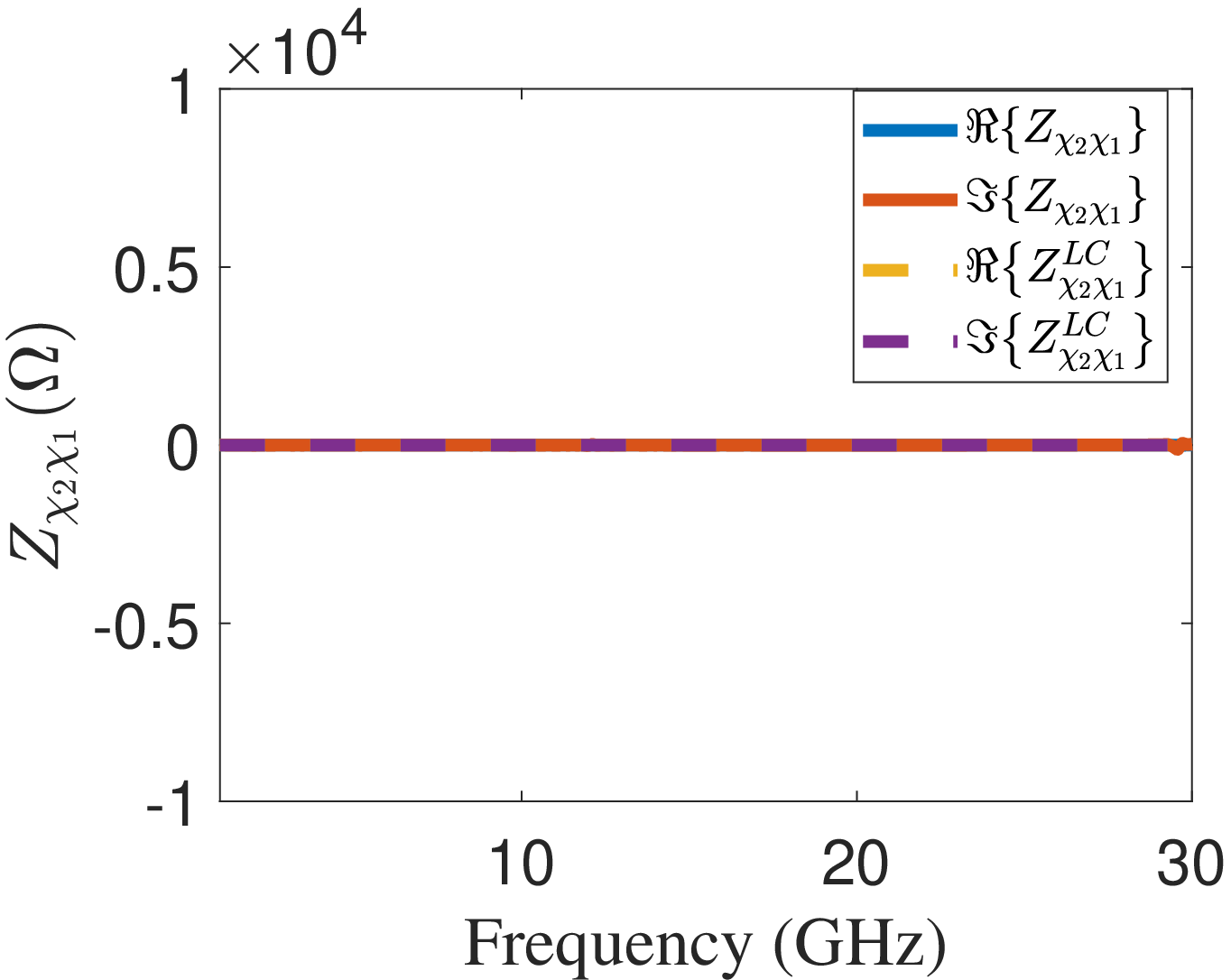}}
    \hfill
  \subfloat[]{%
        \includegraphics[width=0.45\linewidth]{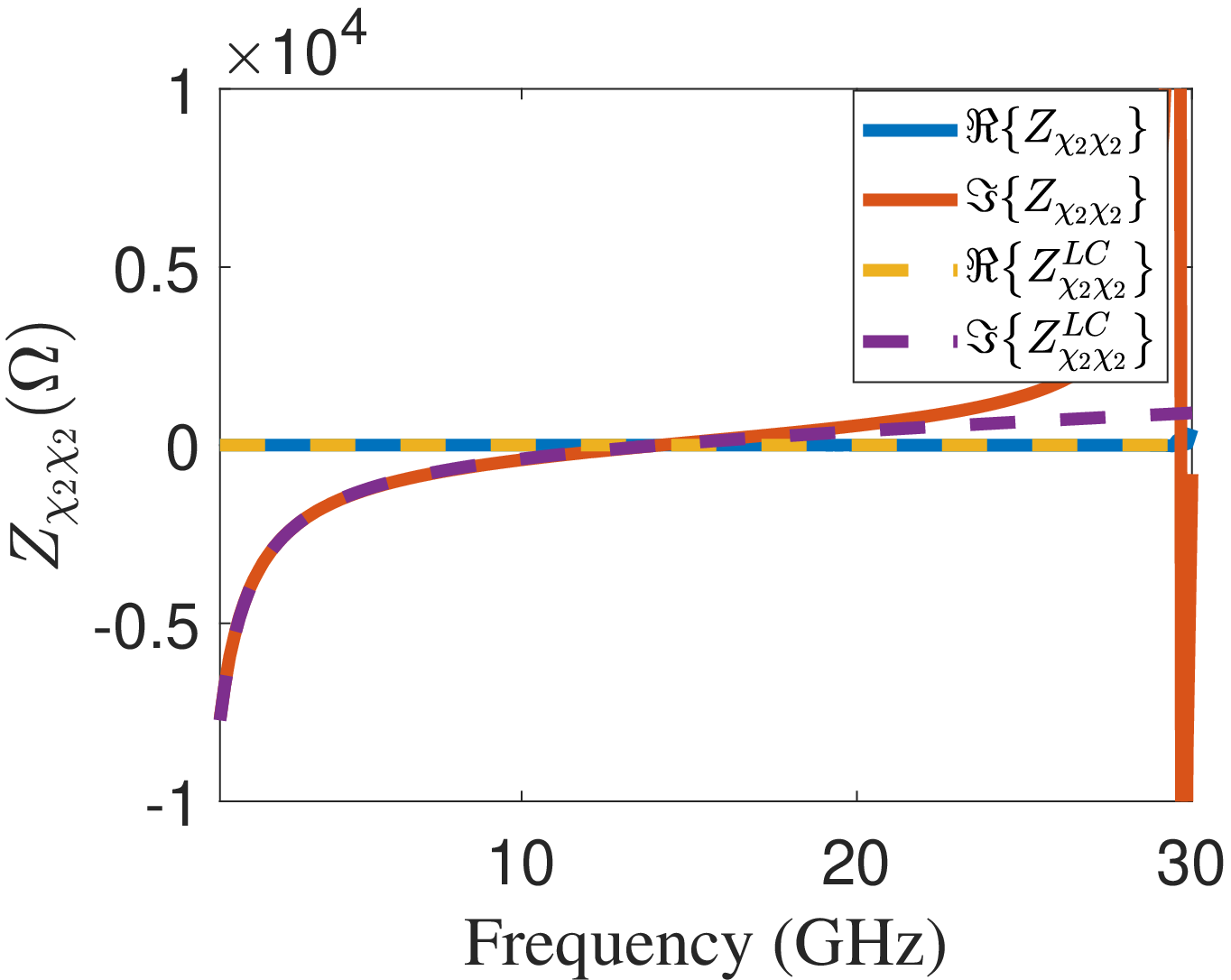}}
      \caption{$ \Re{\lbrace\underline{\underline{Z}}}\rbrace $ and $ \Im{\lbrace\underline{\underline{Z}}}\rbrace $ as a function of the frequency for an FSS based on the element reported in Fig. \ref{fig_UnitCells}(e). The matrix $\underline{\underline{Z}}$ is represented on crystal axes. The direction of the impinging electric field is $ (\theta^{inc} = 0^{\circ},\varphi^{inc} = \varphi^{rot} ) $. (a) $ Z_{\chi_1 \chi_1} $, (b) $ Z_{\chi_1 \chi_2} $, (c) $ Z_{\chi_2 \chi_1} $,(d) $ Z_{\chi_2 \chi_2} $. The periodicity is equal to \SI{1}{\centi\metre} along planar directions.}
  \label{fig_padilla1_impedance} 
\end{figure}

\begin{figure} 
    \centering
  \subfloat[]{%
       \includegraphics[width=0.45\linewidth]{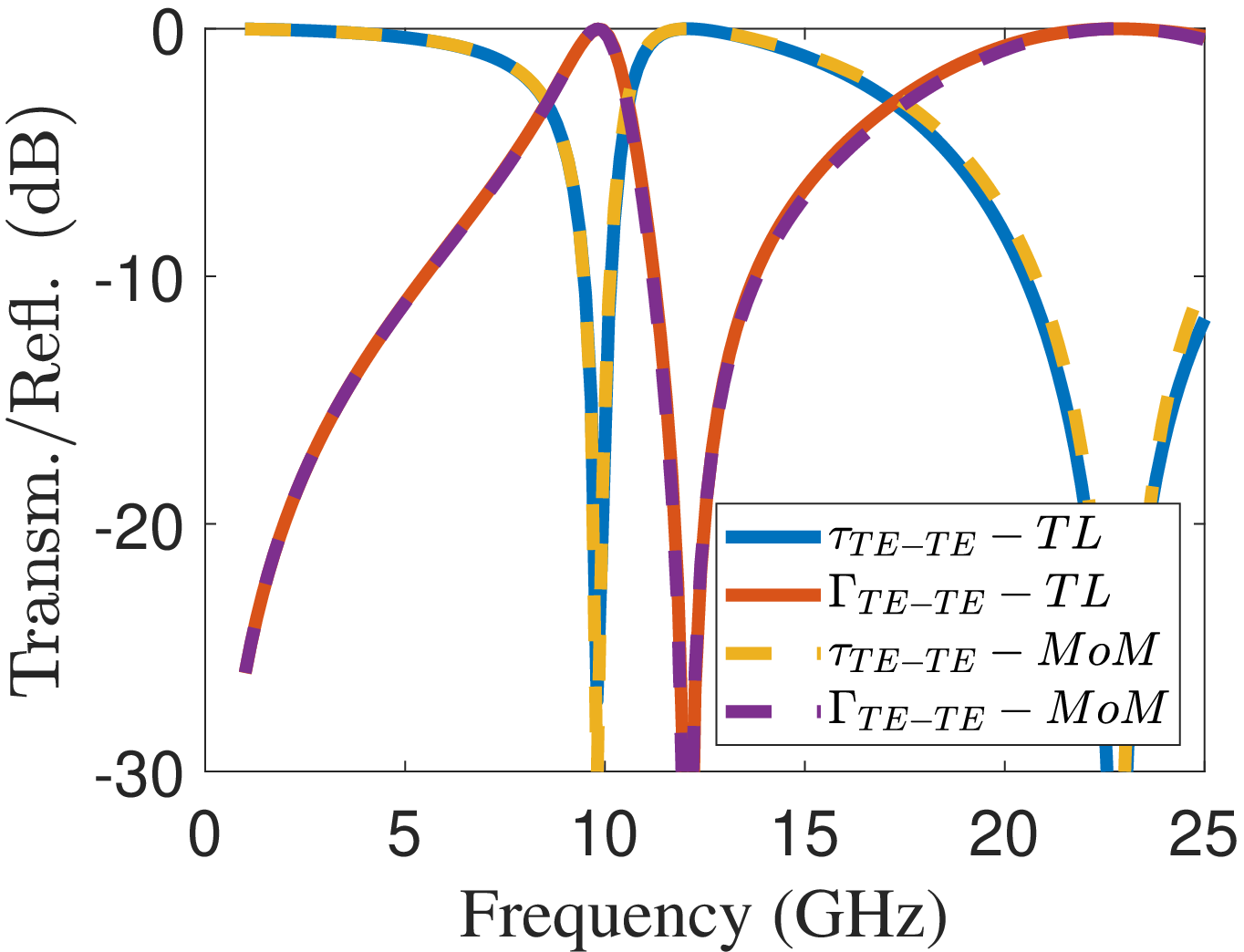}}
    \hfill
  \subfloat[]{%
        \includegraphics[width=0.45\linewidth]{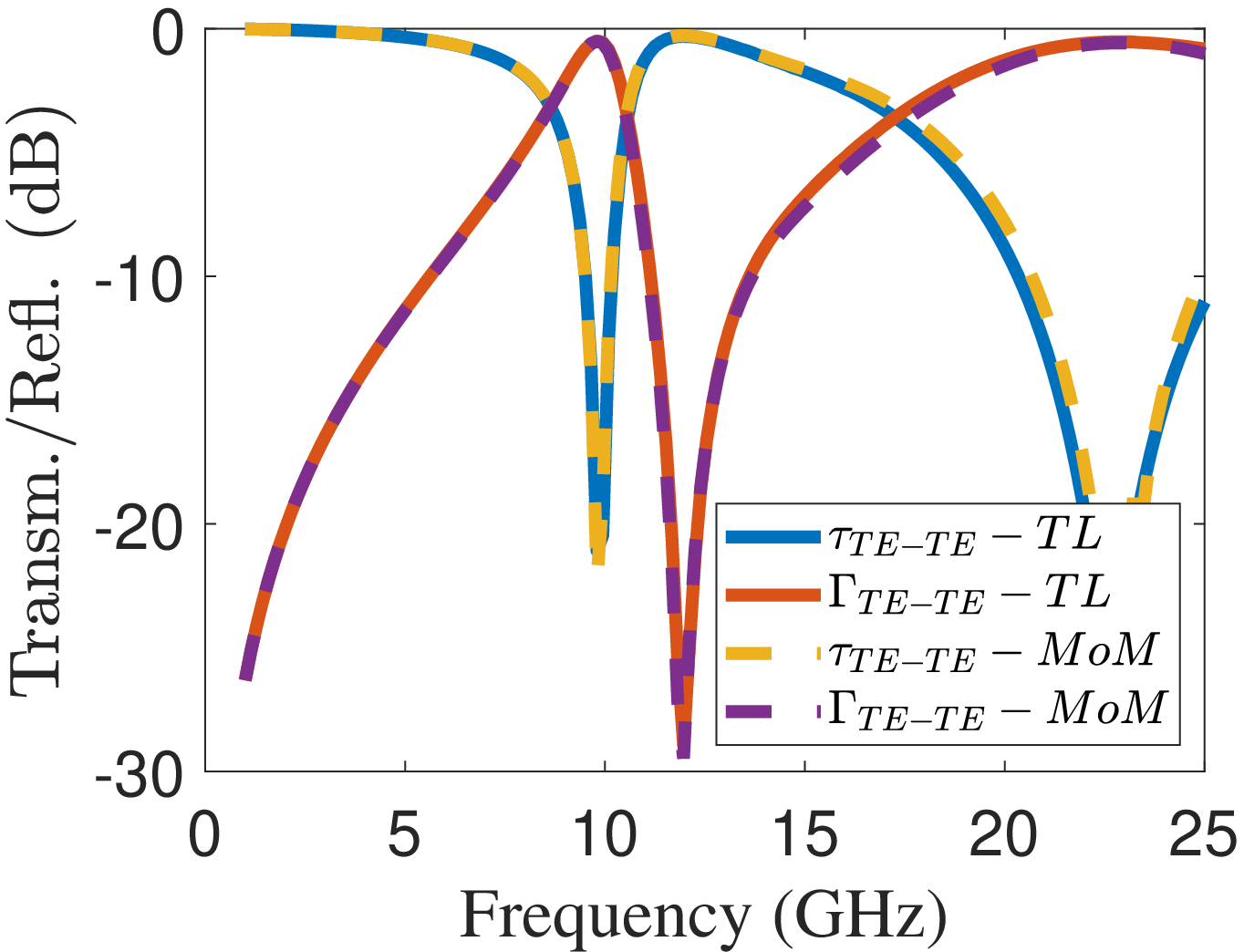}}
   \\
  \subfloat[]{%
        \includegraphics[width=0.45\linewidth]{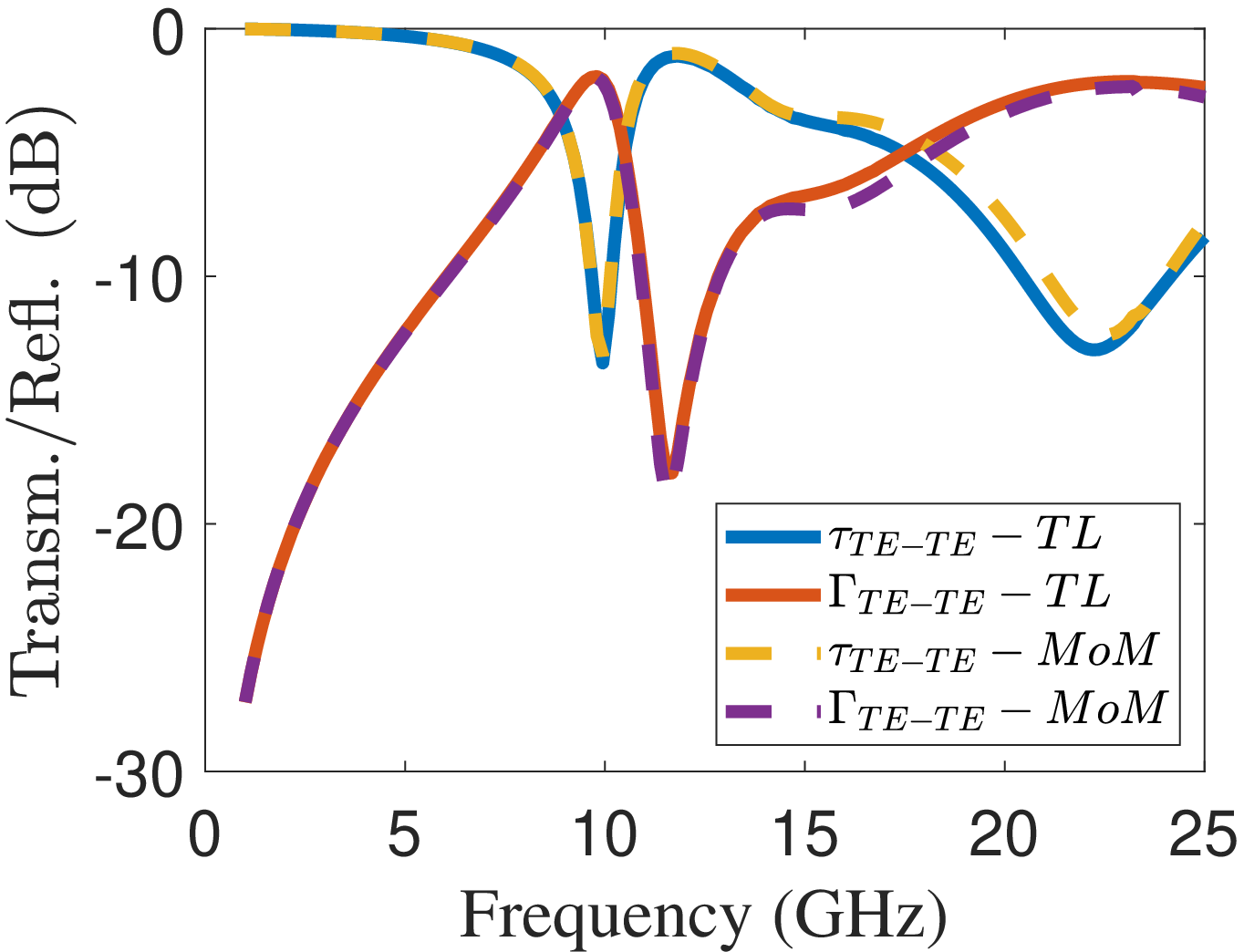}}
    \hfill
  \subfloat[]{%
        \includegraphics[width=0.45\linewidth]{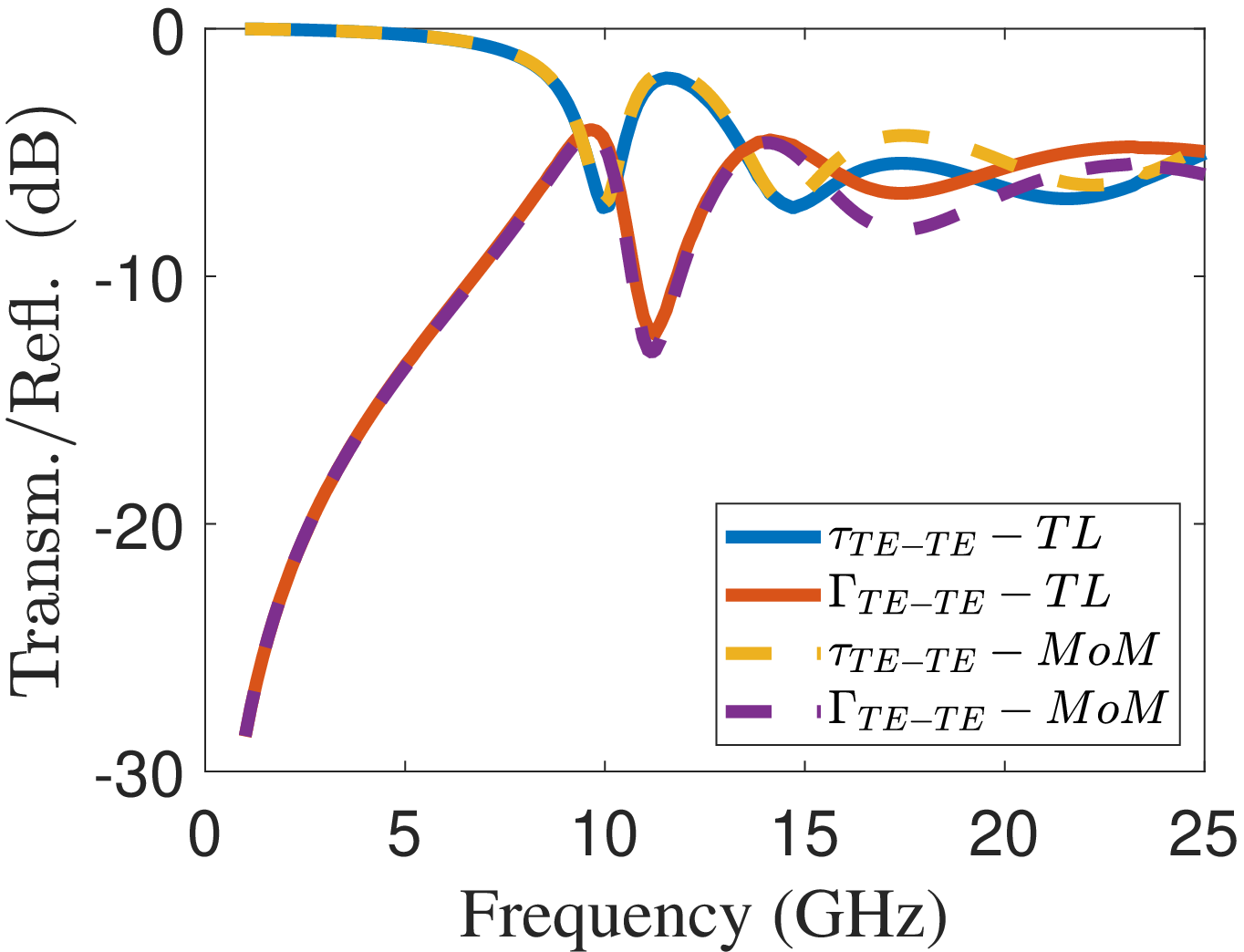}}
      \caption{Reflection and transmission coefficients for an FSS based on the element reported in Fig. \ref{fig_UnitCells}-(e) for different polarization angles ((a) $\varphi^{inc}=0^{\circ}$, (b) $\varphi^{inc}=  15^{\circ}$, (a) $\varphi^{inc} = 30^{\circ}$, (d) $\varphi^{inc} = 45^{\circ}$). The periodicity is equal to \SI{1}{\centi\metre} along planar directions.}
  \label{fig_padilla1_reflection_trasnmission} 
\end{figure}

Considering the generality of the proposed approach, the circuit model analysis can be applied to any unit cell shape. In Fig.~\ref{fig_UnitCells}(e), an interesting unit cell topology employed to design devices such as metamaterials sensors and metamaterials absorbers  \cite{Chen2010} is shown. The behavior of the four terms of the FSS impedance on the two crystal axes for the shape of  Fig.~\ref{fig_UnitCells}(e) is reported in Fig. \ref{fig_padilla1_impedance}. Given the mirror symmetry with respect to $ y $-axis, the crystal axes of this shape coincide with the \textit{x} and \textit{y} axes. The behavior of the impedance on the \textit{x}-axis is not well predicted by a single LC series circuit. For this reason, we used a circuit model comprising two LC series circuits connected in parallel for one of the two crystal axes (for the other crystal axis the model remains a simple LC series circuit). As a consequence, the number of LC parameters for representing the FSS element increases to $7$. This circuit topology allows for the accurate estimation of the impedance extracted from the full wave analysis. Once the four terms of the impedance on the crystal axis are computed, the metasurface can be analysed for an arbitrary azimuth angle $\varphi^{inc}$ by using the spectral rotation of the matrix as shown in (\ref{eq:XaxisRot}) and (\ref{eq:Rot}). This means that the FSS reflection and transmission coefficients for an EM wave impinging with a generic polarization state can be computed without any additional computation effort. For instance, in Fig. \ref{fig_padilla1_reflection_trasnmission} the reflection coefficient of the FSS topology computed with the LC model formulation is compared with MoM simulations for different azimuth angles showing a good agreement. 

\section{Syntesis of Metasurfaces}
Other than an improved visualization and interpretation of the physical phenomena determining the FSS frequency response, the retrieved circuit inductances and capacitances represent a topological representation of a certain shape and can be stored in a database. The circuit topology represents a useful bridge between the element geometry and its impedance matrix representation and can be used for the synthesis of FSS or to apply modifications to already designed elements. Moreover, if a specific FSS element stored in the database should be simulated with a different periodicity or embedded in a dielectric stackup, for instance for optimizing a filter response or its polarization converting properties, there is no need to repeat the full-wave simulation of the periodic surface. It is indeed sufficient to correct the LC parameters with some simple relations which take into account the change of periodicity or the presence of dielectric layers in close proximity of the spatial filter. Subsequently, it is possible to derive the modified FSS response instantaneously using a TL model \cite{costa_efficient_2012}.

\subsection{FSS periodicity}
\begin{figure} 
    \centering
  \subfloat[]{%
       \includegraphics[width=0.45\linewidth]{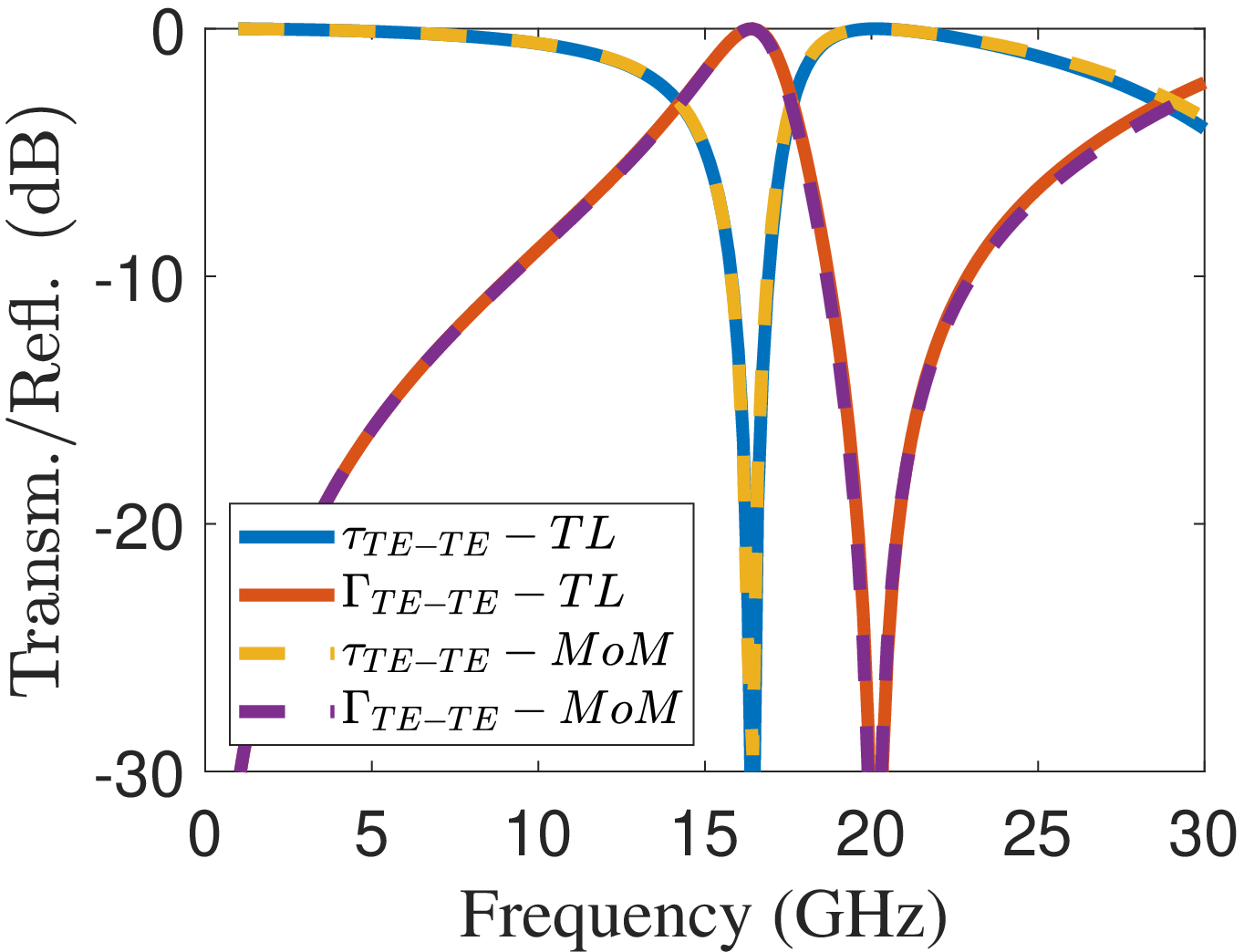}}
    \hfill
  \subfloat[]{%
        \includegraphics[width=0.45\linewidth]{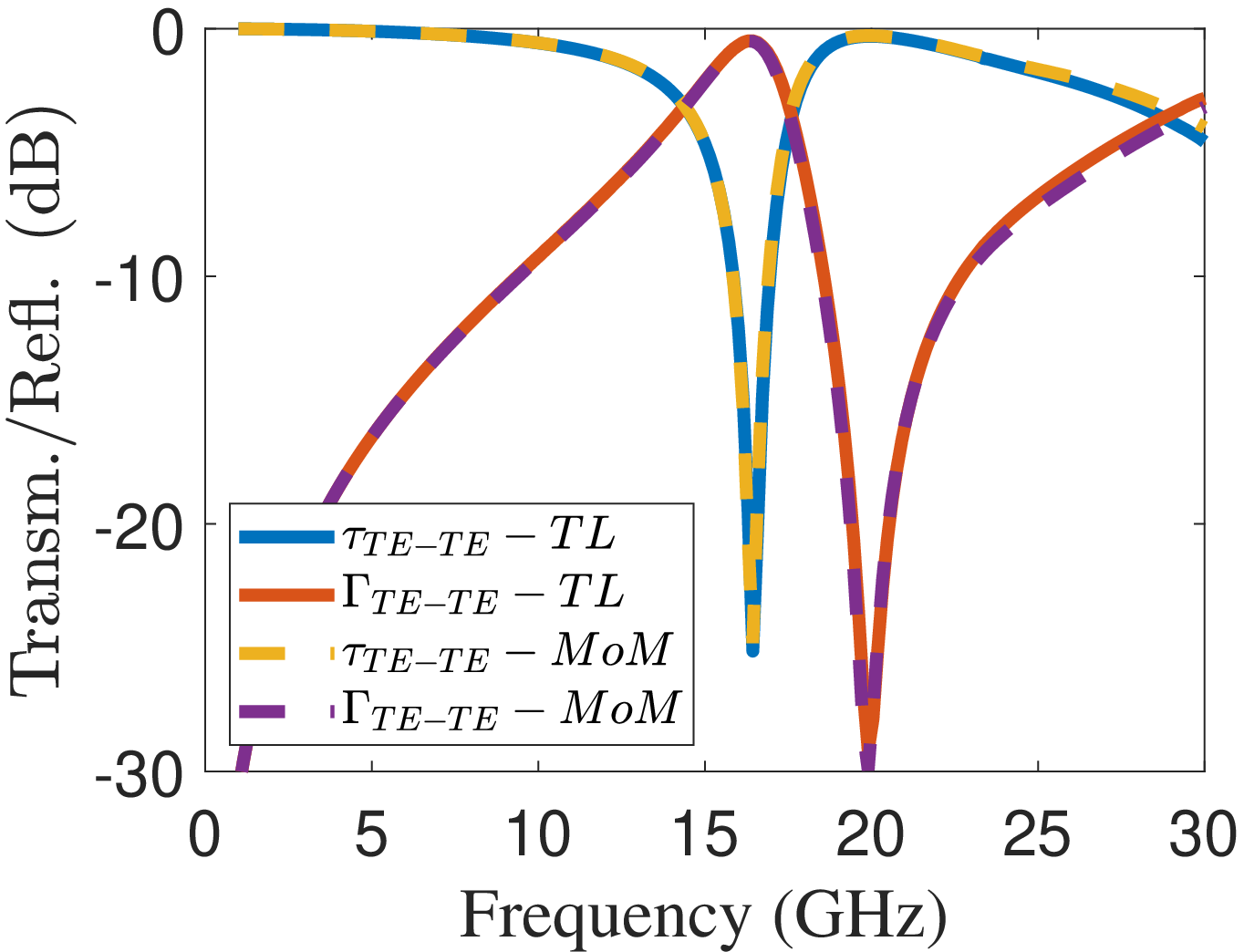}}
   \\
  \subfloat[]{%
        \includegraphics[width=0.45\linewidth]{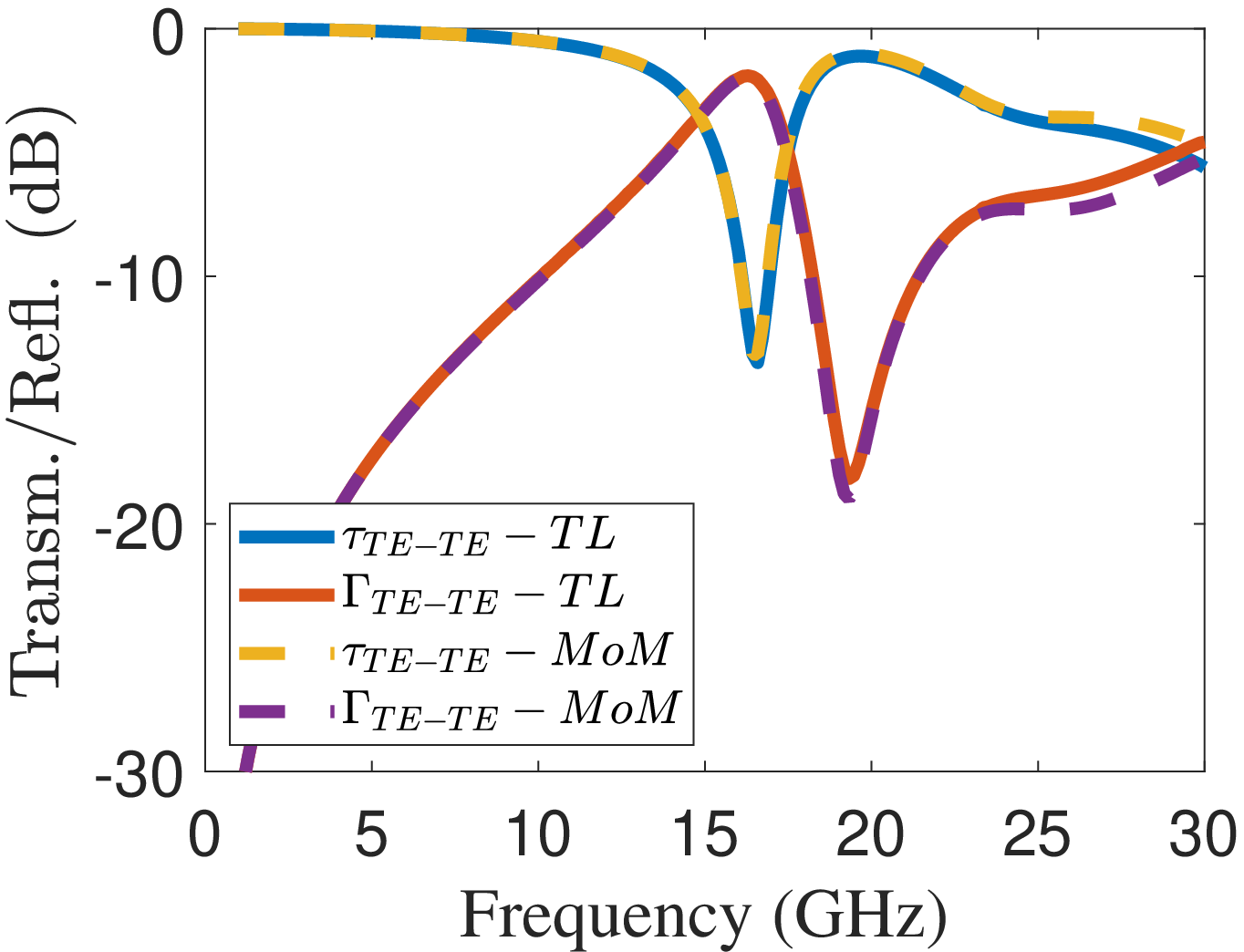}}
    \hfill
  \subfloat[]{%
        \includegraphics[width=0.45\linewidth]{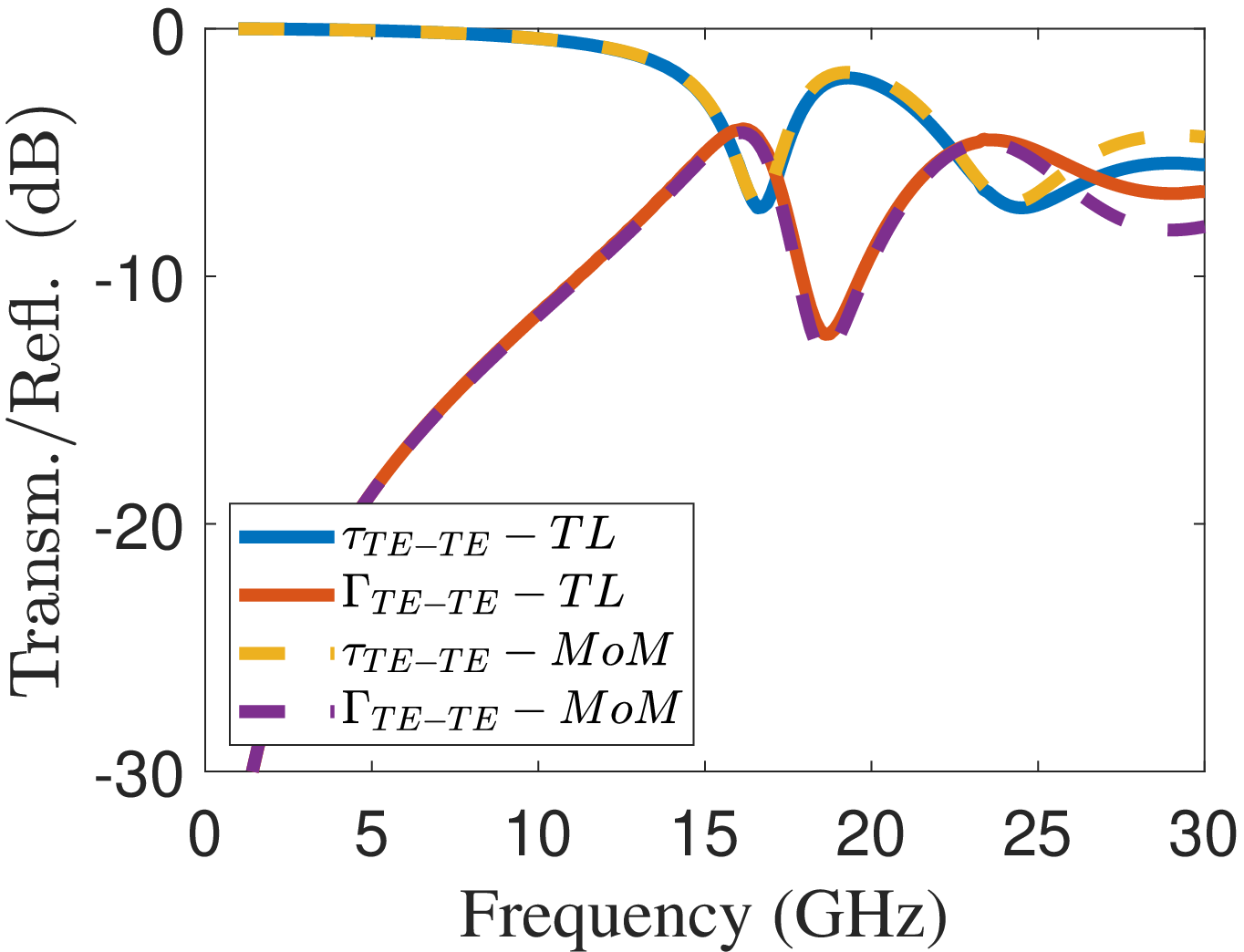}}
      \caption{Reflection and transmission coefficients for an FSS based on the element reported in Fig. \ref{fig_UnitCells}-(e) for different polarization angles ((a) $\varphi^{inc}=0^{\circ}$, (b) $\varphi^{inc}=  15^{\circ}$, (a) $\varphi^{inc} = 30^{\circ}$, (d) $\varphi^{inc} = 45^{\circ}$). The periodicity is equal to \SI{0.6}{\centi\metre} along planar directions.}
  \label{fig_padilla1_reflection_transmission_scaled} 
\end{figure}

Regarding the FSS periodicity, it is sufficient to scale the LC values by the ratio between the new periodicity and the periodicity used to compute the LC model \cite{costa_efficient_2012}. For instance,the FSS inductance is scaled as follows: 

\begin{equation} \label{eq_period_rescale}
 L_{scaled} = L_0
\frac{D_{scaled}}{D_0}
\end{equation}

where $D_{scaled}$ is for the periodicity of the scaled FSS, and $D_0$ and $L_0$ are the periodicity and the inductance of the FSS stored in the database. Similarly, the FSS capacitance is scaled with the same approach.
Clearly, this rescaling cannot be performed without knowing the circuit topology. As an example, in Fig. \ref{fig_padilla1_reflection_transmission_scaled} the reflection and transmission coefficients of the FSS topology reported in Fig. \ref{fig_UnitCells}(e) scaled to $ 0.6 $ with respect to the original period of \SI{1}{\centi\metre} are shown. The circuit simulation is carried out with the parameters of the FSS scaled according to relation (\ref{eq_period_rescale}). The TL model is compared with MoM simulations for different azimuth angles showing a good agreement. 

\subsection{Effect of Dielectric layers}
\label{subsec:Dielectric_layers}
The LC parameters stored in the database can be also used for analysing the FSS embedded within dielectric layers. 
When dielectric substrates are placed in close proximity of FSS, the FSS capacitance can be corrected by multiplying it by the effective permittivity of the stackup \cite{costa_efficient_2012, costa2014overview} ($C_{diel} = C_0 \times \varepsilon_{r_{eff}}$). The effective permittivity can be computed by averaging the effective permittivity towards upper and lower dielectrics:

\begin{equation} \label{eq_eps_eff}
 \varepsilon_{r_{eff}}= \frac{\varepsilon_{r_{eff-up}}+\varepsilon_{r_{eff-down}}}{2}
\end{equation}

If the substrate and superstate are enough thick, the effective permittivity can be computed by simply averaging the dielectric permittivity of the two dielectrics. However, if the substrates are thinner than $0.3 D$ \cite{costa_efficient_2012}, where $D$ is the periodicity of the periodic surface, the effective permittivity depends also on the thickness of the layers. The expression given in \cite{costa_efficient_2012} can used in this case:  

\begin{equation} \label{eq_eps_eff_up}
 \varepsilon_{r_{eff-up}}= \varepsilon_{r-up}+(1-\varepsilon_{r-up}) e^{\frac{-\gamma d_{up}}{D}}
\end{equation}

where $d_{up}$ represents the thickness of the upper layer and $\gamma$ is a coefficient which takes into account the shape of the element \cite{costa_efficient_2012}. $ \varepsilon_{r_{eff-down}}$ is computed with the same approach of  eq. (\ref{eq_eps_eff_up}).

The dispersive effects of dielectric layers are taken into account in the TL model \cite{zhao2012twisted}.
Indeed, in case of two dielectric layers surrounding the metasurface, the ABCD matrix is computed by taking into account the dielectric properties:

\begin{equation}\label{eq:Mdiel}
\begin{bmatrix}
 \underline{\underline {\text{A}}} & \underline{\underline B}  \hfill \\
  \underline{\underline C} & \underline{\underline D}
\end{bmatrix} =  \underline{\underline{M}}^{diel_1} \underline{\underline{M}}^{FSS} \underline{\underline{M}}^{diel_2}
\end{equation}

\setlength{\arraycolsep}{3pt}
	
\begin{equation}\label{eq:ABCD_diel}
         \underline{\underline{M}}^{diel_i} = \begin{bmatrix}
\cos \left( {{k_{zi}}{d_i}} \right)\underline{\underline {\text{I}}} & -j \sin \left( {{k_{zi}}{d_i}} \right)\zeta _i^{TE/TM} \underline{\underline n}  \hfill \\[5pt]
  -j \dfrac{{\sin \left( {{k_{zi}}{d_i}} \right)}}{{\zeta _i^{TE/TM}}}\underline{\underline {\text{n}}} & \cos \left( {{k_{zi}}{d_i}} \right) \underline{\underline I}
                  				\end{bmatrix}
\end{equation}

\setlength{\arraycolsep}{6pt}  

where $\zeta _0^{TE/TM}$ and $\zeta _i^{TE/TM}$ are the impedances, for TE or TM polarization, of the equivalent transmission line for free space and for the $i^{th}$ dielectric layer, respectively. The impedances for TE and TM polarization read:

\begin{equation}\label{eq_imepdance_TE_TM}
\zeta_i^{TE} = \frac{\omega\mu_0\mu_i}{k_{zi}}, \zeta_i^{TM} = \frac{k_{zi}}{\omega\varepsilon_0\varepsilon_i}
\end{equation}

where ${k_{zi}}=\sqrt{{k_0\varepsilon_i\mu_i}^2-{k_t}^2}$ represents the propagation constant along the normal direction inside the medium $i$ with $k_t=k_0sin(\theta^{inc})$. $\varepsilon_0$ and $\mu_0$ represent the dielectric permittivity and the magnetic permeability of free space whereas
$\varepsilon_i$ and $\mu_i$ represent the relative dielectric permittivity and the relative magnetic permeability of the dielectric medium. $d_i$ represents the thickness of the dielectric. Once  the ABCD matrix is computed, the scattering matrix is calculated according to (\ref{eq:S_matrix}). The TL approach can be adopted also for computing the oblique incidence behaviour of the metasurface embedded within dielectrics but, with this approach, the spatial dispersion of metasurface impedance is neglected \cite{luukkonen2009effects,luukkonen2010experimental}. On the contrary, the spatially dispersive effects of the dielectrics are taken into account by the TL model.

\begin{figure} 
    \centering
  \subfloat[]{%
       \includegraphics[width=0.45\linewidth]{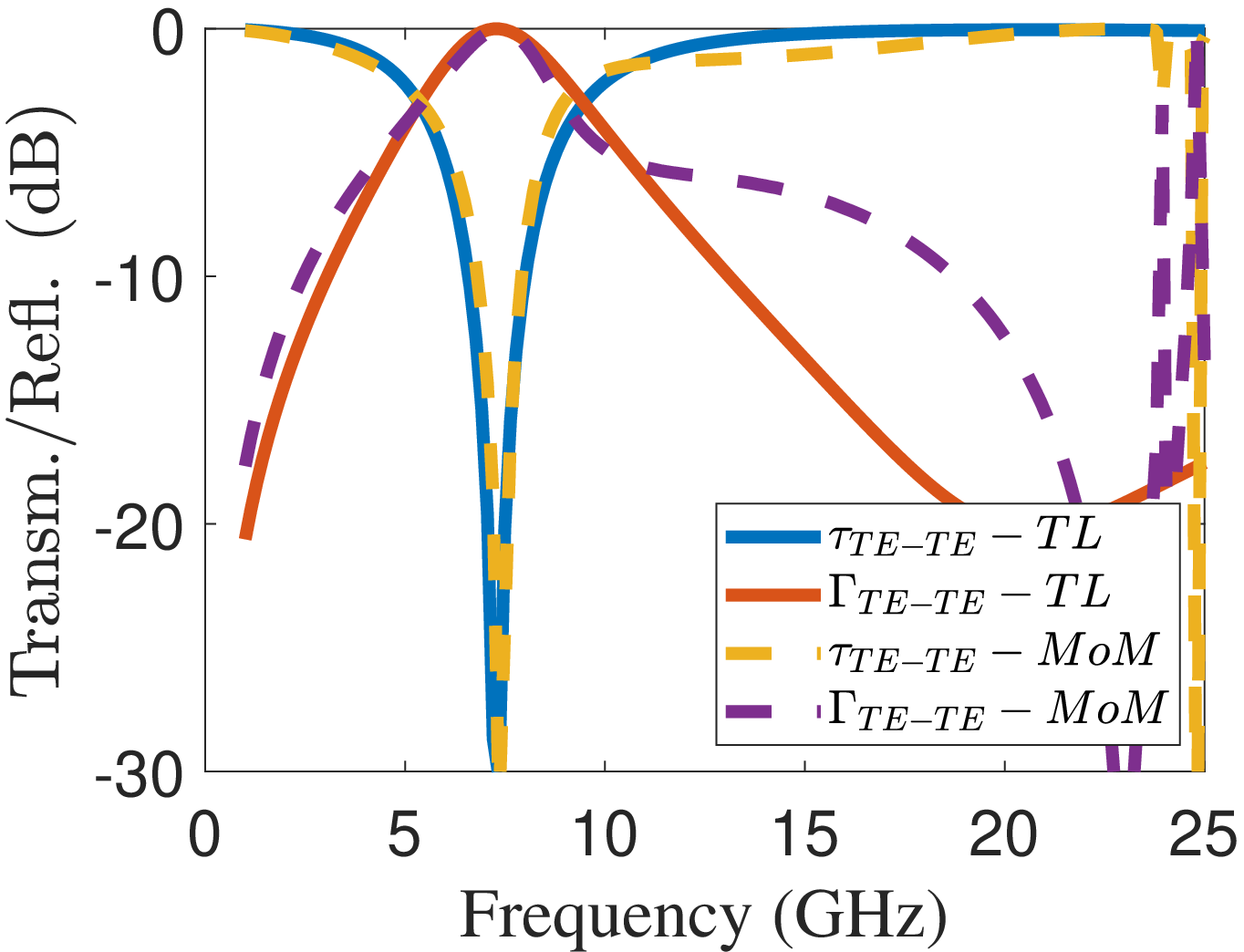}}
    \hfill
  \subfloat[]{%
        \includegraphics[width=0.45\linewidth]{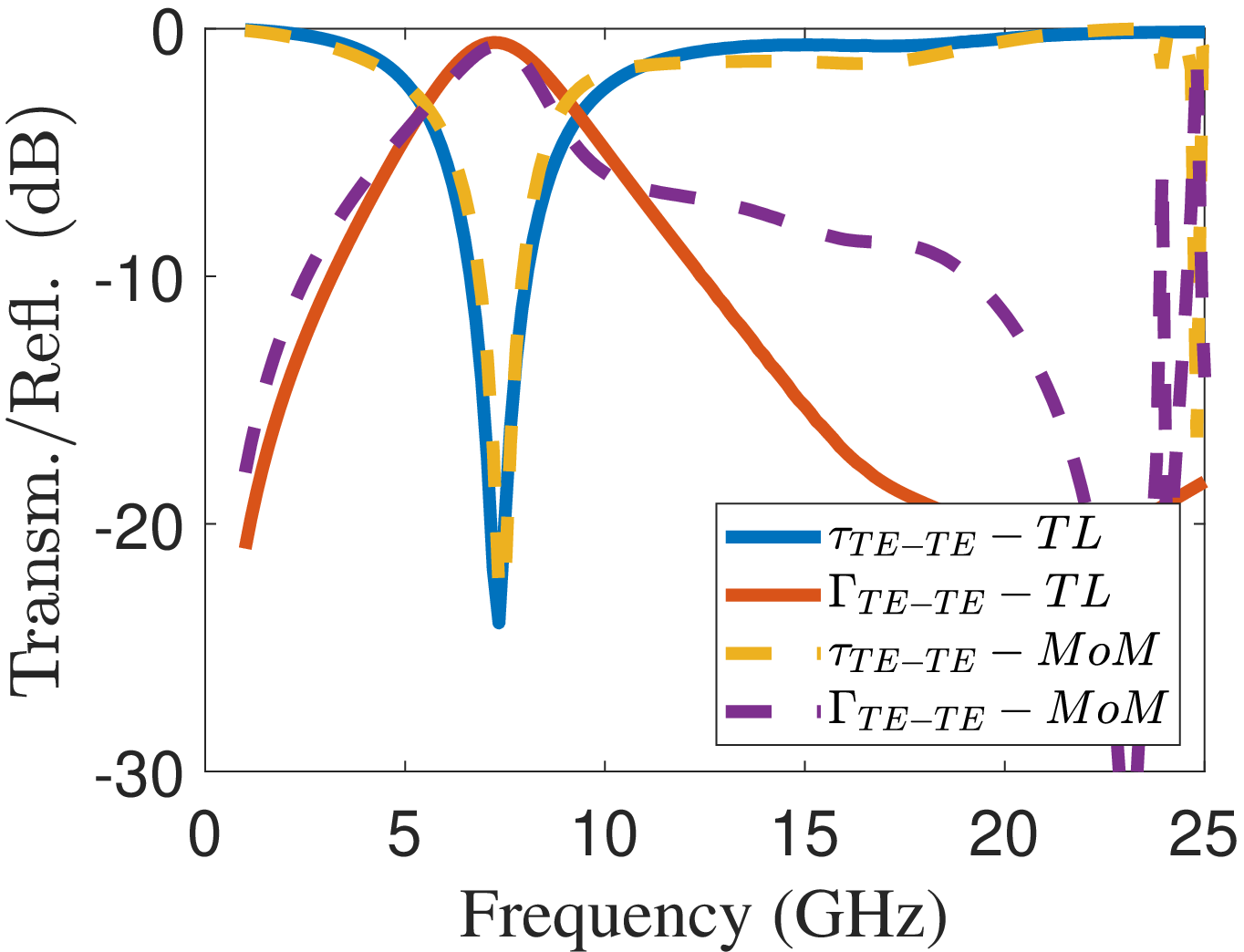}}
   \\
  \subfloat[]{%
        \includegraphics[width=0.45\linewidth]{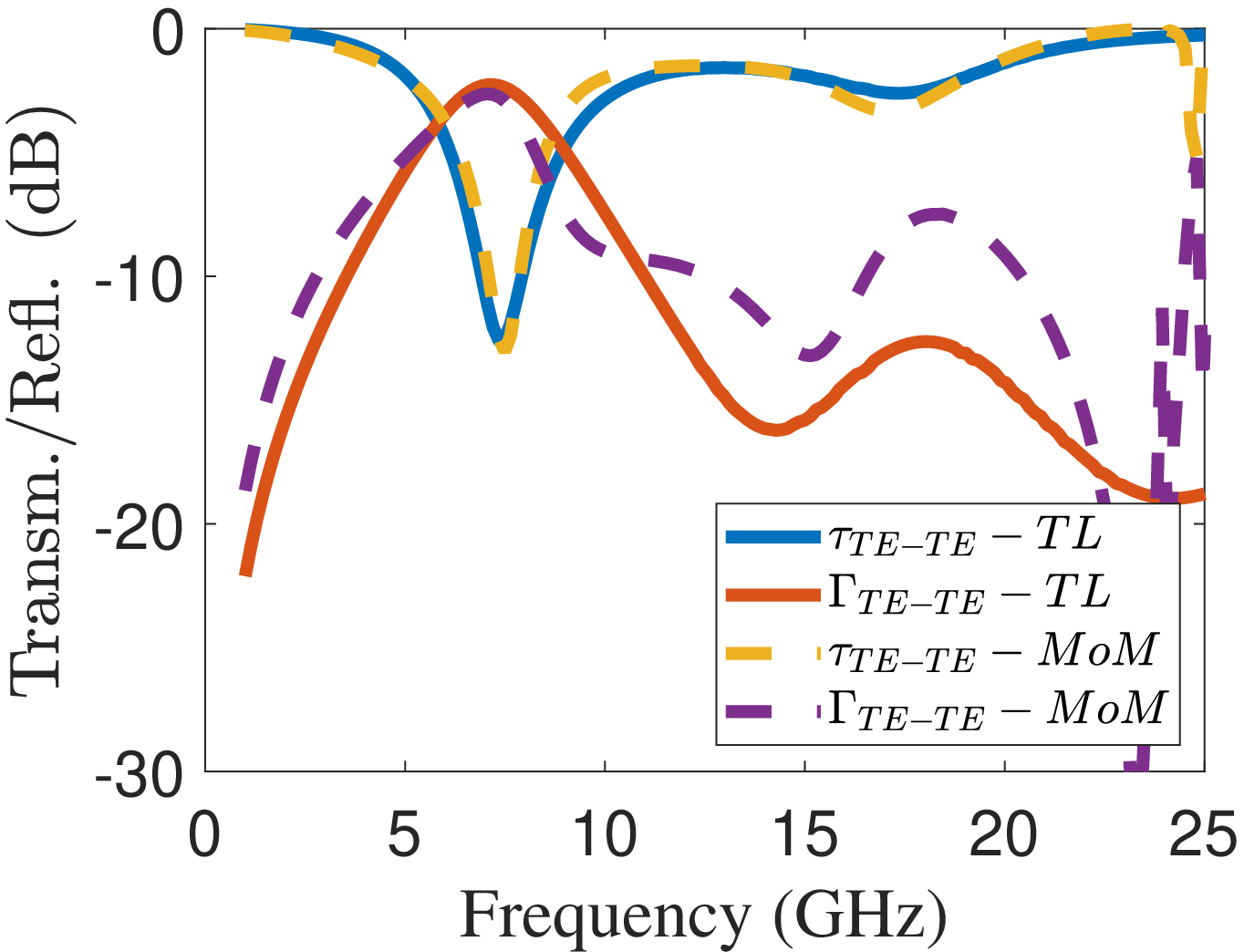}}
    \hfill
  \subfloat[]{%
        \includegraphics[width=0.45\linewidth]{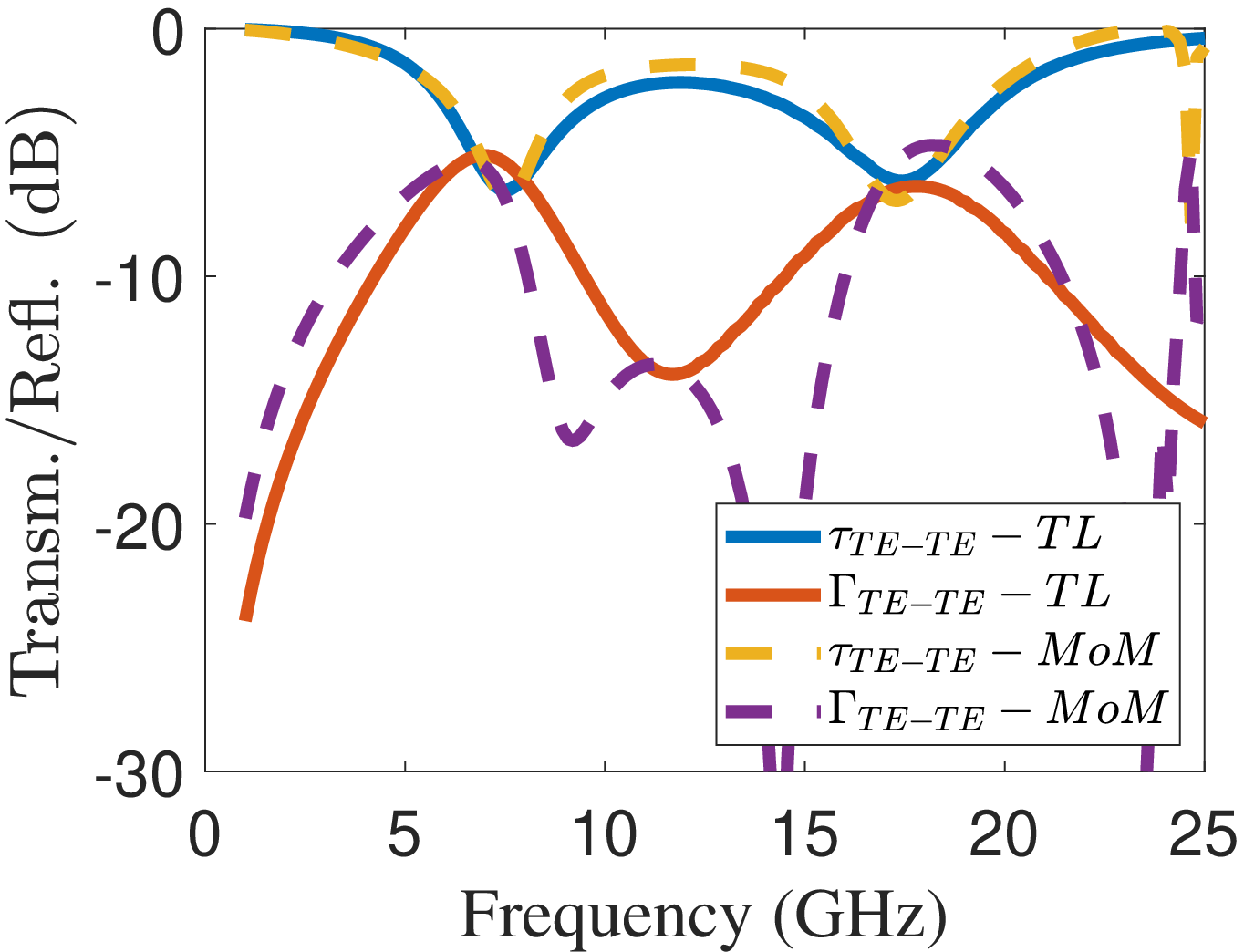}}
      \caption{Reflection and transmission coefficients of a dogbone shaped FSS embedded within two thin dielectric slabs ($t=$ \SI{0.5}{\milli\metre}) with $\varepsilon_r=2$ for different polarization angles within thin substrates: (a) $\varphi^{inc}=0^{\circ}$, (b) $\varphi^{inc}=15^{\circ}$, (c) $\varphi^{inc}=30^{\circ}$, (d) $\varphi^{inc}=45^{\circ}$. The periodicity is equal to \SI{1}{\centi\metre} along planar directions.}
  \label{fig_dogbone_reflection_trasnmission_diel} 
\end{figure}

In order to show the validity of the procedure, the case of a dogbone shaped FSS embedded within two thin dielectric slabs with thickness of \SI{0.5}{\milli\metre} and dielectric permittivity equal to $ 2 $ is shown is Fig.~\ref{fig_dogbone_reflection_trasnmission_diel}. The effective permittivity has been computed according to the transition function presented in eq. (\ref{eq_eps_eff}). The shift of the resonance frequency of the FSS filter with respect to the free standing one is well predicted by the LC model corrected with the effective permittivity. Clearly, is this case, the effect of high order modes is visible at lower frequencies with respect to the freestanding case and the agreement of the LC model with the full-wave MoM simulation is good up to the resonance frequency. 
 
\subsection{Design of Multilayer metasurfaces}

Once the impedance matrix of an FSS element is calculated, the reflection and transmission coefficients of a multilayer structures comprising dielectric layers and generically rotated FSS elements can be computed according to transfer matrix (ABCD) approach \cite{Pfeiffer_PRA}. The analytic approach in computing the response of cascaded metasurfaces is accurate as long as the metasurafaces are at a sufficient distance to avoid the effect of high order Floquet modes \cite{costa2012circuit}. The ABCD matrix of the multilayer structure can be obtained by simply modifying relation (\ref{eq:Mdiel}).

The equivalent circuit approach is typically used for the design of devices based on multilayer metasurfaces both in microwave and optic regimes	\cite{lerner1965wave,abadi2015wideband,pisano2008metal}. Usually, these approaches analyze the TE and TM polarization separately. On the other hand, the circuit model presented in this work, is based on a tensorial analysis of TE and TM modes thus taking account both polarization simultaneously.

\begin{figure} 
    \centering
  \subfloat[]{%
       \includegraphics[width=0.45\linewidth]{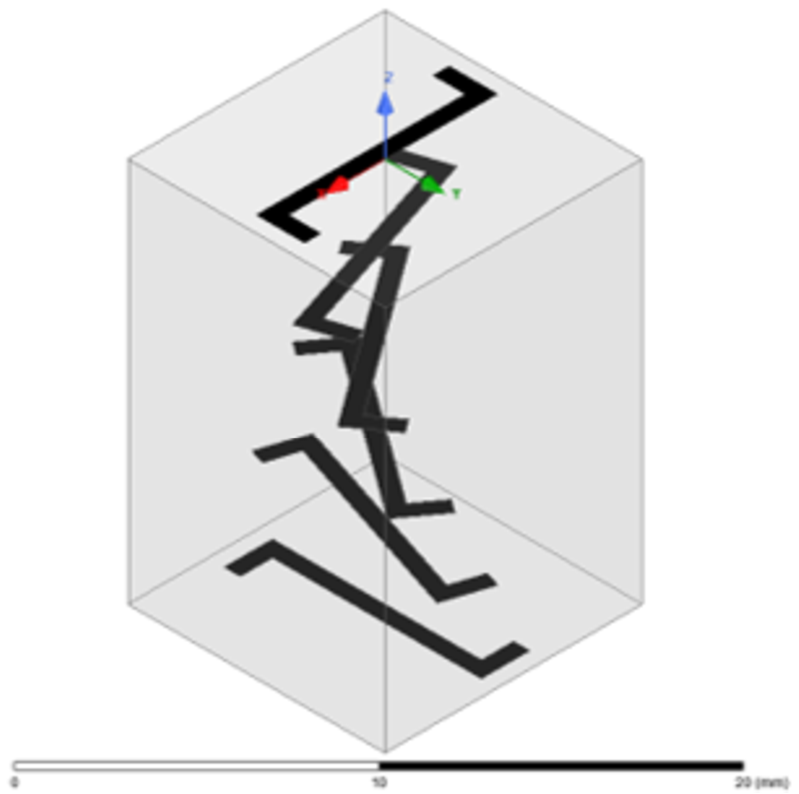}}
    \hfill
  \subfloat[]{%
        \includegraphics[width=0.45\linewidth]{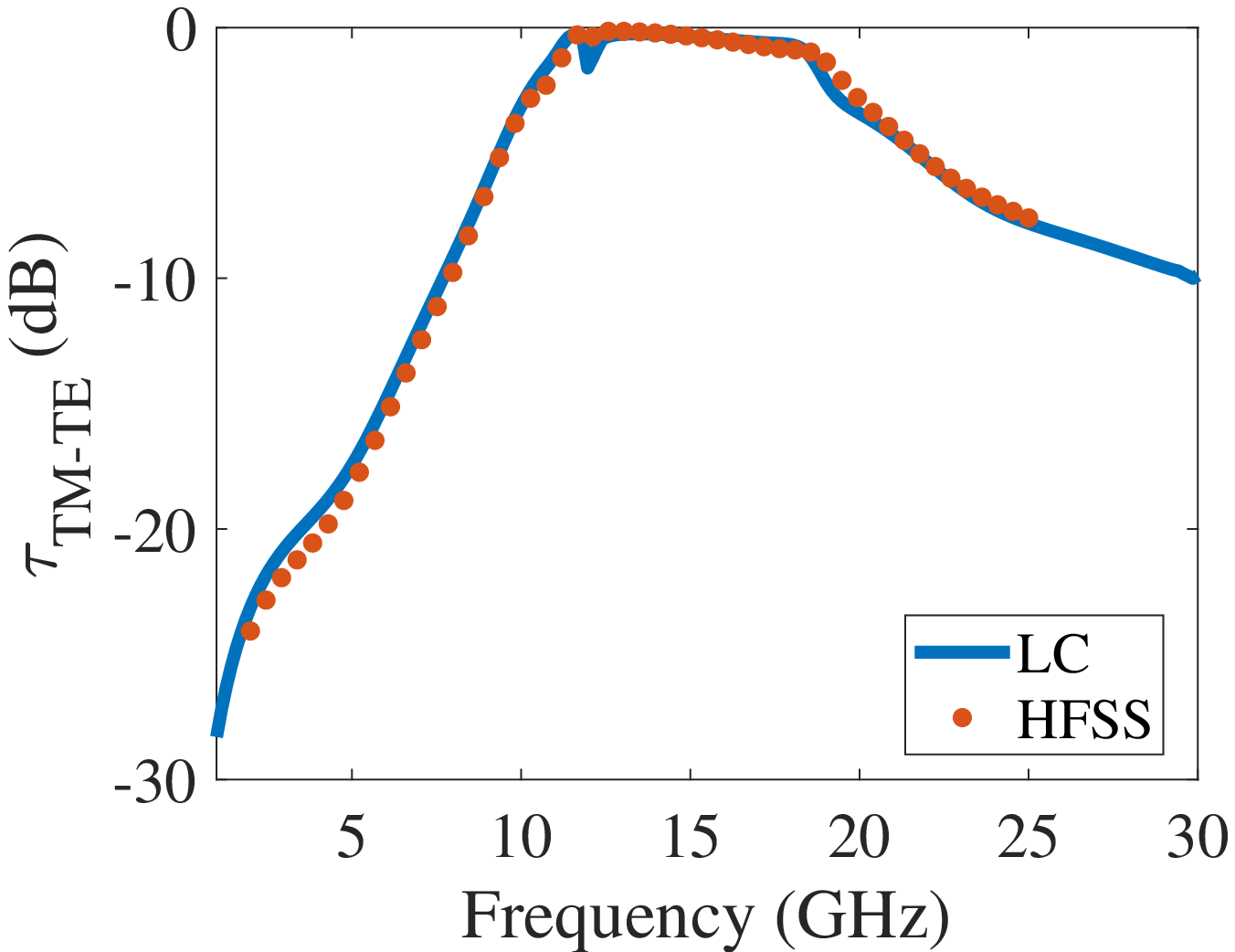}}
      \caption{Linear-to-linear polarization converted designed with the proposed method: (a) 3D view; (b) transmission coefficient $\tau_{{TM-TE}}$ calculated with the proposed method (LC) and with the full-wave simulation performed with HFSS.}
  \label{fig:design_lin} 
\end{figure}

\begin{figure} 
    \centering
  \subfloat[]{%
       \includegraphics[width=0.45\linewidth]{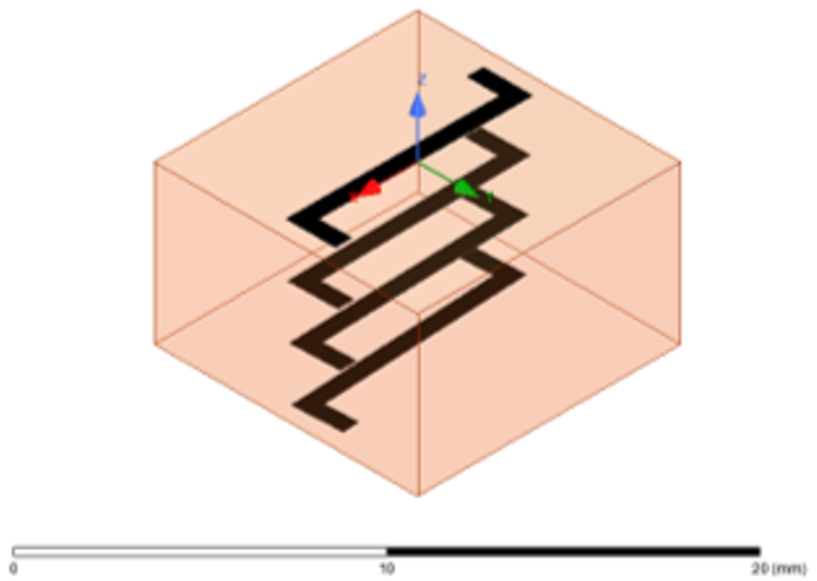}}
    \hfill
  \subfloat[]{%
        \includegraphics[width=0.45\linewidth]{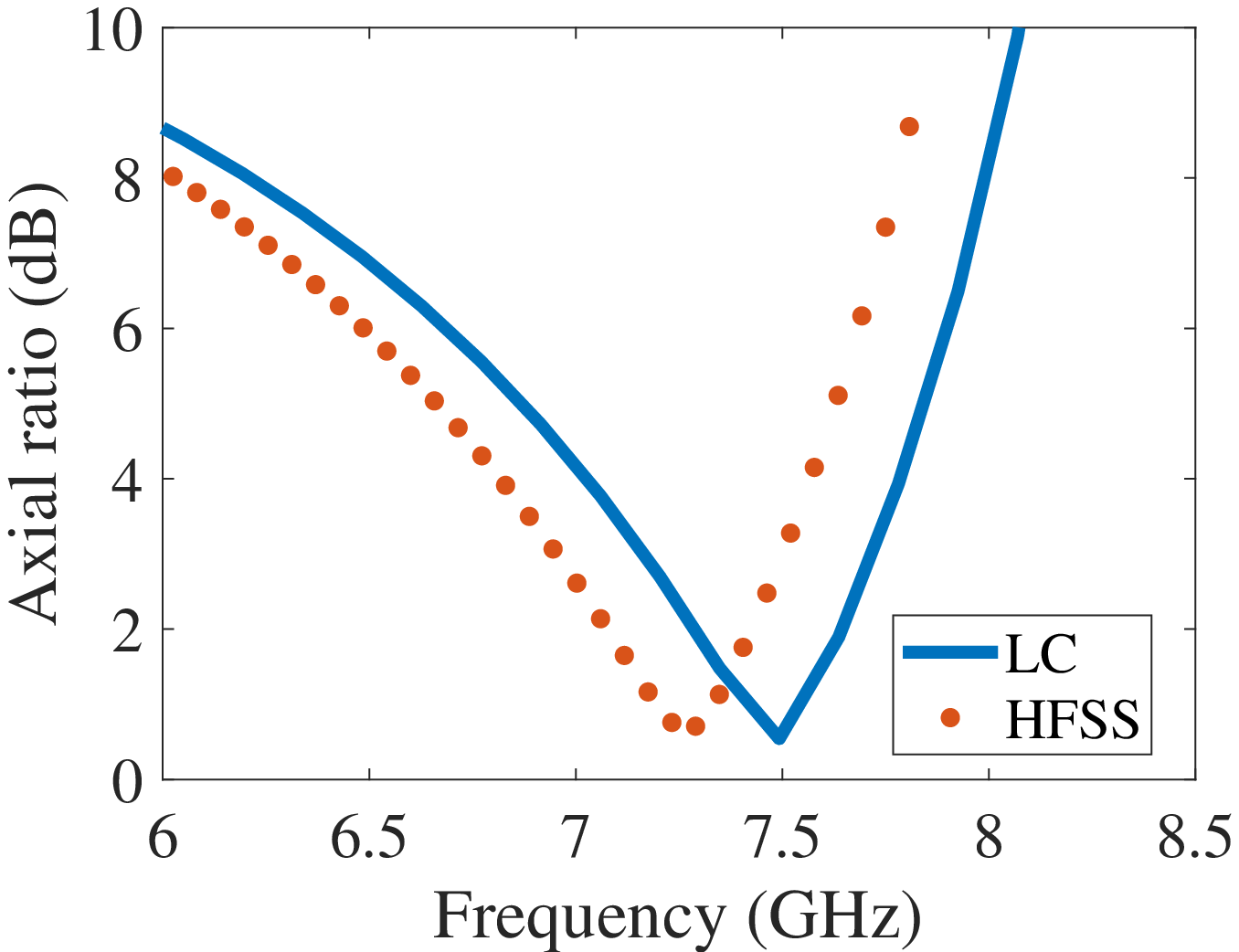}}
   \\
  \subfloat[]{%
        \includegraphics[width=0.45\linewidth]{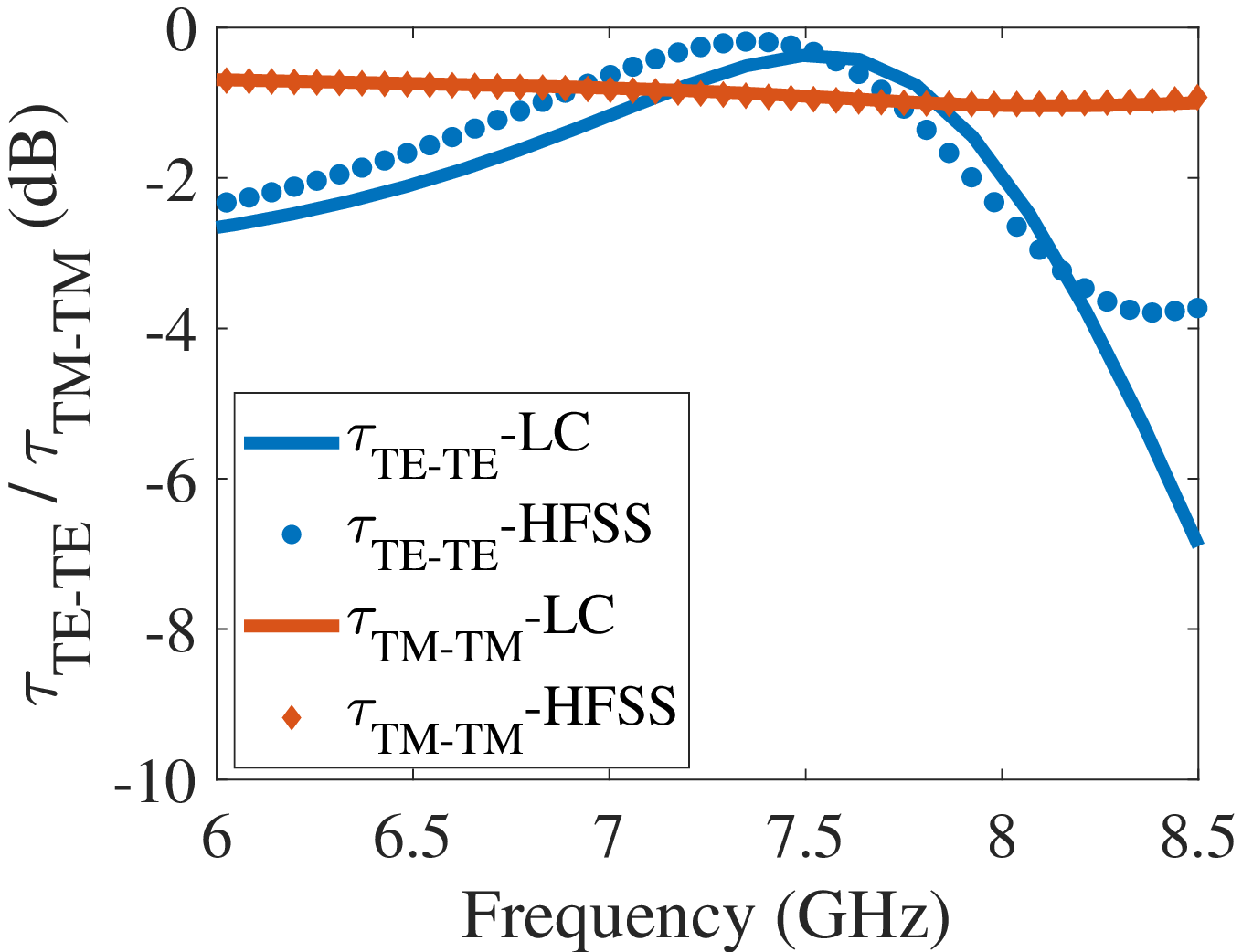}}
    \hfill
  \subfloat[]{%
        \includegraphics[width=0.45\linewidth]{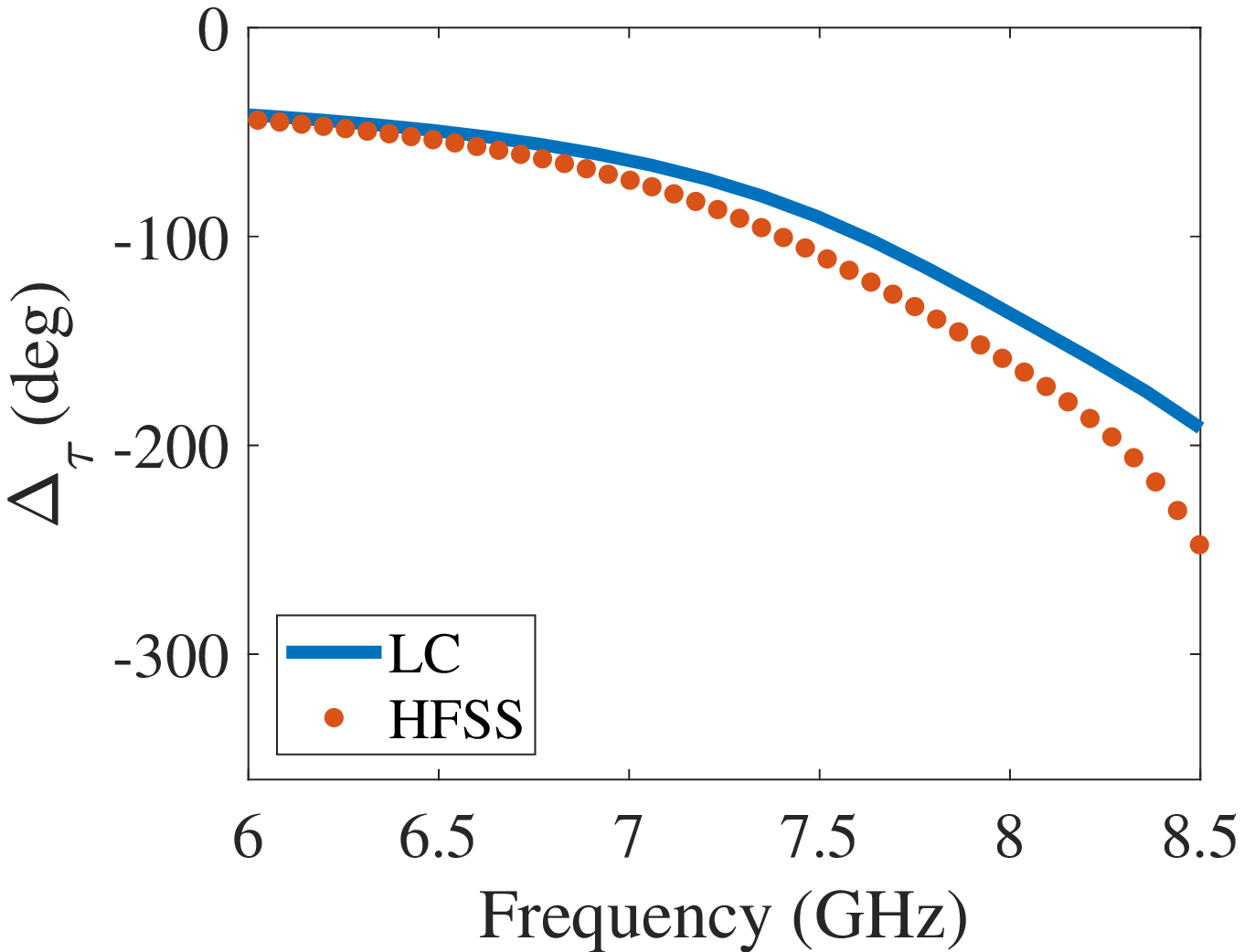}}
      \caption{Linear-to-circular polarization converter based on multilayer impedance surfaces. (a) Layout of the LP-CP converter: the loaded dipoles are gradually rotated of $1^\circ$, (b) Axial Ratio of the transmitted field, (c) Transmission coefficient for TE and TM polarizations, (d) Phase displacement between TE and TM transmitted fields.}
  \label{fig:design_CP} 
\end{figure}

A structure composed of cascaded metasurfaces can be employed to synthesize transmission-type polarization converters. A couple of examples are provided in this section in order to show the potentialities of the proposed circuit model approach in the synthesis of devices based on metasurfaces \cite{costa2019design}. The first example is a wideband linear polarization converter comprising a six layer metasurface where a loaded dipole unit cell is employed on every layer. The dipole element is partially and gradually rotated layer by layer. The spacer thickness and its permittivity, together with the rotation angle of the dipole resonator, is optimized through the proposed circuit approach which allows to test a large number of configurations in a few minutes. In Fig.~\ref{fig:design_lin} the layout of the six layers polarization converter comprising loaded dipole resonators is shown together with the cross-polarized transmission coefficient. As is evident, the structure is able to completely transmit and convert a linear polarized field with TE polarization into a TM polarized field over a considerable frequency band. The rotation angle which is progressively applied layer by layer is $18^\circ $, the spacer thickness between the layers is \SI{2.5}{\milli\metre} and the permittivity is equal to $1$. 
The second example consists of a transmission type linear-to-circular polarization converter. The structure comprises four metasurfaces separated by \SI{2}{\milli\metre} of Teflon ($\varepsilon_r=2.2$) and the unit cell formed by the same loaded dipole of the previous example is partially rotated layer by layer by  $1^\circ$. When the field impinges with $\varphi^{inc}=45^\circ$ with respect to the \textit{x}-axis, the structure is able to transform the transmitted wave into a circular polarized one around 7.5 GHz. To achieve these results, the optimization of the structure is performed both on the amplitude and phase of the TE and TM co-polarized transmission coefficients. Indeed, in order to achieve a good polarization converter, the TE and TM components of the field should be completely transmitted and, at the same time, subject to a relative phase delay of  $90^\circ$. It is worth to remark that, as is the previous example, the tested cell shapes are described in terms of the LC parameters computed in freestanding configuration and their values are corrected according to the procedure described in paragraph \ref{subsec:Dielectric_layers}. In Fig.~\ref{fig:design_CP} the layout of the four layers polarization converter comprising loaded dipole resonators is shown together with the axial ratio of the polarization converter, and the co-polarized transmission coefficient (amplitude and phase) for TE and TM polarizations at $\varphi^{inc}=0^\circ$ . The results obtained with the circuit approach, used for the fast optimization process, exhibit a good agreement with full-wave simulations performed with HFSS.

\bigskip

\section{Conclusion}
A simple approach for deriving the equivalent circuit model of anisotropic FSS has been presented. The methodology relies on the spectral theorem stating that every impedance matrix of a passive system can be diagonalized. After an initial simulation of the anisotropic element, the crystal angles are computed frequency by frequency and a second simulation is carried out on the crystal axis where the FSS impedance matrix is diagonal. The two terms of this diagonal matrix are then fitted by using an LC model. Every FSS cell can be described by using only $ 5 $ parameters ($L_{\chi_1}$, $C_{\chi_1}$, $L_{\chi_2}$, $C_{\chi_2}$ and the rotation angle $\varphi^{rot}$). Once computed the $ 5 $ parameters, the FSS cell can be simulated for a generic azimuth angle or with different unit cell periodicity. Moreover, the effect of the dielectric substrate can be taken into account by scaling the capacitance value for the effective dielectric permittivity of the surrounding medium. Finally, 
the potentialities of the proposed circuit model approach for the synthesis of devices based on cascade metasurfaces is demonstrated. In particular, a linear-to-linear and a linear-to-circular transmission-type polarization converter designed with the proposed circuit model approach have been validated with full-wave simulations performed with HFSS.

\appendices
  
  \section{Calculation of the Rotation Angle}
  \label{Appendix-B}
\begin{figure} 
    \centering
  \subfloat[][${\underline E ^r} = \underline{\underline{\Gamma}} \,{\underline E ^i}$]{%
       \includegraphics[width=.29\linewidth]{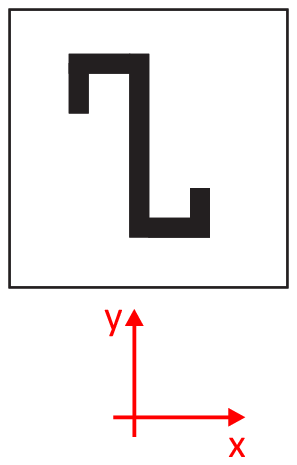}}
\hfill
  \subfloat[][${\underline E ^{r^\prime}} = \underline{\underline{\Gamma}}^\prime\,{\underline E ^{i^{\prime}}}$]{%
        \includegraphics[width=.35\linewidth]{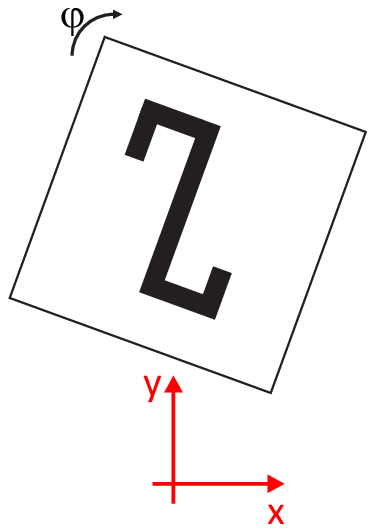}}
  \subfloat[][${\underline E ^{r^{\prime\prime}}} = \underline{\underline {\Gamma}}^{\prime}\,{\underline E ^{i^{\prime\prime}}}$]{%
        \includegraphics[width=.29\linewidth]{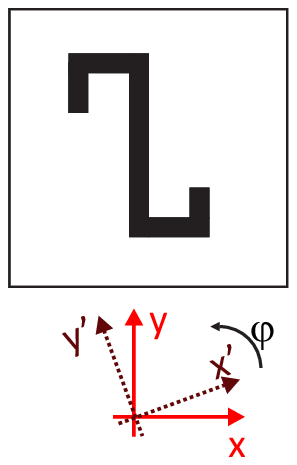}}
  \caption{(a) FSS in the $ (x,y) $ plane; (b) FSS in the $ (x,y) $ rotated by $ \varphi $ in the clockwise direction, (c) counter-clockwise rotation of the impinging electric field by an angle $ \varphi $. }
  \label{fig:Rotation} 
\end{figure}

Let us suppose to have a generic FSS element in the coordinate system $(x,y)$ as depicted in Fig.~\ref{fig:Rotation}~(a). The incident electric fields lays on the $(x,y)$ plane with a generic polarization. The scattered electric field $ \underline E ^{r}  $ is obtained from the scattering matrix $ \underline{\underline{\Gamma}}. $
If the FSS is rotated in the $(x,y)$ plane by an angle $\varphi$ in the clockwise direction (Fig.~\ref{fig:Rotation}~(b)), the reflected field ${\underline E ^{r^{\prime}}}$ is obtained from the reflection matrix $\underline{\underline{\Gamma}}^{\prime}$.
Similarly, it is possible to consider the FSS in the coordinate system $(x,y)$ and to perform a counter-clockwise rotation of the impinging electric field by an angle $\varphi$ as depicted in Fig.~\ref{fig:Rotation}~(c). In this case the reflected field is obtained as follows:
\begin{equation}
{\underline E ^{r^{\prime\prime}}} = \underline{\underline{\Gamma}}^{\prime}\,{\underline E ^{i^{\prime\prime}}} = \underline{\underline{\Gamma}}^{\prime}\,\underline{\underline R}^{T}\,\underline{E}^{i}
\end{equation}

Alternatively, the scattered field can be calculated in the following form:
\begin{equation}
{\underline E ^{r^{\prime\prime}}} = \underline{\underline R}^{T}\,\underline{E}^{r} = \underline{\underline R}^{T}\, \underline{\underline \Gamma}\,\underline{E}^{i}
\end{equation}
Consequently, the scattering matrix $ \underline{\underline{\Gamma}}^{\prime} $ can be calculated from $ \underline{\underline{\Gamma}} $ with the following relation:
\begin{equation}\label{eq:RotGamma} 
\underline{\underline{\Gamma}}^{\prime} = \underline{\underline R}^{T}\, \underline{\underline{\Gamma}}\,\underline{\underline R}
\end{equation}
Being an FSS a passive system, according to the spectral theorem, there is always a particular rotation for which the scattering matrix of the FSS is diagonal ($ \underline{\underline{D}} $). In light of this consideration, using the equation \eqref{eq:RotGamma}, it is possible to write the following relation:
\begin{equation}
{\underline{\underline R} ^{ - 1}}\underline{\underline{\Gamma}} \;\underline{\underline R}  = \underline{\underline D}
\end{equation}
It is convenient to write the rotation matrix $ \underline{\underline{R}} $ as a function of the real parameter $ m $:
\begin{equation}\label{eq:RotMatrixparam}
\underline{\underline R}  = {1 \over {\sqrt {1 + {m^2}} }} \begin{bmatrix}
                   1 & -m \\[3pt]
                   m &  1  \\
                  				\end{bmatrix} \\ 
\end{equation}
It is worth noticing that the parametric form of $ \underline{\underline{R}} $ reported in equation \eqref{eq:RotMatrixparam} exhibits the typical properties of the rotation matrix:
\begin{equation}
\underline{\underline R}^{T}\underline{\underline R} = \underline{\underline{I}} \,\,\,\, \text{and} \,\,\,\, det(\underline{\underline{R}})=1
\end{equation}
Using relations \eqref{eq:RotGamma} and \eqref{eq:RotMatrixparam} the following equation is obtained:
\begin{equation}
{\underline{\underline R} ^{ - 1}}\underline{\underline{\Gamma}} \;\underline{\underline R}  =  {\underline{\underline R} ^T}\underline{\underline{\Gamma}} \;\underline{\underline R} = \underline{\underline{D}}
\end{equation}
Imposing that $ \underline{\underline{D}} $ is diagonal, the following relation is obtained:
\begin{equation}\label{eq:DiagZero}
\Gamma_{xy}-m\,\Gamma_{xx}+m\, \Gamma_{yy}- m^2\, \Gamma_{xy}= 0
\end{equation}
From equation \eqref{eq:DiagZero}, the parameter $ m $ can be calculated:
\begin{equation}
{m} = {{\Gamma _{yy} \pm \Gamma _{xx} + \sqrt {{{\left( {\Gamma _{yy} - \Gamma _{xx}} \right)}^2} + 4{\Gamma _{yx}^2}} } \over {2\Gamma _{yx}}}
\end{equation}

\begin{equation}
\varphi^{rot} = \arcsin\left( m \over{\sqrt{1+m^2}} \right) 
\end{equation}

  \section{anisotropic FSS Impedance}
  \label{Appendix-A}
The impedance matrix $ {{\underline{\underline Z} }}  $ is calculated from the reflection coefficient $ \underline{\underline{\Gamma}} $ computed trough a full-wave simulation. The reflection coefficient can be written as a function of the input impedance ${{\underline{\underline Z} }^V}$ and of the free space impedance $\zeta_0$: 

\begin{equation} \label{eq:6}
\underline{\underline \Gamma }  = {{{{\underline{\underline Z} }^V} - {\zeta _0}} \over {{{\underline{\underline Z} }^V} + {\zeta _0}}}
\end{equation}

where ${{\underline{\underline Z} }^V}=[{{\underline{\underline Y} }^V}]^{-1}$ is the parallel connection of the FSS impedance $ ({{\underline{\underline Z} }})  $ and the free space impedance $\zeta _0$  $ ({{\underline{\underline Y} }^V}~=~{{\underline{\underline Y} }}~+~\underline{\underline {{Y_0}}})$, according to the transmission line model shown in Fig.~\ref{fig_eq_circuit}(b), 
with $ Y_0 = diag \left( {1 \over{\zeta_0}}\right) $.

Considering a Cartesian reference system, the matrix $ \underline{\underline{\Gamma}} $ can be written as:
\begin{equation} \label{eq:GammaInv}
\scalebox{0.95}[1]{$\underline{\underline \Gamma }  = {\begin{bmatrix}
                    Z_{xx}^V + \zeta _0 & Z_{xy}^V \\[3pt]
                    Z_{yx}^V & Z_{yy}^V + \zeta _0  \\

                  				\end{bmatrix}}^{-1} \begin{bmatrix}
                    Z_{xx}^V - \zeta _0 & Z_{xy}^V \\[3pt]
                    Z_{yx}^V & Z_{yy}^V - \zeta _0  \\

                  				\end{bmatrix} \\$}                				
\end{equation}

Indicating with $\Delta$ the determinant of the matrix in \eqref{eq:GammaInv} that needs to be inverted:
\begin{equation}
\Delta  = \left( {Z_{xx}^V + {\zeta _0}} \right)\left( {Z_{yy}^V + {\zeta _0}} \right) - Z_{xy}^VZ_{yx}^V
\end{equation}

With some mathematical manipulations, each term of the reflection coefficient can be written in the following form:

\begin{equation}
\left\{ \,
\begin{IEEEeqnarraybox}[
\IEEEeqnarraystrutmode
\IEEEeqnarraystrutsizeadd{7pt}
{7pt}][c]{rCl}
\Gamma _{xx} & = & {{\left( {Z_{yy}^V + {\zeta _0}} \right)\left( {Z_{xx}^V - {\zeta _0}} \right) - Z_{xy}^VZ_{yx}^V} \over \Delta }
\\
\Gamma _{xy} & = & {{2Z_{yx}^V{\zeta _0}} \over \Delta }
\\
\Gamma _{yx} & = & {{2Z_{xy}^V{\zeta _0}} \over \Delta }
\\
\Gamma _{yy} & = & {{{ - Z_{xy}^V} Z_{yx}^V + \left( {Z_{xx}^V + {\zeta _0}} \right)\left( {Z_{yy}^V - {\zeta _0}} \right)} \over \Delta }
\end{IEEEeqnarraybox}
\right.
\label{eq:Gammasystem}
\end{equation}

In order to obtain the impedance matrix $\underline{\underline{Z}}^{V} $,  the system reported in \eqref{eq:Gammasystem} needs to be solved. The four terms of the input impedance matrix read:
\begin{IEEEeqnarray}{rCl}
\label{eq:block1}
Z_{xx}^V  & = & - {{{\zeta_0}\left( {{\Gamma_{xx}}\left( {{\Gamma_{yy}} - 1} \right) - {\Gamma_{xy}}{\Gamma_{yx}} + {\Gamma_{yy}} - 1} \right)} \over {{\Gamma_{xx}}\left( {{\Gamma_{yy}} - 1} \right) - {\Gamma_{xy}}{\Gamma_{yx}} - {\Gamma_{yy}} + 1}}
\\[5pt] \label{eq:block2}
Z_{xy}^V  & = & {{2{\zeta _0}{\Gamma_{yx}}} \over {{\Gamma_{xx}}\left( {{\Gamma_{yy}} - 1} \right) - {\Gamma_{xy}}{\Gamma_{yx}} - {\Gamma_{yy}} + 1}}\\[5pt]\label{eq:block3}
Z_{yx}^V  & = & {{2{\zeta _0}{\Gamma_{xy}}} \over {{\Gamma_{xx}}\left( {{\Gamma_{yy}} - 1} \right) - {\Gamma_{xy}}{\Gamma_{yx}} - {\Gamma_{yy}} + 1}}\\[5pt]\label{eq:block4}
Z_{yy}^V  & = & - {{{\zeta_0}\left( {{\Gamma_{xx}}\left( {{\Gamma_{yy}} + 1} \right) - {\Gamma_{xy}}{\Gamma_{yx}} - {\Gamma_{yy}} - 1} \right)} \over {{\Gamma_{xx}}\left( {{\Gamma_{yy}} - 1} \right) - {\Gamma_{xy}}{\Gamma_{yx}} - {\Gamma_{yy}} + 1}}
\end{IEEEeqnarray}

The solutions reported in equations (\eqref{eq:block1}-\eqref{eq:block4}) are valid if the following condition is met:
\begin{equation} \label{eq:DispersionEq} 
Z_{xx}^V (Z_{yy}^V + \zeta_0) - Z_{xy}^V Z_{yx}^V + \zeta_0 (Z_{yy}^V + \zeta_0) \neq 0 
\end{equation}

The relation \eqref{eq:DispersionEq} is the dispersion equation as a function of the input impedance $\underline{\underline{Z}}^{V}$. The input admittance matrix $\underline{\underline{Y}}^{V}$ can be calculated as the inverse of the input impedance matrix impedance $\underline{\underline{Z}}^{V}$.  
At this stage, the admittance of the FSS can be extracted as follows:
\begin{equation}
 \underline{\underline{Y}}={\underline{\underline Y} ^V} - \underline{\underline {{Y_0}}} = \begin{bmatrix}
                    Y_{xx}^{V}- {1 \over {\zeta_0}} & Y_{xy}^{V} \\[5pt]
                    Y_{yx}^{V} &  Y_{yy}^{V}- {1 \over {\zeta_0}}  \\

                  				\end{bmatrix} \\ 
\end{equation}

Finally, the FSS impedance is computed by inverting the admittance matrix $\underline{\underline{Y}}$.

\bibliographystyle{IEEEtran}
\bibliography{references}

\end{document}